\pdfoutput=1

\documentclass[%
 reprint,
 amsmath,amssymb,
 aps,
]{revtex4-2}

\usepackage[pdftex]{graphicx}
\graphicspath{{./Figure/}}
\usepackage{braket}
\usepackage{dcolumn}
\usepackage{bm}

\makeatletter
\let\MYcaption\@makecaption
\makeatother

\usepackage{subcaption}
\makeatletter
\let\@makecaption\MYcaption
\makeatother

\usepackage{booktabs}
\usepackage{threeparttable}

\begin{document}

\preprint{APS/123-QED}

\title{Densest binary sphere packings and phase diagram : revisited}

\author{Ryotaro Koshoji}
\email{cosaji@issp.u-tokyo.ac.jp}
\affiliation{Institute for Solid State Physics, The University of Tokyo, Kashiwa 277-8581, Japan}
\author{Mitsuaki Kawamura}%
\email{mkawamura@issp.u-tokyo.ac.jp}
\affiliation{Institute for Solid State Physics, The University of Tokyo, Kashiwa 277-8581, Japan}%
\author{Masahiro Fukuda}%
 \email{masahiro.fukuda@issp.u-tokyo.ac.jp}
\affiliation{Institute for Solid State Physics, The University of Tokyo, Kashiwa 277-8581, Japan}
\author{Taisuke Ozaki}%
 \email{t-ozaki@issp.u-tokyo.ac.jp}
\affiliation{Institute for Solid State Physics, The University of Tokyo, Kashiwa 277-8581, Japan}

\date{\today}

\date{\today}

\begin{abstract}

We revisit the densest binary sphere packings (DBSP) under the periodic boundary conditions and present an updated phase diagram, including newly found 12 putative densest structures over the $x - \alpha$ plane, where $x$ is the relative concentration and $\alpha$ is the radius ratio of the small and large spheres. To efficiently explore the DBSP, we develop an unbiased random search approach based on both the piling up method to generate initial structures in an unbiased way and the iterative balance method to optimize the volume of a unit cell while keeping the overlap of hard spheres minimized. With those two methods, we have discovered 12 putative DBSP and thereby the phase diagram is updated, while our results are consistent with those of the previous study [Hopkins et al., Phys. Rev. E 85, 021130 (2012)] with a small correction for the case of 12 or fewer spheres in the unit cell. The 5 of the new 12 densest packings are discovered in the small radius range of $0.42 \le \alpha \le 0.50$ where several structures are competitive to each other with respect to packing fraction. Through the exhaustive search, diverse dense packings are discovered and accordingly we find that packing structures achieve high packing fractions by introducing distortion and/or combining a few local dense structural units. Furthermore, we investigate the correspondence of the DBSP with crystals based on the space group. The result shows that many structural units in real crystals, e.g., $\mathrm{LaH_{10}}$ and $\mathrm{SrGe_{2-\delta}}$ being high-pressure phases, can be understood as DBSP. The correspondence implies that the densest sphere packings can be used effectively as structural prototypes for searching complex crystal structures, especially for high-pressure phases.

\end{abstract}

\maketitle

\section{INTRODUCTION}

The densest sphere packings can be used as structural models for many physical systems, e.g., crystals, colloids~\cite{doi:10.1021/acsnano.9b04274, doi:10.1021/acs.chemrev.6b00196}, and glasses~\cite{doi:10.1063/1.4769981}. Atoms in crystals are sometimes approximated as spheres: In ion-bonded materials such as NaCl~\cite{villars}, atoms are often spherically symmetrical because atoms have closed-shell structures due to the charge transfer. In intermetallic compounds such as AgCu~\cite{10.1007/BF02652162}, atoms are also sometimes spherically symmetrical due to the bonds formed by electrons populating in \textit{s}-orbitals. In materials under high pressure, distances between atoms become so close that the directional orientation of the bond is weakened due to the strong repulsive force by Pauli's exclusion principle~\cite{doi:10.1021/jacs.9b02634}. Accordingly, many structural units in crystals can be understood as sphere packings~\cite{Derivativestructuresbasedonthespherepacking}. The correspondences indicate that the densest sphere packings may be used effectively as structural prototypes for searching complex crystal structures, especially for high-pressure phases.

Identifying the densest sphere packings is one of the most difficult mathematical problems. It was proved only in the 2000s that the Barlow packing is the densest packing of monodisperse spheres in $\mathbb{R}^3$~\cite{10.2307/20159940}. There seems no general way to determine the densest binary sphere packings (DBSP) in a mathematically rigorous way despite considerable efforts under limited conditions~\cite{PhysRevE.76.041304, PhysRevE.78.011303, PhysRevE.87.032202}. However, there have been several studies that attempt to estimate the DBSP by numerical calculations~\cite{doi:10.1021/jp804953r, PhysRevE.79.046714, doi:10.1021/jp206115p, doi:10.1021/jp1045639, Hudson_2011, doi:10.1063/1.5052478, de_laat_de_oliveira_filho_vallentin_2014}. Recently, Hopkins and coworkers explored the DBSP under the restriction that the number of spheres in the unit cell is less than or equal to 12. They used the original method to generate initial structures~\cite{torquato2013random} and the structural optimization algorithms~\cite{PhysRevE.82.061302}. Accordingly, they constructed the phase diagram for DBSP~\cite{PhysRevLett.107.125501, PhysRevE.85.021130}; hereafter we call the diagram as the HST phase diagram. The 18 distinct putative DBSP were identified on the HST phase diagram. The phase diagram for the densest ternary sphere packings has not yet been constructed.
 
An exhaustive search for packing structures becomes more difficult with an increase in the number of spheres. The considerable increase in the number of local minima is inferred from the explosive increase in the number of the permutation of lining up spheres in a row. The number of cases with 6 large and 6 small spheres is 924, whereas the number of cases with 12 large and 12 small spheres is 2,704,156. In general, the number of cases $R$ with $N$ large and $N$ small spheres can be calculated as
\begin{equation}
R = \frac{\left(2N \right)!}{N! N!}.
\end{equation}
With Stirling's formula, $R$ can be estimated as
\begin{equation}
R \sim \frac{4^{N}}{\sqrt{\pi N}}. \label{eq:lining-up}
\end{equation}
Equation (\ref{eq:lining-up}) indicates that the number of local minima in three-dimensional structures is getting larger exponentially with an increase in the number of spheres.

Prediction of crystal structures faces the same difficulties. Many algorithms to explore effectively in coordination spaces have been devised for crystal structure prediction, e.g., the evolutionary algorithm~\cite{GLASS2006713, doi:10.1063/1.2210932, Oganov_2008, LYAKHOV20101623, LYAKHOV20131172} and the particle-swarm optimization method~\cite{PhysRevB.82.094116, WANG20122063, Wang_2015, WANG2016406}. It is also known that symmetry constraints enhance the efficiency of finding the most stable structure~\cite{LYAKHOV20131172, PhysRevB.82.094116, WANG20122063, WANG2016406}. Those methods have successfully predicted many materials, followed by experimental confirmations~\cite{GLASS2006713, doi:10.1063/1.2210932, Oganov_2008, LYAKHOV20101623, LYAKHOV20131172, PhysRevB.82.094116, WANG20122063, Wang_2015, WANG2016406, PhysRevLett.119.107001}.

The random structure searching method~\cite{Pickard_2011} is also a powerful method for structure prediction. We consider that the densest sphere packings should be searched by the random structure searching method because we have no a priori knowledge about what kind of structures are the densest multinary sphere packings. For example, the densest structures may not be highly symmetric; therefore those structures may not be found under symmetry constraints. On the other hand, the random structure searching method has a drawback that it is impossible to explore exhaustively in coordination space unless structural optimization is efficient. However, in the case of hard spheres, structural optimization is expected to be efficient because repulsive forces occur only when two spheres overlap each other. Therefore, the random structure searching method is to be regarded as an effective way to search for the densest packings.

In the present research, we revisit the densest binary sphere packings under the periodic boundary conditions. To efficiently explore the densest sphere packings, firstly we invent the \textit{piling up method} to generate initial structures in an unbiased way; secondly, we develop the \textit{iterative balance method} to optimize the volume of a unit cell while keeping the overlap of hard spheres minimized. The piling up method is developed based on the idea of stacking spheres randomly one by one on top of a randomly generated first layer. It enables us to search the densest packings unbiasedly from the vast coordination space. The iterative balance method is developed based on the idea of repeating collision and repulsion among spheres under pressure while the maximum displacement in position vectors and lattice vectors is gradually decreased. The method not only generates a dense, periodic packing of nonoverlapping spheres but also predicts the maximum packing fraction with high accuracy. Those two methods are implemented in our open-source program package \textit{\textbf{SAMLAI}} (Structure search Alchemy for MateriaL Artificial Invention). With the \textbf{\textit{SAMLAI}}, we exhaustively search the DBSP with extending the unit cell compared to the previous study \cite{PhysRevE.85.021130} and as a result, we update the phase diagram, including newly found 12 putative densest structures over the $x - \alpha$ plane, where $x$ is the relative concentration and $\alpha$ is the radius ratio of the small and large spheres. $x$ is defined as
\begin{equation}
x \equiv \frac{s}{S + s},
\end{equation}
where $S$ ($s$) is the number of large (small) spheres in the unit cell. For the case of 12 or fewer spheres in the unit cell, our phase diagram is consistent with that of the previous study~\cite{PhysRevE.85.021130} with a small correction. Through the exhaustive search, diverse mono-phase densest binary sphere packings are discovered; accordingly, we find that high packing fractions are achieved by introducing distortion and/or combining a few local dense structural units. Furthermore, we investigate the correspondence of the DBSP with crystals based on the space group. The result shows that many structural units in real crystals, e.g., $\mathrm{LaH_{10}}$~\cite{doi:10.1002/anie.201709970} and $\mathrm{SrGe_{2-\delta}}$ \cite{doi:10.1021/acs.inorgchem.7b01446} synthesized under high pressure, can be understood as DBSP. The correspondence implies that the densest sphere packings can be used effectively as structural prototypes for searching complex crystal structures, especially for high-pressure phases.

The paper is organized as follows: Sec.~\ref{sec:our_implementation} describes our method to explore densest sphere packings; Sec.~\ref{sec:numeric_aspect_of_our_method} discusses the numerical aspects of our method, e.g., the accuracy of the packing fraction, the distribution of generated packing fractions; Sec.~\ref{sec:method_for_phase_diagram} details the methodology for constructing the phase diagram for densest binary sphere packings; Sec.~\ref{sec:result} presents the phase diagram for DBSP and the 12 newly discovered putative densest binary sphere packings; Sec.~\ref{sec:discussion} discusses the effectiveness of our method and the geometry of densest binary sphere packings. In the Sec.~\ref{sec:conclusions}, we summarize this study.

\section{our implementation}
\label{sec:our_implementation}

To efficiently explore the densest packings, we develop an unbiased random search method. The method is implemented in our open-source program package \textit{\textbf{SAMLAI}} (Structure search Alchemy for MateriaL Artificial Invention). It consists of two steps: (A) random generation of multi-layered structures; (B) structural optimization. Step A is aimed at generating initial structures; step B is aimed at optimizing the initial structures for high packing fractions.

\subsection{Generation of initial structures : \textit{piling up method}}

As discussed by Pickard and coworkers~\cite{Pickard_2011}, randomly generated structures may contain spheres that are very close together, or spheres may be crowded in a direction of a short lattice vector instead of a long lattice vector. To avoid generating such an abnormal initial structure, we develop the \textit{piling up method} to generate appropriate initial structures randomly.  The method is based on an idea of stacking spheres randomly one by one on top of a randomly generated first layer, since any periodic structure can be understood as a multi-layered structure if it is extended in a direction perpendicular to a chosen base plane.

The piling up method consists of two steps. At first, an initial structure is generated randomly. Next, a multi-layered structure is constructed by expanding initial structures.

\subsubsection{Generation of seed structure}

To generate an initial structure, firstly the largest sphere is placed at $\left(0,0,0 \right)$. Hereafter, the position of sphere $\left(q_{i1}, q_{i2}, q_{i3} \right)$ is represented in the fractional coordinates and the maximum radius of spheres is set to 1. All values we discuss in the paper are dimensionless. The initial lattice vectors are set to
\begin{align}
\bm{a}_1 &= \left(a_{11}, 0, 0 \right), \\
\bm{a}_2 &= \left(a_{21}, a_{22}, 0 \right), \\
\bm{a}_3 &= \left(a_{31}, a_{32}, a_{33} \right),
\end{align}
where we set $a_{11} = 2$ and $a_{22} = 2$; the $a_{21}$ is set to a random value in the range of $-1 \le a_{21} < 1$, in order to allow $\bm{a}_2$ to have freedom of angle of $60^{\circ}$ to $120 ^{\circ}$ relative to $\bm{a}_1$. Next, zero or more spheres are randomly selected to place on the first layer spanned by $\bm{a}_1$ and $\bm{a}_2$. The selected spheres are placed at $\left(r, r^{\prime}, 0 \right)$, where $r$ and $r^{\prime}$ are random values in the range of $0 \le r,r^{\prime} < 1$. These operations correspond to the random generation of the first layer. The initial lattice vectors have a very strong restriction as $a_{11} = a_{22}$; furthermore, there can be a large overlap between spheres placed on the first layer. However, as discussed in the next subsection, these structural features can be relaxed simultaneously by expanding the unit cell with the steepest descent method. At this stage, we do not expand the cell, but continue to generate the initial structure.

Next, the unselected spheres are stacked one by one. As the first step in the second stage, $\bm{a}_3$ is set to
\begin{equation}
\bm{a}_3 = r \bm{a}_1 + r^{\prime} \bm{a}_2 + \left(0,0,a_{33} \right),
\end{equation}
where $r$ and $r^{\prime}$ are random values in the range of $-0.5 \le r,r^{\prime} < 0.5$; $a_{33}$ is set to
\begin{equation}
a_{33} = \sum_{I} c_I + 1,
\end{equation}
where $c_i$ is defined as the radius of the sphere $i$ and $I$ is the labels of unselected spheres. Finally, all unselected spheres are picked up one by one and they are placed at $\left(r, r^{\prime}, q_{i3} \right)$ where $r$ and $r^{\prime}$ are random values within $0 \le r,r^{\prime} < 1$ and the $q_{i3}$ is set so that the $z$ coordinate will be increased by the radius of the sphere to be placed every time a sphere is placed.

\subsubsection{Expansion}

The generated unit cell seems to be very biased, and generally there can be a large overlap between spheres. If we simply scale the unit cell so that the largest overlap can become zero, the cell may expand explosively and the initial constraint of $a_{11} = a_{22}$ would nearly hold. However, if the initial structure is expanded with the steepest descent method, the lattice vectors are adjusted to the optimal length and angle depending on the sphere arrangement; the optimized lattice vectors have a large variation in length and angle. In fact, if many spheres are piled up in a particular direction, the lattice vectors are greatly expanded in that direction.

In order to apply the steepest decent method for the expansion, we introduce a two body interaction potential between spheres $U \left(\left|\bm{r}_{j} + \bm{T} - \bm{r}_{i} \right| \right)$, where $\bm{r}_{i}$ is the position of sphere $i$ and $\bm{T}$ is the translational lattice vector. We also define the total energy $E$ per unit cell as
\begin{equation}
E = \frac{1}{2} \sum _{\bm{T}} \sum _{i = 1}^{N} \sum _{j = 1}^{N} U \left(\left|\bm{r} _{j} + \bm{T} - \bm{r}_{i} \right| \right), \label{eq:energy}
\end{equation}
where $N$ is the number of spheres in the unit cell. The two body interaction potential $U \left(\left|\bm{r}_{j} + \bm{T} - \bm{r}_{i} \right| \right)$ is defined to be
\begin{equation}
U \left(\left|\bm{r} _{j} + \bm{T} - \bm{r}_{i} \right| \right) \equiv \begin{cases}
-z _{i j} ^{\left(\bm{T}\right)} & \text{$z _{i j} ^{\left(\bm{T}\right)} \le 0$} \\
0 & \text{$0 < z _{i j} ^{\left(\bm{T}\right)}$} \end{cases} \label{eq:potential}
\end{equation}
with
\begin{equation}
z _{i j} ^{\left(\bm{T}\right)} \equiv \left|\bm{r} _{j} + \bm{T} - \bm{r}_{i} \right| - \left(c_i + c_j \right),
\end{equation}
where $c_i$ is the radius of sphere $i$. The characteristic features of the potential are two folds: one is that the potential is nonzero only when two spheres overlap with each other, and the other is that the repulsive force is constant when overlapping. If the potential $U \left(\left|\bm{r}_{j} + \bm{T} - \bm{r}_{i} \right| \right)$ is smoothly connected with $z _{i j} ^{\left(\bm{T}\right)} = 0$, it becomes a pseudo hard-sphere potential. The smoothness causes an undesirable overlap between spheres when a finite pressure is applied to increase the packing fraction as discussed later on.

To expand the unit cell, the extended coordinates defined with
\begin{equation}
\bm{u} \equiv \left(q _{11}, q _{12}, q _{13} \cdots q _{N3}, a _{11}, a _{12}, a _{13}, \cdots a _{33} \right)
\end{equation}
is updated by the steepest descent method as
\begin{equation}
\Delta \bm{u} = - k_1 \, \frac{\partial E}{\partial \bm{u}},
\end{equation}
where $k_1$ is the steepest descent prefactor. The steepest descent method is repeated until the size of the gradient is less than a threshold value. As a result, a multi-layered structure is generated. We calculate the derivatives analytically in Appendix \ref{sec:force-and-stress}.

If the size of the gradient is more than a threshold value, $\Delta \bm{u}$ is scaled so that the maximum displacement in the position vectors and lattice vectors will be equal to the predetermined value. The prefactor $k_1$ is not set directly. The expansion step is aimed at not optimizing structures, but producing diverse multi-layered structures; lattice vectors have large variations in the length and angle. Therefore, it makes sense to scale $\Delta \bm{u}$ to an optimal size that is neither too large nor too small. The value of the maximum displacement is presented in Sec.~\ref{sec:structural-optimization-parameters}.

The piling up method is able to create diverse multi-layered structures with appropriate lattice vectors depending on the sphere arrangement. With the iterative balance method discussed in Sec.~\ref{sec:local_optimization}, the generated structures can be optimized to diverse packing structures: e.g., towerlike structures and symmetric structures such as the fcc structure. The piling up method is very effective for searching densest sphere packings as discussed in Sec.~\ref{sec:Distribution_and_update_history_of_packing_fractions}.

\subsection{Global optimization}
\label{sec:global_optimization}

The generated multi-layered structure contains a large gap, which makes the packing fraction reduced. The large gap can be filled by repeating collision and repulsion between spheres under pressure. The steepest descent method with the hard-sphere potential defined as Eq.~(\ref{eq:potential}) can cause a repetition of collision and repulsion. If the constant repulsive force is large enough, the pressure and the repulsive force cannot be balanced. The repulsion is dominant when overlapping, leading to a structure without any overlap as a result of the optimization. Once the optimization reaches a structure without any overlap, the pressure is only the driving force to change the structure because of the characteristic feature of the two-body interaction defined by Eq.~(\ref{eq:potential}), and the optimization leads to a structure with overlap again. This means that the optimization cannot finish, but enable collision and repulsion between spheres to be repeated as long as we continue. The pressure is necessary to minimize the volume of the unit cell. In each steepest descent step, the extended coordinates $\bm{u}$ is changed to minimize the enthalpy per unit cell $H$ as
\begin{equation}
\Delta \bm{u} = - k_2 \, \frac{\partial H}{\partial \bm{u}},
\end{equation}
where $k_2$ is the steepest descent prefactor and the enthalpy is defined as $H \equiv E + PV$; $P$ is the pressure and $V$ is the volume of the unit cell. The operation is not aimed at converging the enthalpy to a local minimum but transforming significantly an initial structure to a dense structure with a large number of the repetition of collision and repulsion. In fact, we have confirmed that the linear minimization method does not work. A lot of iteration is essential for sufficient structural transformation. In the global optimization step, the steepest descent method is repeated several thousand times.

In the optimization, no structure stops transforming, because the forces and the pressure are never balanced. Any structure can transform into a dense structure. The effectiveness comes from the hard-sphere potential defined as Eq.~(\ref{eq:potential}).

As with the expansion process discussed before, the steepest descent prefactor $k_2$ is not set directly, but $\Delta \bm{u}$ is scaled so that the maximum displacement in the position vectors and the lattice vectors will be equal to the predetermined value. The purpose of the global optimization is only to fill the wasteful gaps, so it is the most important that how much the position vectors and lattice vectors are allowed to displace. Setting the maximum displacement has the advantage of allowing each sphere to move enough even when the number of spheres in the unit cell is large. Therefore, it makes sense to set $\Delta \bm{u}$ to an appropriate size so that the wasted gaps can be filled. During the global optimization, the maximum displacement is kept constant. The pressure $P$ is also kept constant as $P = 0.1$. The other values are presented in the Sec.~\ref{sec:structural-optimization-parameters}.

Even after a large number of optimization steps, the overlap never converges to zero, because $z _{i j} ^{\left(\bm{T}\right)}$ between neighboring spheres oscillate around zero. The overlap converges to zero with the local optimization, which we discuss the next.

\subsection{Local optimization: \textit{iterative balance method}}
\label{sec:local_optimization}

\begin{figure}
\centering
\includegraphics[width=\columnwidth]{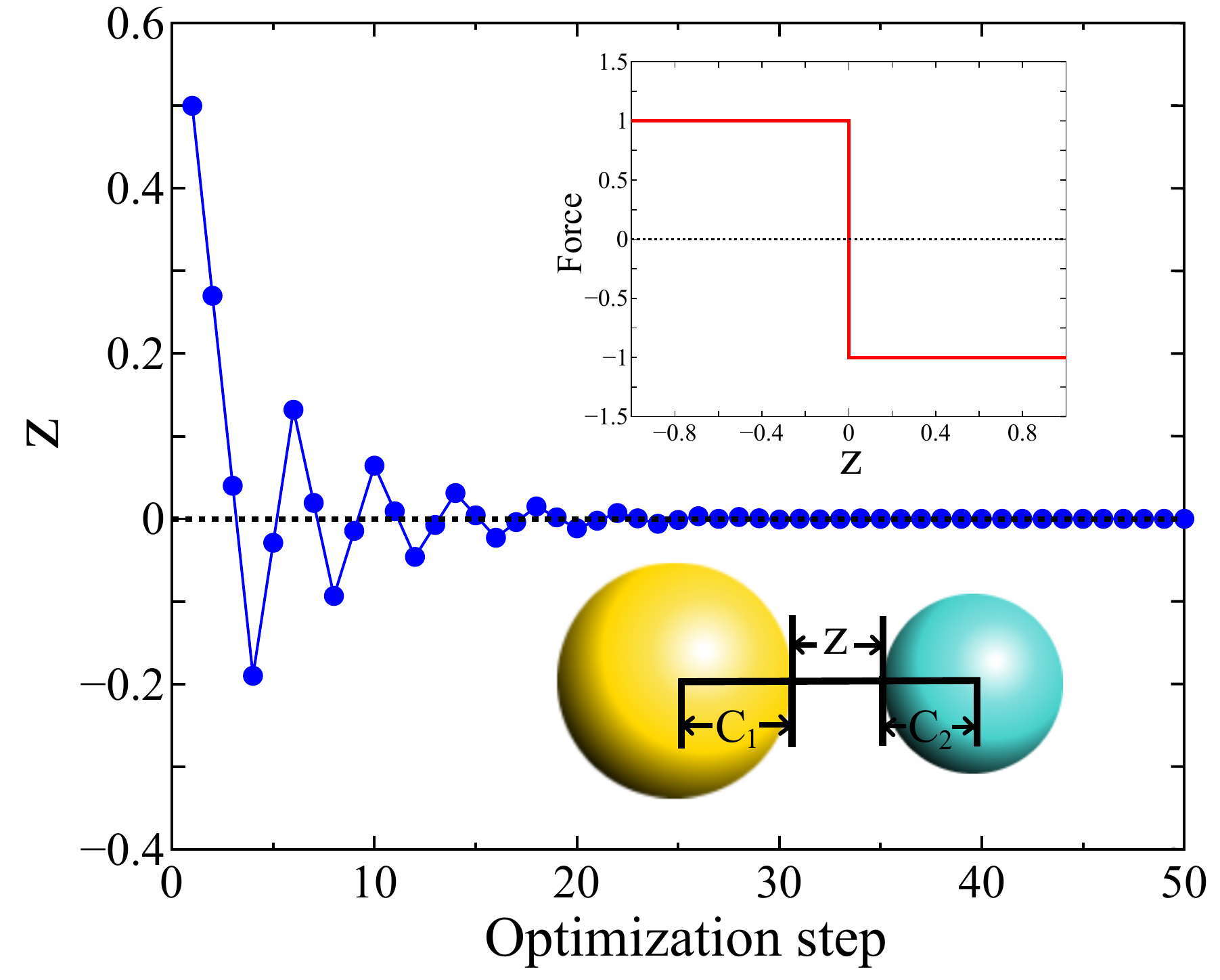}
\caption{The iterative balance method in the one-dimensional case. Two spheres interact with the discontinuous force. With decrease in the displacement, $z$ eventually converges to zero.}
\label{fig:binaryOverlap}
\end{figure}%

To optimize a structure to a periodic packing of nonoverlapping spheres, all of the overlaps have to be converged to zero. In other words, all of the $z _{i j} ^{\left(\bm{T}\right)}$ have to be converged to more than or equal to zero. To achieve the condition, we develop the \textit{iterative balance method}. The iterative balance method optimizes the volume of a unit cell while keeping the overlap of hard spheres minimized. The method is developed based on the idea of repeating collision and repulsion among spheres under pressure while the maximum displacement in position vectors and lattice vectors is gradually decreased. The pressure $P$ is kept constant as $P = 0.1$ through the iterative balance method.

In the iterative balance method, collision and repulsion between spheres are also repeated with the steepest descent method. As in the global optimization, the steepest descent prefactor $k_2$ is not set directly, but $\Delta \bm{u}$ is scaled so that the largest displacement in the position vectors and lattice vectors will be equal to a certain value. To converge all of $z _{i j} ^{\left(\bm{T}\right)}$ to more than or equal to zero, the maximum displacement is gradually decreased for each step.

To illustrate the convergence of $z _{i j} ^{\left(\bm{T}\right)}$, we consider the structural optimization in the one-dimensional case. Here, as shown in Fig.~\ref{fig:binaryOverlap}, we assume that two spheres interact with the discontinuous force defined to be
\begin{equation}
f = \begin{cases}
1 & \text{$z \le 0$} \\
-1 & \text{$0 < z $} \end{cases}, \label{eq:model-force}
\end{equation}
with
\begin{equation}
z \equiv r - \left(c_1 + c_2 \right),
\end{equation}
where $r$ is the distance between the two spheres. If the model is optimized with the steepest descent method with a constant prefactor, $z$ oscillates around $z=0$, and never reaches to $z=0$. On the other hand, if the displacement is gradually reduced, $z$ eventually converges to zero. The convergence behavior is shown in Fig.~\ref{fig:binaryOverlap} together with the one-dimensional model and the  discontinuous force defined by Eq.~(\ref{eq:model-force}) as in the inset.

The idea works even for the three-dimensional case we are interested in. If the repulsive force is much larger than the applied pressure $P$, some of $z _{i j} ^{\left(\bm{T}\right)}$ between neighboring spheres oscillate around zero when the steepest descent method with a constant prefactor is applied. Once we control the prefactor so that the maximum displacement can be gradually reduced, the structure eventually reaches a packing structure with almost zero overlaps. In this sense, the $z_{i j} ^{\left(\bm{T}\right)} = 0$ can be regarded as a \textit{balanced} point. This is the reason why we call it the iterative balance method. The size of the maximum displacement is the same order as the maximum overlap which is defined as the maximum value in $-z_{i j} ^{\left(\bm{T}\right)}$, as discussed in Sec.~\ref{maximum-overlap}. It should be clearly noted that the discontinuous sign change in forces produced by Eq.~(\ref{eq:potential}) is crucial to realize the balanced point. Other potentials such as the squared potential of $z$ do not provide the interesting feature.

The iterative balance method can find optimal distortion. Due to the pressure, the volume of a unit cell is decreased as much as possible. Therefore, the method makes as many $z_{i j} ^{\left(\bm{T}\right)}$ converge to zero as possible, because the void between spheres causes an increase in the volume. Some of $z_{i j} ^{\left(\bm{T}\right)}$ between neighboring spheres converge to zero. The convergence causes distortion in the structure. Accordingly, the generated packing structure has a high packing fraction, because the volume of the unit cell is minimized. The feature indicates that the iterative balance method can find the local maximum of packing fractions despite the fact that a large number of optimization steps is necessary to find the maximum packing fraction. The optimization parameters such as the optimization step number are presented in the Sec.~\ref{sec:structural-optimization-parameters}.

Just like the Torquato-Jiao sphere-packing algorithm~\cite{PhysRevE.82.061302}, in principle the iterative balance method allows us to evaluate the packing fraction of the dense sphere packings while avoiding overlaps of spheres. The implementation is straightforward and the computational time for the structural optimization is very short as discussed later on, since it requires only simple calculations of forces and stress at each steepest decent step. We will demonstrate the efficiency and ability to find putative densest structures in Sec.~\ref{sec:numeric_aspect_of_our_method}.

\subsection{Examples of structure generation}

\begin{figure}
\centering
\begin{subfigure}{0.32\columnwidth}
\centering
\includegraphics[width=\columnwidth]{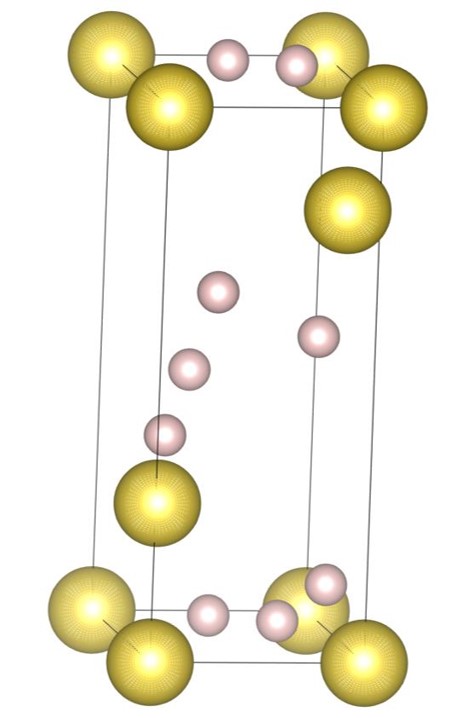}
\caption{Generation of seed structure}
\label{fig:7-3-generate}
\end{subfigure}
\begin{subfigure}{0.32\columnwidth}
\centering
\includegraphics[width=\columnwidth]{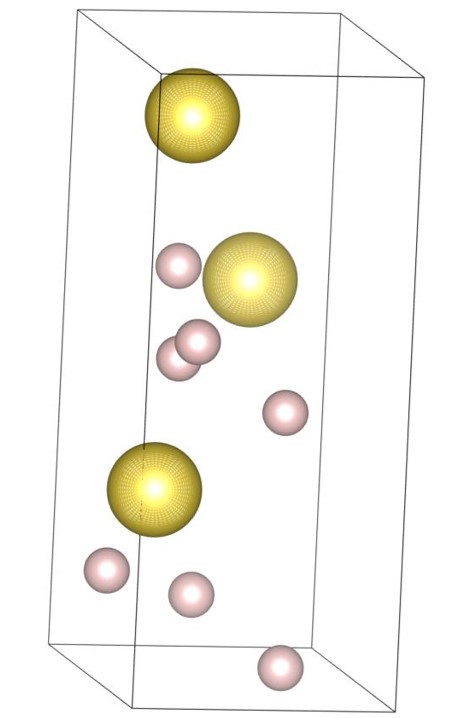}
\caption{Expansion}
\label{fig:7-3-expand}
\end{subfigure}
\begin{subfigure}{0.32\columnwidth}
\centering
\includegraphics[width=\columnwidth]{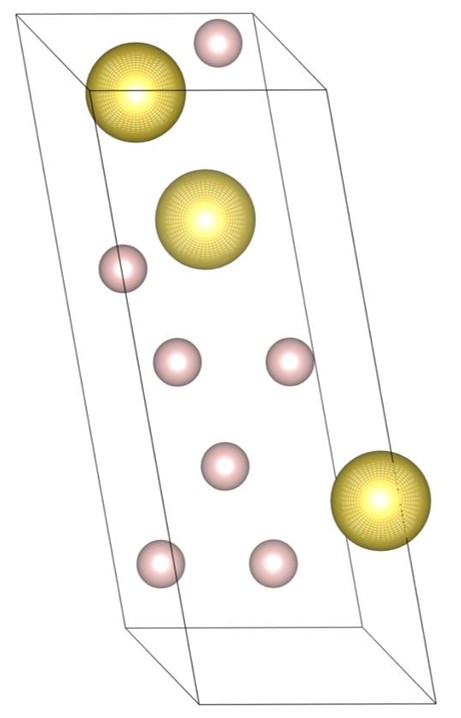}
\caption{Result}
\label{fig:7-3-result}
\end{subfigure}
\caption{Generation process of the (7-3) structure~\cite{PhysRevE.85.021130}. (a) Generation of a seed structure with the piling-up method. (b) Expansion of the seed structure with the steepest decent method. (c) The final structure obtained with the iterative balance method, where the bottom side is narrow and the vertical direction is long, corresponding to the the feature of the seed structure.}
\label{fig:7-3-production}
\end{figure}
\begin{figure}
\centering
\begin{subfigure}{0.32\columnwidth}
\centering
\includegraphics[width=\columnwidth]{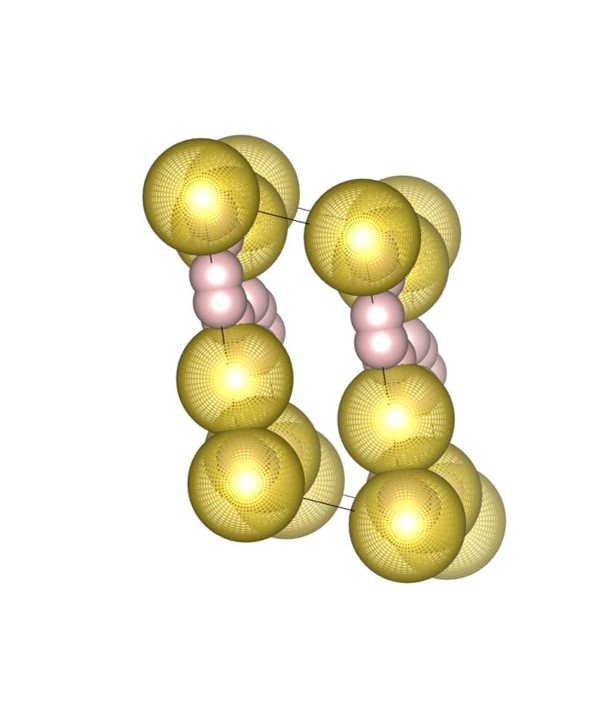}
\caption{Generation of seed structure}
\label{fig:16-4-generate}
\end{subfigure}
\begin{subfigure}{0.32\columnwidth}
\centering
\includegraphics[width=\columnwidth]{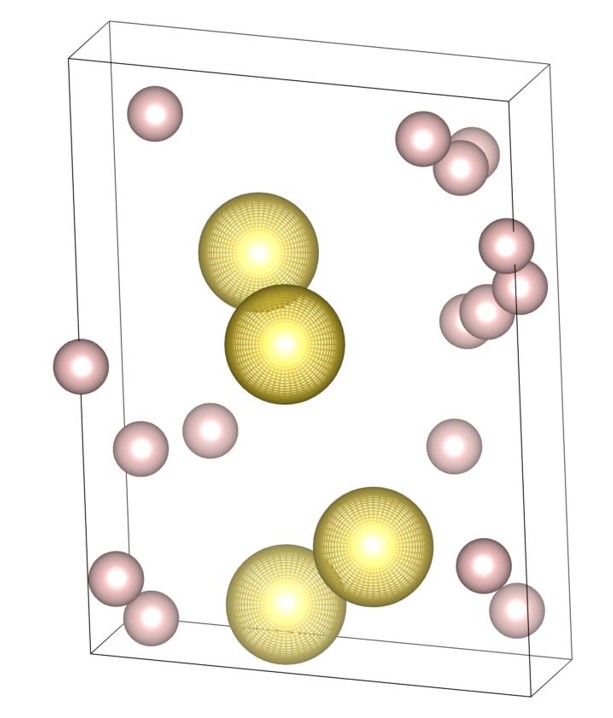}
\caption{Expansion}
\label{fig:16-4-expand}
\end{subfigure}
\begin{subfigure}{0.32\columnwidth}
\centering
\includegraphics[width=\columnwidth]{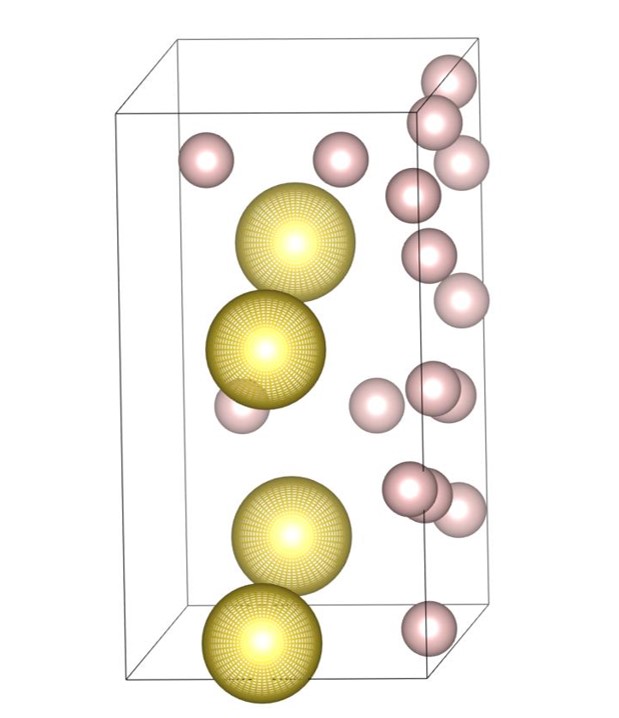}
\caption{Result}
\label{fig:16-4-result}
\end{subfigure}
\caption{Generation process of the (16-4) structure, where for visibility of the figures, the bottom side, which is the first layer generated by the piling up method, is arranged at the front side. (a) Generation of a seed structure with the piling-up method, where all the spheres are placed in the first layer. (b) Expansion of the seed structure with the steepest decent method. (c) The final structure obtained with the iterative balance method. }
\label{fig:16-4-production}
\end{figure}

The (7-3) structure~\cite{PhysRevE.85.021130} is the tower-like DBSP with seven small and three large spheres in the unit cell. The formation process of the structure is shown in Fig.~\ref{fig:7-3-production}. Firstly, the tower-like structure is generated by stacking many spheres on top of the small bottom plane as shown in Fig.~\ref{fig:7-3-production}(a), and then the structure is expanded as shown in Fig.~\ref{fig:7-3-production}(b) with the steepest descent method. The bottom plane corresponds to the first layer generated by the piling up method. The final structure, shown in Fig.~\ref{fig:7-3-production}(c), also has a small bottom plane and is long in the vertical direction.

The (16-4) structure is the DBSP with sixteen small and four large spheres in the unit cell. The structure can be understood as a two-layered structure except for four small spheres. The formation process of the structure is shown in Fig.~\ref{fig:16-4-production}. For visibility of the figures, the bottom plane, which is the first layer generated by the piling up method, is arranged at the front side. Firstly, all the spheres are placed on the first layer as shown in Fig.~\ref{fig:16-4-production}(a), and then the structure is expanded as shown in Fig.~\ref{fig:16-4-production}(b). An overlap in the direction perpendicular to the first layer causes a distribution of spheres' positions in the direction. The final structure, shown in Fig.~\ref{fig:16-4-production}(c), can also be regarded as a layer-by-layer structure with a small height.

\subsection{Neighboring spheres}
\label{sec:neighbor-sphere}

If a packing structure sufficiently converges to a dense packing, few spheres' positions displace significantly even after many optimization steps. Besides, repulsive forces occur only when two spheres overlap each other. Therefore, the computational cost can be reduced significantly by predetermining the neighboring spheres. Of course, we must reset the neighboring list periodically, because accidentally position vectors may displace largely. In our code, the neighboring list is reset once every 100 times in the global optimization step and once every 200 times in the local optimization step.

\subsection{Treatment of similar initial multilayer structures}

One may consider that it is futile to optimize all similar structures which converge to an isomorphic structure. Hereafter, the isomorphic is defined that two structures are equal when the distortion is corrected. The distortion is necessary to achieve a high packing fraction. In some cases, the number of local minima is huge due to a large number of distortion patterns and accordingly many structures are trapped at local minima. Therefore, optimizing all of the similar structures is necessary to determine the highest packing fraction. Hence, all of the generated structures are optimized with our method.

\section{numeric aspect of our method}
\label{sec:numeric_aspect_of_our_method}

In this section, we discuss the numerical aspects of our method including the accuracy of the packing fraction, structural optimization parameters, and the efficiency of our method.

\subsection{Structural optimization parameters}
\label{sec:structural-optimization-parameters}

The efficiency of structural optimization depends on the choice of parameters used in the optimization, and even the reachable maximum packing fraction is varied depending on the parameters in case that many local minima are having competitive packing fractions. These parameters we have are listed below:
\begin{itemize}
\setlength{\parskip}{0pt} 
\item $\Delta_{\mathrm{expand, max}}$ : The maximum displacement in the expansion step
\item $\Delta_{\mathrm{global, max}}$ : The maximum displacement in the global optimization step
\item $N_{\mathrm{global}}$ : The maximum step number in the global optimization step
\item $\Delta_{\mathrm{local, max}}$ : The maximum displacement in the local optimization step
\item $N_{\mathrm{local}}$ : The maximum step number in the local optimization step
\item $d$ : The decreasing factor of the maximum displacement in the local optimization step
\end{itemize}
The default values used in our study are given below.
\begin{align*}
&0.03 \le \Delta_{\mathrm{global, max}} \le  0.15 \\
&0.02 \le \Delta_{\mathrm{local, max}} \le 0.10 \\
&\Delta_{\mathrm{expand, max}} = 0.3 \\
&N_{\mathrm{global}} = 2000 \\
&N_{\mathrm{local}} = 40000 \\
&d = 0.9997
\end{align*}
Here, we note that $\Delta_{\mathrm{global, max}}$ ($\Delta_{\mathrm{local, max}}$) is chosen randomly in the range of $0.03 \le \Delta_{\mathrm{global, max}} \le  0.15$ ($0.02 \le \Delta_{\mathrm{local, max}} \le 0.05$) for each structural optimization. The random choice of them enables us to obtain various local minima. The default parameters lead to a convergence that almost all the minimum value of $z _{i j} ^{\left(\bm{T}\right)}$ becomes the same order as $-10^{-7}$ to $-10^{-6}$.

\subsection{Maximum overlap}
\label{maximum-overlap}

\begin{figure}
\centering
\includegraphics[width=\columnwidth]{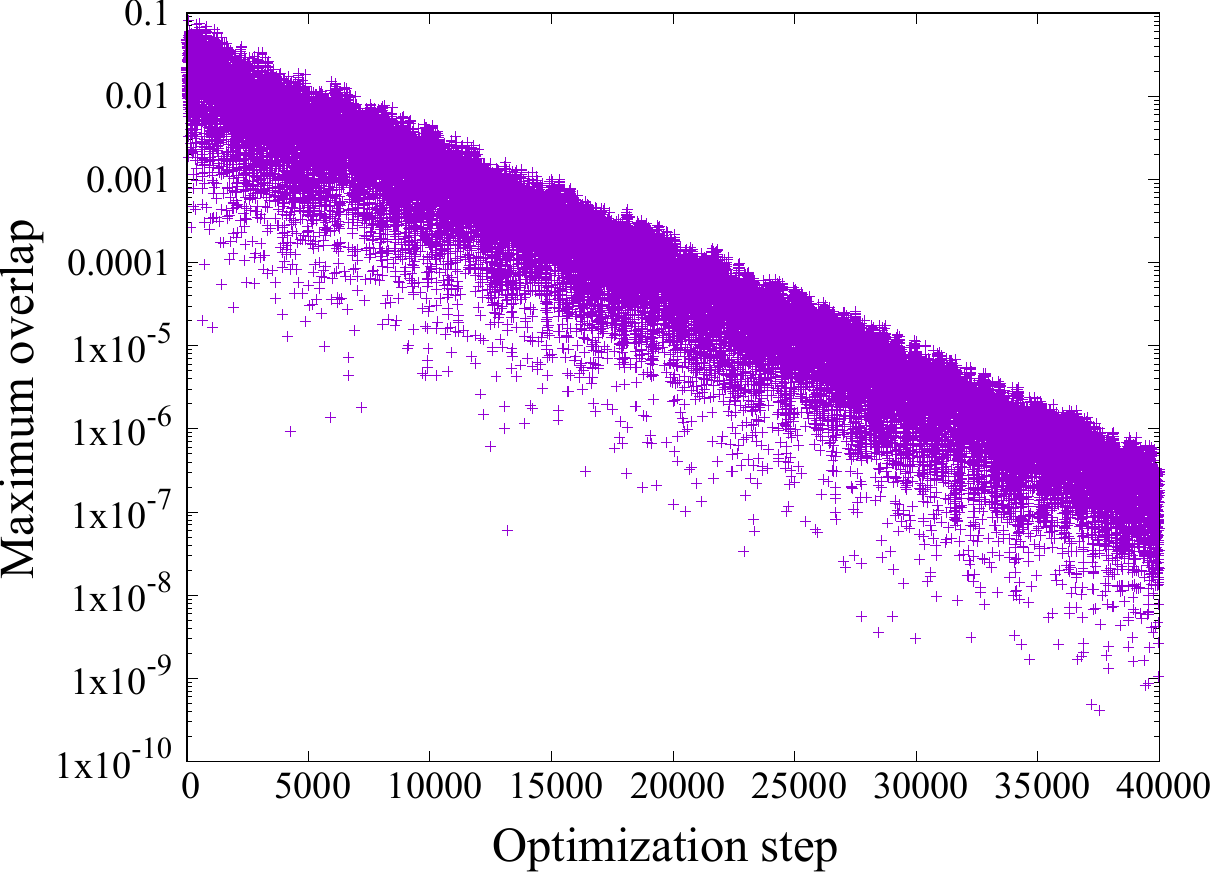}
\caption{The maximum overlap at each optimization step where the maximum overlap is defined as the maximum value of $-z _{i j} ^{\left(\bm{T}\right)}$. The maximum overlap decreases exponentially as the maximum displacement is decreased exponentially. }
\label{fig:overlap}
\end{figure}

In the iterative balance method, the maximum displacement of the position vectors and the lattice vectors are gradually reduced. Figure~\ref{fig:overlap} shows the maximum overlaps at each optimization step. $\Delta_{\mathrm{local, max}}$ is set to $0.0546375$. The maximum overlap is defined as the maximum value of $-z _{i j} ^{\left(\bm{T}\right)}$. It is confirmed from Fig.~\ref{fig:overlap} that the maximum overlap decreases exponentially as the maximum displacement is decreased exponentially, and those values are the same order. The relation can be understood from the fact that some of $z _{i j} ^{\left(\bm{T}\right)}$ between neighboring spheres oscillate around zero.

\subsection{Accuracy of packing fraction}
\label{sec:accuracy-of-filling-rate}

In the radius ratio of $\alpha \le \sqrt{2}-1$, the $\mathrm{XY_n}$ structures appear on the phase diagram. The $\mathrm{XY_n}$ structures are defined as packing structures in which large spheres X constitute the fcc densest structure and the small spheres penetrate into the tetrahedral and octahedral sites constituted by X. This definition is the same as that of Hopkins \textit{et al.} \cite{PhysRevE.85.021130}. The packing fractions of the $\mathrm{XY_n}$ structures can be calculated analytically as
\begin{equation}
\phi = \frac{\pi}{3 \sqrt{2}} \left(1 + n \alpha ^3 \right). \label{eq:analytical-solution-of-XYn}
\end{equation}
In the radius ratio of $\alpha \le \sqrt{2}-1$, as discussed by Hopkins and coworkers, six $\mathrm{XY_n}$ structures of $\mathrm{XY}$, $\mathrm{XY_2}$, $\mathrm{XY_4}$, $\mathrm{XY_8}$, $\mathrm{XY_{10}}$, and $\mathrm{XY_{11}}$ \cite{PhysRevE.85.021130} appear on the phase diagram. An octahedral site of $\mathrm{XY}$ structure is occupied by one small sphere, an octahedral site of $\mathrm{XY_2}$ structure is occupied by two small spheres, an octahedral site of $\mathrm{XY_4}$ structure is occupied by a tetrahedron consisting of four small spheres, an octahedral site of $\mathrm{XY_8}$ is occupied by an cubic consisting of eight small spheres, a tetrahedral site of $\mathrm{XY_{10}}$ structure is occupied by one small sphere and an octahedral site of $\mathrm{XY_{10}}$ is occupied by a cubic consisting of eight small spheres, and a tetrahedral site of $\mathrm{XY_{11}}$ structure is occupied by one small sphere and an octahedral site of $\mathrm{XY_{11}}$ is occupied by a bcc structure consisting of nine small spheres, respectively. 

It was relatively easy to calculate the packing fractions of the $\mathrm{XY_n}$ structures with our method because there is no distortion in the fcc structure constituted by large spheres. However, when the optimization parameters are set to the default values, in some cases an error of the packing fraction becomes the same order as $10^{-6}$ due to overlap which is also the same order as $10^{-6}$. As discussed in the next section, by optimizing again with changing the optimization parameters, the maximum overlap can be decreased to the same order as $10^{-13}$. In that case, the packing fractions agree by more than 10 decimal points with the analytical solution of Eq.~(\ref{eq:analytical-solution-of-XYn}).

As discussed in Sec.~\ref{sec:local_optimization}, the iterative balance method can minimize the volume of a unit cell. Therefore, high-precision re-optimization enables us to find the highest packing fractions with high accuracy because of a small overlap and a minimized volume of a unit cell.

\subsection{High-precision structural optimization}
\label{sec:high-precision-structural-optimization}

As discussed in Sec.~\ref{sec:accuracy-of-filling-rate}, the local optimization with default values presented in Sec.~\ref{sec:structural-optimization-parameters} cannot reduce an overlap enough. The overlap causes an error in packing fraction after the six decimal points. The error can be reduced by a high-precision re-optimization. With an increase in the number of local optimization steps, the overlap can be decreased because the maximum displacement is decreased as the number of optimization steps is increased. As discussed in Sec.~\ref{maximum-overlap}, the maximum overlap and the size of the maximum displacement is the same order.

When the number of local minima is getting larger, more structures are trapped at local minima, so the global minimum might not be found during an exhaustive search. Therefore, it is necessary to re-optimize the putative densest structure many times in order to identify the global minimum. The optimization results depend on the initial structure and the optimization parameters such as $\Delta_{\mathrm{global, max}}$ and $\Delta_{\mathrm{local, max}}$. Therefore, slight fluctuations are given to the structure before re-optimization and the range of $\Delta_{\mathrm{global, max}}$ and $\Delta_{\mathrm{local, max}}$ is set to be large. When the number of local minima is small, almost all the fluctuated structures converge to the densest packing. On the other hand, when the number of local minima is large, we obtain many packing fractions corresponding to diverse distortion patterns. In some cases, tens of thousand re-optimizations are necessary to determine the maximum packing fraction. As discussed in the next subsection, the computational time for the structural optimization is very short, so the necessity of the several thousand re-optimization is not a serious problem.

The default parameters for re-optimization are given below.
\begin{align*}
&0.003 \le \Delta_{\mathrm{global, max}} \le  0.05 \\
&0.001 \le \Delta_{\mathrm{local, max}} \le 0.02 \\
&N_{\mathrm{global}} = 4000 \\
&N_{\mathrm{local}} = 80000 \\
&d = 0.9997
\end{align*}
If we use the default parameters, in almost all case, the minimum values of $z _{i j} ^{\left(\bm{T}\right)}$ become the same order as $-10^{-13}$ to $-10^{-12}$.

\subsection{Speed of structure generation}
\label{sec:speed-of-structure-generation}

The computational cost for calculating forces with the hard-sphere potential is very low because repulsive forces occur only when two spheres overlap each other. Besides, if a packing structure converges to a dense packing, the positions of spheres do not change significantly even after many optimization steps. Therefore, as discussed in Sec.~\ref{sec:neighbor-sphere}, the computational cost can be significantly reduced by pre-determining the neighboring spheres. In fact, if the number of spheres in the unit cell is about ten, it takes less than 0.1 seconds from generation to optimization using a single core of Intel(R) Xeon(R) CPU E5-1650 v4 @ 3.60GHz. Therefore, our method can generate a large number of packing structures. The efficiency enables us to find the densest packing from the vast coordination space.

In addition, since the number of optimization steps is set externally, the computational cost can be estimated as $O \left(N \right)$, where $N$ is the number of spheres. Hence, it is possible to conduct an exhaustive search for the long periodic densest packings.

\subsection{Distribution and update history of packing fractions}
\label{sec:Distribution_and_update_history_of_packing_fractions}

\begin{figure}
\centering
\includegraphics[width=\columnwidth]{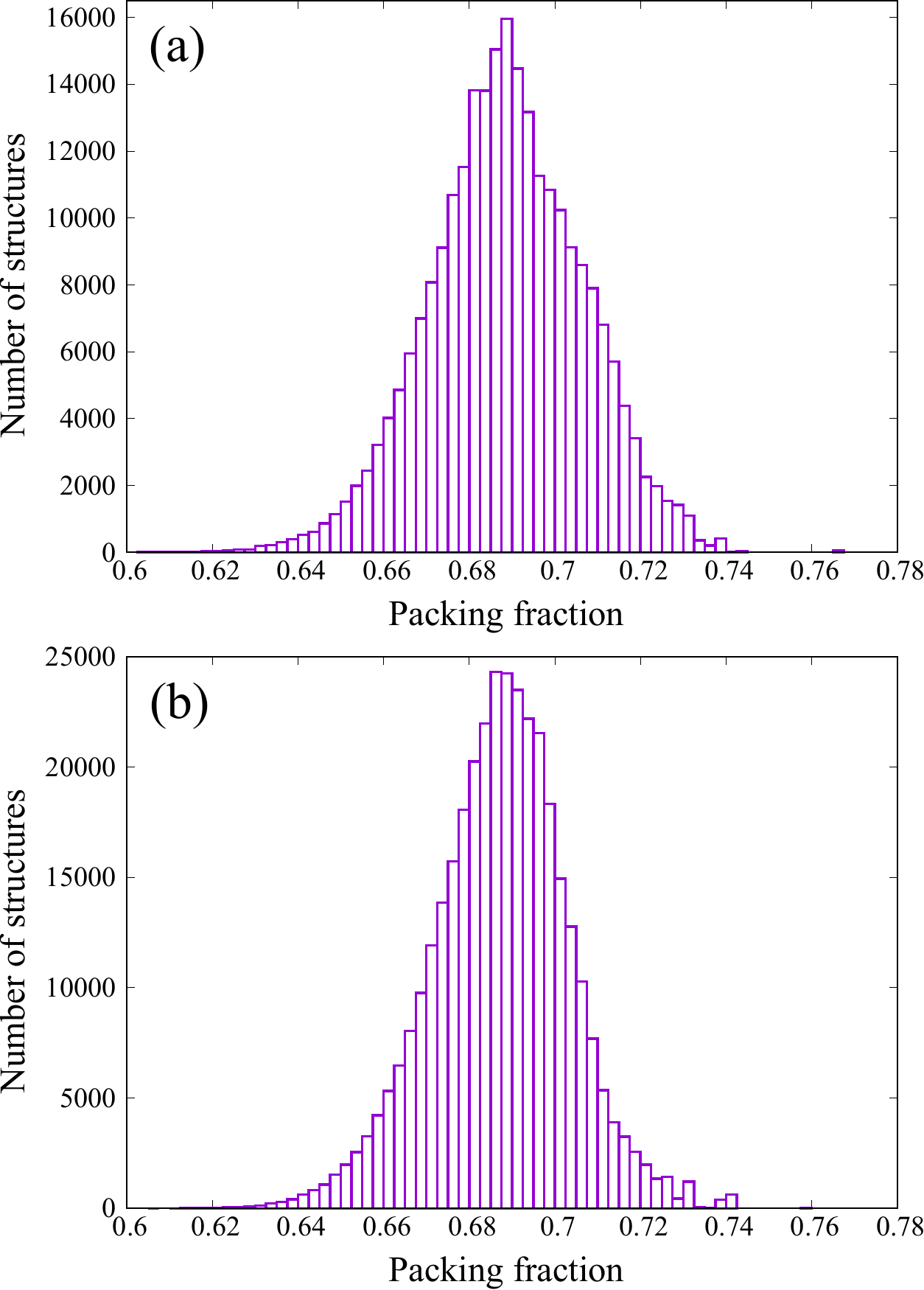}
\caption{The distribution of the packing fractions during the exhaustive search. (a) The distribution for 14-5 system; the unit cell contains 14 small spheres and 5 large spheres. The radius ratio is $\alpha = 0.445$. (b) The distribution for 16-4 system; the unit cell contains 16 small spheres and 4 large spheres. The radius ratio is $\alpha = 0.455$.}
\label{fig:distribution}
\end{figure}
\begin{figure}
\centering
\includegraphics[width=\columnwidth]{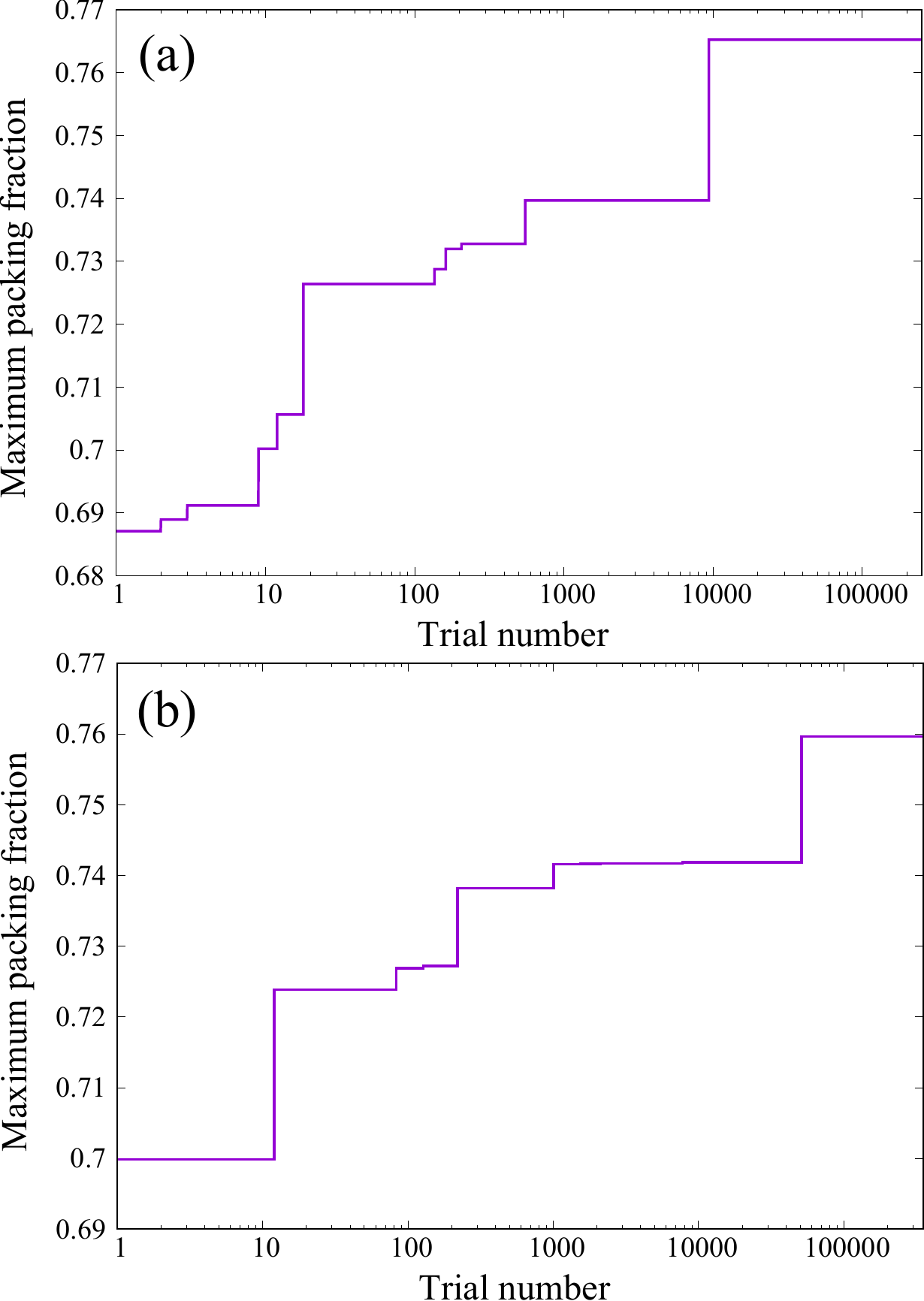}
\caption{The updating history of packing fractions during the exhaustive search. (a) The updating history for 14-5 system; the unit cell contains 14 small spheres and 5 large spheres. The radius ratio is $\alpha = 0.445$. (b) The updating history for 16-4 system; the unit cell contains 16 small spheres and 4 large spheres. The radius ratio is $\alpha = 0.455$.}
\label{fig:history}
\end{figure}

In this subsection, we discuss the distribution and updating history of packing fractions during the exhaustive search, to show that our method can find the densest packings from the vast coordination space.

First, we conducted an exhaustive search for the mono-phase densest binary sphere packing at the radius ratio of $\alpha = 0.445$ and the composition ratio of $x=14/19$. The unit cell contains 14 small spheres and 5 large spheres. In that case, the (14-5) structure with the packing fraction of $0.765259$ is the densest. The number of structures generated during the exhaustive search is 250,182. The distribution of the packing fractions is shown in Fig.~\ref{fig:distribution}(a). The number of structures having the packing fraction between 0.6875 and 0.6900 is 15959 and the largest. On the other hand, the number of structures having the packing fraction between 0.7650 and 0.7675 is only 58. The result indicates that our method can find the densest packings from the vast coordination space. Figure \ref{fig:history}(a) shows the updating history of the highest packing fraction. The structure with a packing fraction of $0.765256$ is generated at the 9372nd steps, while the densest structure with the packing fraction of $0.765259$ is generated at the 50155th steps.

Similarly, we conducted an exhaustive search for the mono-phase densest binary sphere packing at the radius ratio of $\alpha = 0.455$ and the composition ratio of $x=16/20$. The unit cell contains 16 small spheres and 4 large spheres. In that case, the (16-4) structure with the packing fraction of $0.759629$ is the densest. The number of structures generated during the exhaustive search is 351,124. The distribution of the packing fractions is shown in Fig.~\ref{fig:distribution}(b). The number of structures having the packing fraction between 0.6850 and 0.6875 is 24303 and the largest. On the other hand, the number of structures having the packing fraction between 0.7575 and 0.7600 is only 8. The result also indicates that our method can find the densest packings from the vast coordination space. Figure \ref{fig:history}(b) shows the updating history of the packing fraction. The densest structure with the packing fraction of $0.765259$ is generated at the 51097th steps.

\section{method for phase diagram}
\label{sec:method_for_phase_diagram}

In the previous sections, we have already discussed the details of our methods implemented in \textit{\textbf{SAMLAI}} and the numerical aspects. In this section, we discuss how the phase diagram for DBSP is constructed with \textit{\textbf{SAMLAI}}.

\subsection{Exhaustive search conditions}

To determine the densest phase separation at each composition ratio $x$ and each radius ratio $\alpha$, firstly the mono-phase densest binary sphere packings have to be identified at each $\left(x, \alpha \right)$. 

To identify the mono-phase densest binary sphere packings at each $\left(x, \alpha \right)$, up to one million structures are generated with \textit{\textbf{SAMLAI}}. We terminate the exhaustive search when the highest packing fraction is not updated 200,000 or 300,000 times, and regard the structure with the highest packing fraction as putative densest packing.   

The radius ratio $\alpha$ is changed by a step of 0.02 in the range of $0.20 \le \alpha \le 0.64$ while the radius ratio is changed by a step of 0.005 in the radius range of $0.42 \le \alpha \le 0.50$. In total, the 35 radius ratios are investigated in the construction of the phase diagram. 

The number of spheres in the unit cell is set between 6 and 24 while the number of spheres in the unit cell is set between 12 and 32 in the radius range of $0.42 \le \alpha \le 0.50$. Under the condition, all possible compositions are investigated with the constraint that the number of small spheres is equal to or larger than that of large spheres. In the former case, there are 138 compositions and in the latter case, there are 220 compositions, respectively. In general, the number of possible compositions is $n^2/4$ if the maximum number of spheres is set to $n$, where $n$ is an even number and the minimum number of spheres is set to 2.

The optimization parameters are set to the default as discussed in Sec.~\ref{sec:structural-optimization-parameters}.

\subsection{Re-optimization}

Almost all of the generated structures have an overlap of about $10^{-7}$ to $10^{-6}$. Therefore, packing fractions may have an error on the sixth digit. In addition, the highest packing fraction may not be found due to a large number of local minima. Therefore, putative densest packings have to be re-optimized for determining the highest packing fractions with high accuracy. In some cases, tens of thousand steps for the re-optimization is necessary in order to find the highest packing fraction from the huge number of local minima. The re-optimization parameters are set to the defaults discussed in Sec.~\ref{sec:high-precision-structural-optimization} in most cases. The re-optimization updates the fourth digit and beyond of the packing fraction for cases that many local minima exist, while the packing fraction is unchanged by the re-optimization in most of the cases.

\subsection{Phase separation}

As Hopkins and coworkers have shown, the highest packing fractions for the binary system are achieved by phase separation into two or fewer densest packings~\cite{PhysRevE.85.021130}. We also proved it in a way that seems more intuitive to us. The proof is given in Appendix \ref{sec:phase_separation}. The phase separations with the highest packing fraction are determined from all possible phase separations.

\section{result}
\label{sec:result}

\begin{figure*}
\centering
\includegraphics[width=2\columnwidth]{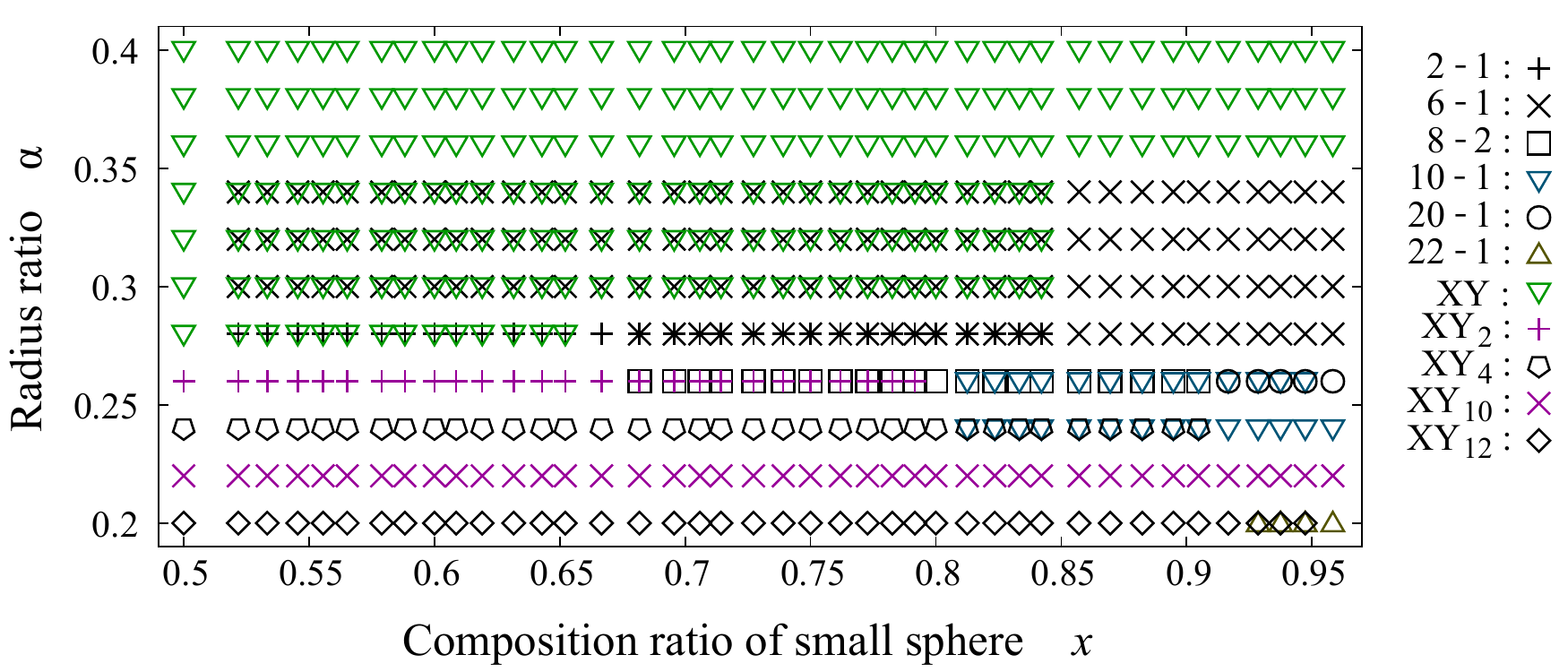}
\caption{The phase diagram in the radius ratio of $0.20 \le \alpha \le 0.40$. When the highest packing fraction is achieved by phase separation into a densest binary sphere packing $\Phi$ and the fcc densest packing consisting of small spheres, only the symbol of the densest binary packing $\Phi$ is plotted. On the other hand, when the highest packing fraction is achieved by phase separation into two densest binary sphere packings, those two symbols are plotted together.}
\label{fig:smallPhaseDiagram}
\end{figure*}
\begin{figure*}
\centering
\includegraphics[width=2\columnwidth]{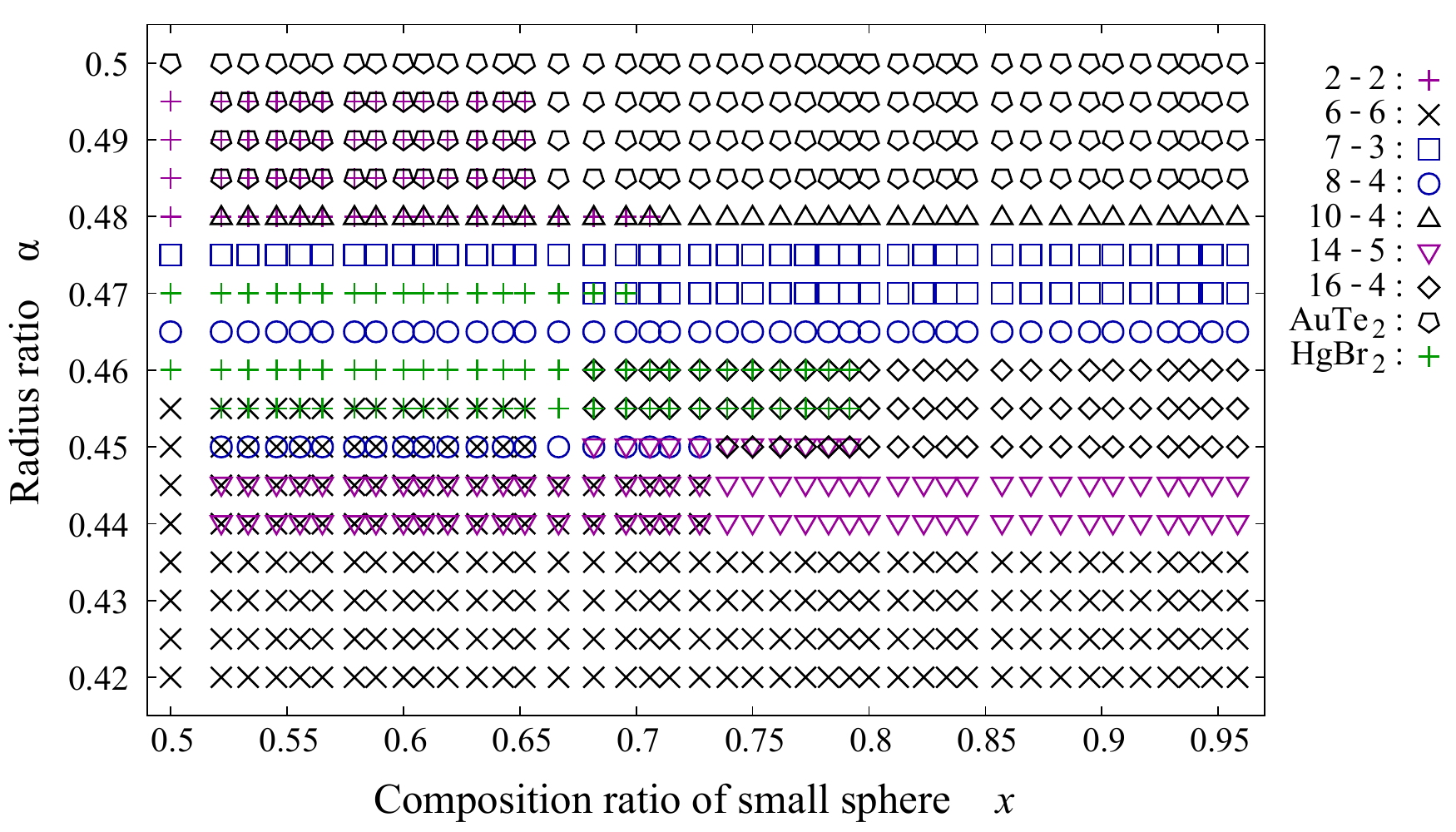}
\caption{The phase diagram in the radius ratio of $0.420 \le \alpha \le 0.500$. The rule in plotting symbols follows that explained in the caption of Fig.~\ref{fig:smallPhaseDiagram}.}
\label{fig:mediumPhaseDiagram}
\end{figure*}
\begin{figure*}
\centering
\includegraphics[width=2\columnwidth]{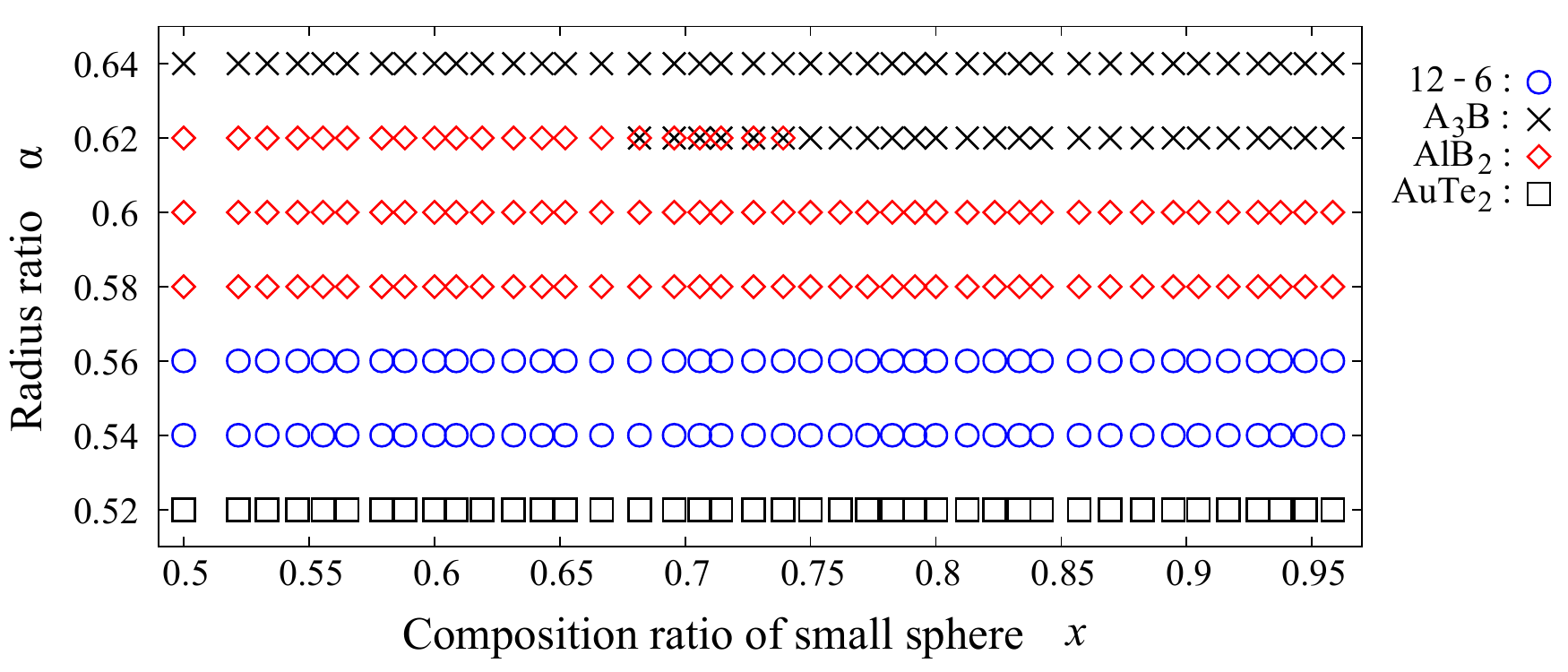}
\caption{The phase diagram in the radius ratio of $0.52 \le \alpha \le 0.64$. The rule in plotting symbols follows that explained in the caption of Fig.~\ref{fig:smallPhaseDiagram}.}
\label{fig:largePhaseDiagram}
\end{figure*}

In this section, we describe the newly found 12 putative DBSP and the updated phase diagram. Next, we detail the geometry of binary densest packings. Finally, we discuss geometric features of the mono-phase densest binary sphere packings.

\subsection{Overview}

We have discovered 12 putative densest packings and accordingly updated the phase diagram over the $x-\alpha$ plane, where $x$ is the relative concentration and $\alpha$ is the radius ratio of the small sphere relative to the large sphere. The phase diagram is shown in Fig.~\ref{fig:smallPhaseDiagram} for $0.20 \le \alpha \le 0.40$, Fig.~\ref{fig:mediumPhaseDiagram} for $0.420 \le \alpha \le 0.500$, and Fig.~\ref{fig:largePhaseDiagram} for $0.52 \le \alpha \le 0.64$, respectively. On the phase diagrams, 24 DBSP are plotted. Some DBSP are not plotted, e.g., newly found (12-1) structure and the (9-4) structure, because they appear in a very narrow region. For the case of 12 or fewer spheres in the unit cell, our phase diagram is consistent with the HST phase diagram~\cite{PhysRevE.85.021130} with a small correction.

The $\mathrm{XY_n}$ structures are defined as DBSP in which the large spheres X constitute the fcc densest structure and the small spheres penetrate into the tetrahedral and octahedral sites constituted by X. If the $\mathrm{XY_n}$ structures are excluded, there are 21 putative DBSP. Most of those structures are named as ($m$-$n$) structure. A ($m$-$n$) structure contains $m$ small spheres and $n$ large spheres. Their packing fractions are shown in Tables~\ref{table:small-filling-factors}, \ref{table:medium-filling-factors}, and \ref{table:large-filling-factors}.

For binary systems, the highest packing fraction is generally achieved by phase separation into two or fewer densest packings. When the highest packing fraction is achieved by phase separation into a structure $\Phi$ and the fcc densest packing consisting of small spheres, only the symbol of structure $\Phi$ is plotted on the phase diagram. On the other hand, when the highest packing fraction is achieved by phase separation into two structures, those two symbols are plotted together on the phase diagram. For example, in the area of $0.30 \le \alpha \le 0.34$ and $6/7 < x$ the highest packing fraction is achieved by the phase separation into the (6-1) structure and the fcc densest structure consisting of small spheres, and in the area of $0.30 \le \alpha \le 0.34$ and $1/2 < x < 6/7$ the highest packing fraction is achieved by the phase separation into the $\mathrm{XY}$ structure~\cite{PhysRevE.85.021130} and the (6-1) structure~\cite{PhysRevE.85.021130}.

For $0.20 \le \alpha \le 0.40$, the exhaustive search is conducted for 11 radius ratios of 0.20, 0.22, 0.24, 0.26, 0.28, 0.30, 0.32, 0.34, 0.36, 0.38, and 0.40. The number of spheres in the unit cell is set between 6 and 24. By the exhaustive search, we identify 6 new putative densest packings: the $\mathrm{XY_{12}}$ structure, the (22-1) structure, the (12-1) structure, the (20-1) structure, the (8-2) structure, and the (2-1) structure. The $\mathrm{XY_{12}}$ structure and the (22-1) structure appear on the phase diagram at $\alpha=0.20$ and $0.203$. The (12-1) structure appears on the phase diagram at $\alpha=0.203$; it can be understood as an extended $\mathrm{XY_{12}}$ structure. The (12-1) structure is not plotted on the phase diagram of Fig.~\ref{fig:smallPhaseDiagram}, because it appears in a very narrow region. The (8-2) structure and the (20-1) structure appear on the phase diagram at $\alpha=0.26$; it is distorted $\mathrm{XY_4}$ structure. The (2-1) structure appears on the phase diagram at $\alpha = 0.28$; it is distorted $\mathrm{XY_2}$ structure.

For $0.420 \le \alpha \le 0.500$, the exhaustive search is conducted for 17 radius ratios of 0.420, 0.425, 0.430, 0.435, 0.440, 0.445, 0.450, 0.455, 0.460, 0.465, 0.470, 0.475, 0.480, 0.485, 0.490, 0.495, and 0.500. The number of spheres in the unit cell is set between 12 and 32. By the exhaustive search, we identify 5 new putative densest packings: the (14-5) structure, the (16-4) structure, the (8-4) structure, the (10-4) structure, and the (9-4) structure. The (14-5) structure appears on the phase diagram at the three radius ratios of $\alpha = 0.440$, $0.445$, and $0.450$. The (16-4) structure appears on the phase diagram at the three radius ratios of $\alpha = 0.450$, $0.455$, and $0.460$. The (8-4) structure appeares on the phase diagram at $\alpha=0.450$ and $0.465$ and it is isomorphic to the $\mathrm{HgBr_2}$ structure. The (10-4) structure appears on the phase diagram at $\alpha = 0.480$ and $0.481$; it is isomorphic to the (5-2) structure. The (9-4) structure appears on the phase diagram at the three radius ratios of $\alpha=0.481$, $0.482$, and $0.483$. In addition, in contrast to the HST phase diagram, there is a phase separation into the (7-3) structure~\cite{PhysRevE.85.021130} and the $\mathrm{HgBr_2}$ structure at $\alpha=0.470$. Finally, the packing fractions of the $\mathrm{AuTe_2}$ structure~\cite{PhysRevE.79.046714, PhysRevE.85.021130} are found to be consistent with those of the (4-2) structure~\cite{PhysRevE.85.021130}. Otherwise, the results are consistent with the HST phase diagram~\cite{PhysRevE.85.021130}.

For $0.52 \le \alpha \le 0.64$, the exhaustive search is conducted for 7 radius ratios of 0.52, 0.54, 0.56, 0.58, 0.60, 0.62, and 0.64. The number of spheres in the unit cell is set between 6 and 24. By the exhaustive search, we identify one new putative densest packing: the (12-6) structure. It is isomorphic to the $\mathrm{AlB_2}$ structure~\cite{PhysRevE.79.046714, PhysRevE.85.021130}. Otherwise, our phase diagram is consistent with the HST phase diagram \cite{PhysRevE.85.021130}.

In Secs.~\ref{sec:small-densest-packings}, \ref{sec:medium-densest-packings}, and \ref{sec:large-densest-packings}, the densest packings in each radius range are detailed.

\subsection{Densest Packings for $0.20 \le \alpha \le 0.40$}
\label{sec:small-densest-packings}

\begin{figure}
\centering
\begin{subfigure}{0.49\columnwidth}
\centering
\includegraphics[width=\columnwidth]{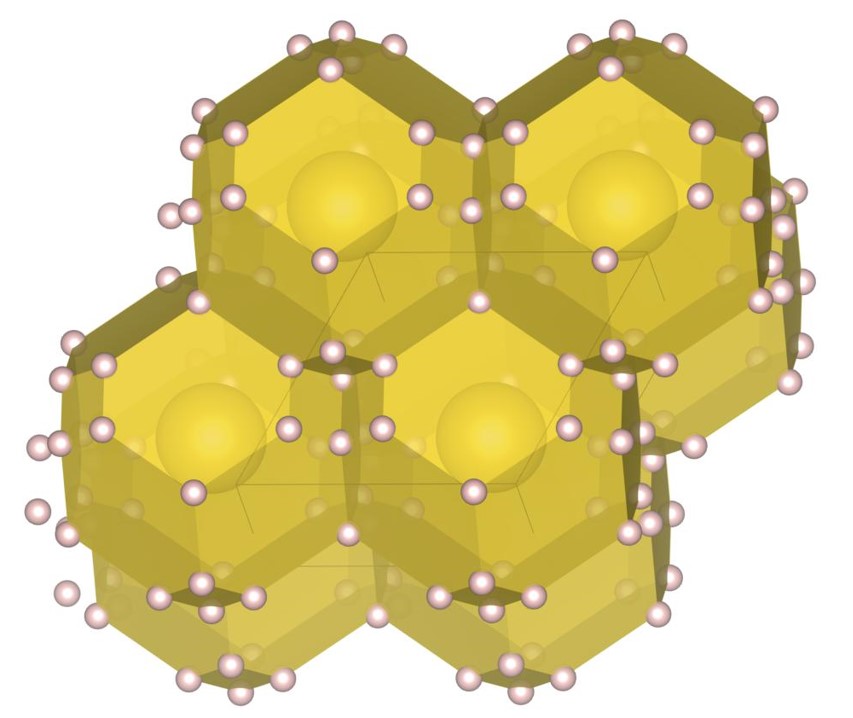}
\caption{(10-1) structure}
\label{fig:10-1}
\end{subfigure}
\begin{subfigure}{0.49\columnwidth}
\centering
\includegraphics[width=\columnwidth]{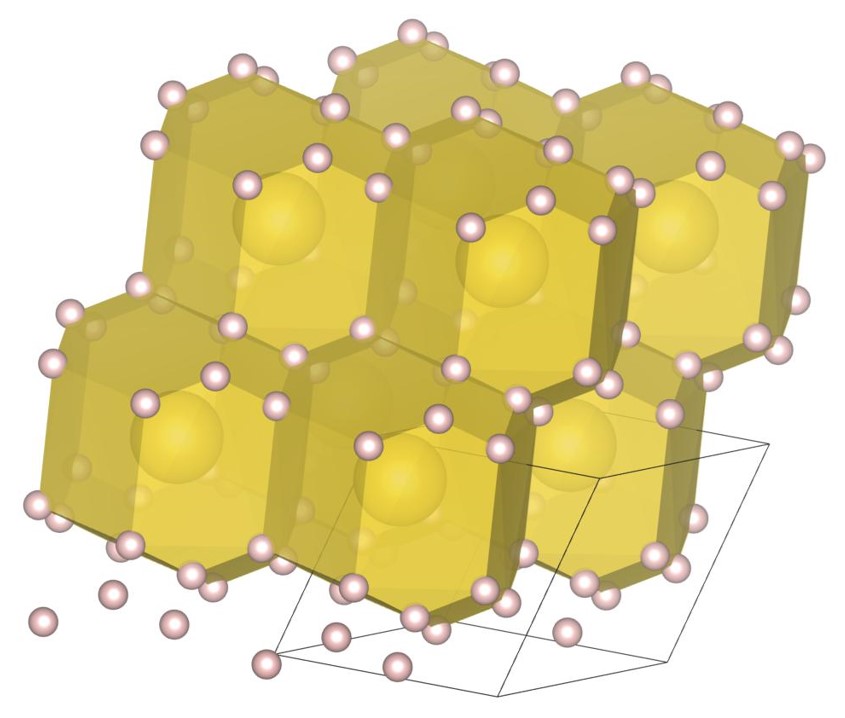}
\caption{(6-1) structure}
\label{fig:6-1}
\end{subfigure}
\caption{Two examples of putative DBSP already identified on the HST phase diagram in the radius ratio of $0.20 \le \alpha \le 0.40$~\cite{PhysRevE.85.021130}. (a) The (10-1) structure; it is a distorted $\mathrm{XY_{10}}$ structure. (b) The (6-1) structure; it can be understood as a clathrate structure. It is an enlarged and distorted bcc structure consisting of the large spheres surrounded by small spheres.}
\end{figure}

\begin{figure*}
\centering
\begin{subfigure}{0.66\columnwidth}
\centering
\includegraphics[width=\columnwidth]{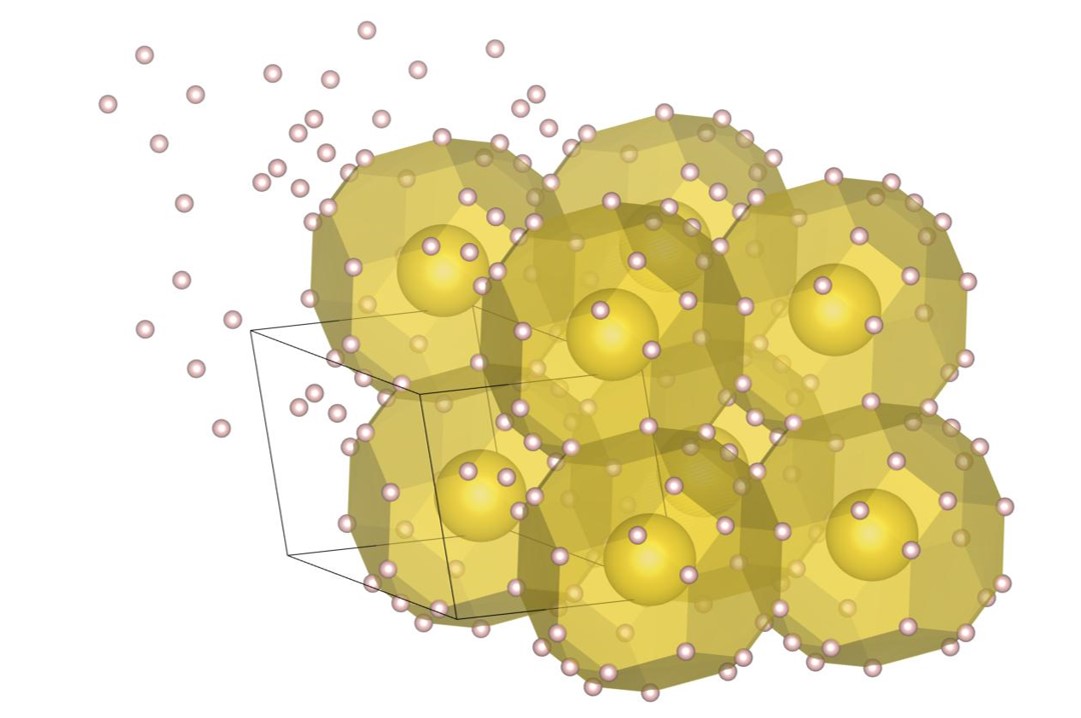}
\caption{$\mathrm{XY_{12}}$ structure}
\label{fig:XY12}
\end{subfigure}
\begin{subfigure}{0.66\columnwidth}
\centering
\includegraphics[width=\columnwidth]{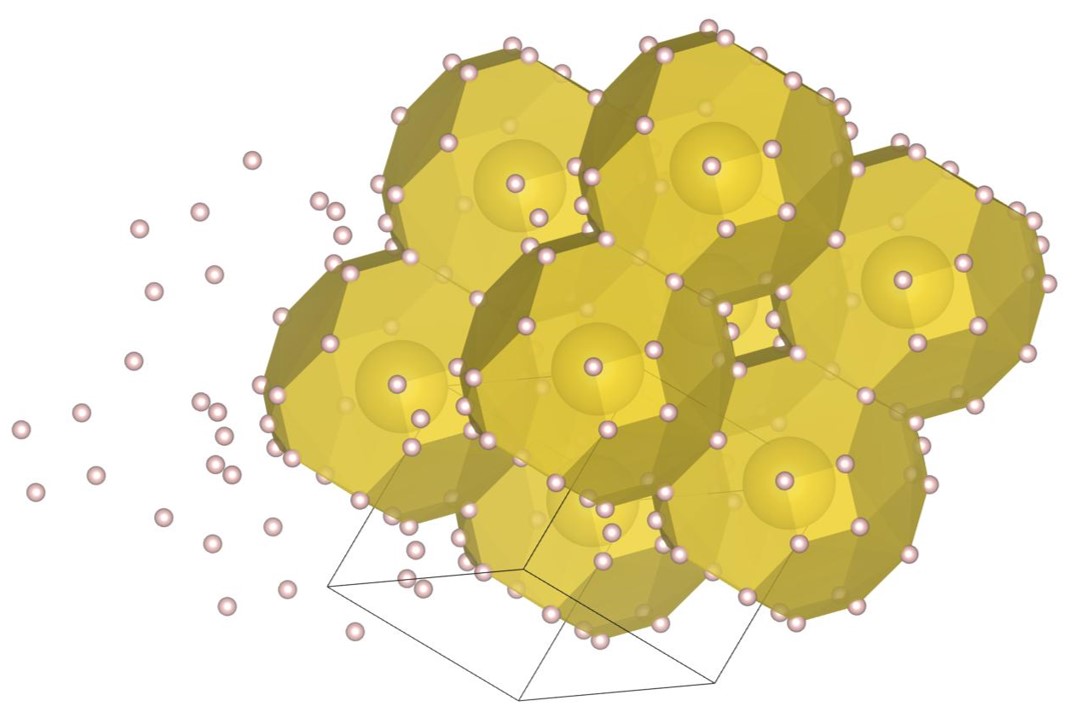}
\caption{(12-1) structure}
\label{fig:12-1}
\end{subfigure}
\begin{subfigure}{0.66\columnwidth}
\centering
\includegraphics[width=\columnwidth]{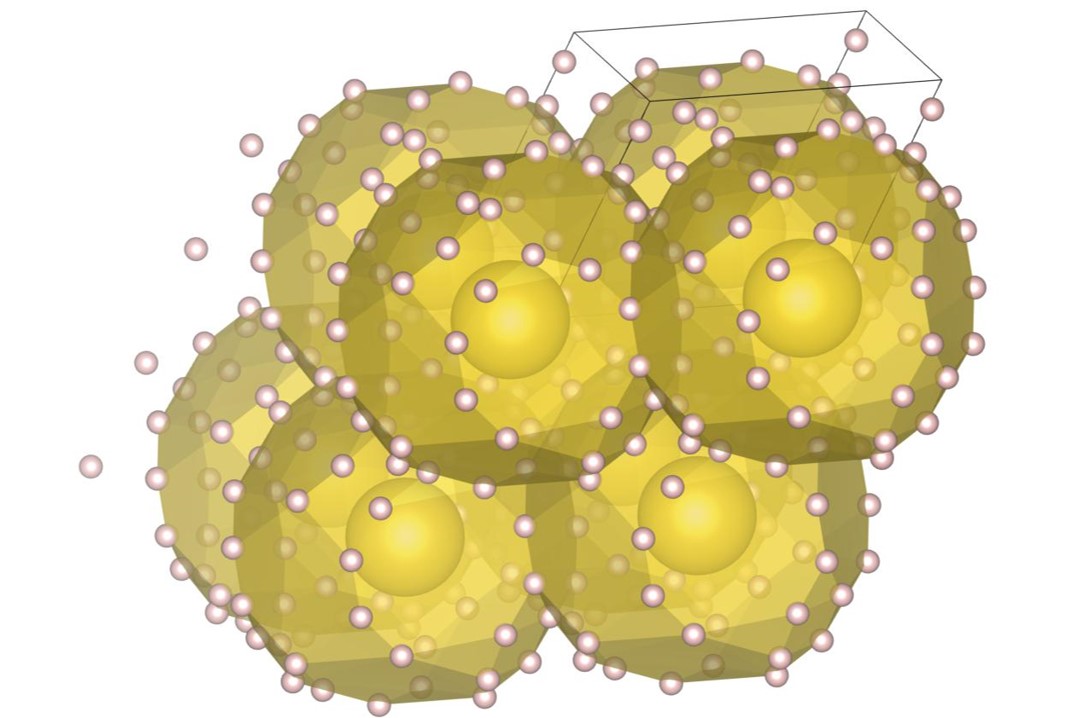}
\caption{(22-1) structure}
\label{fig:22-1}
\end{subfigure}
\begin{subfigure}{0.66\columnwidth}
\centering
\includegraphics[width=\columnwidth]{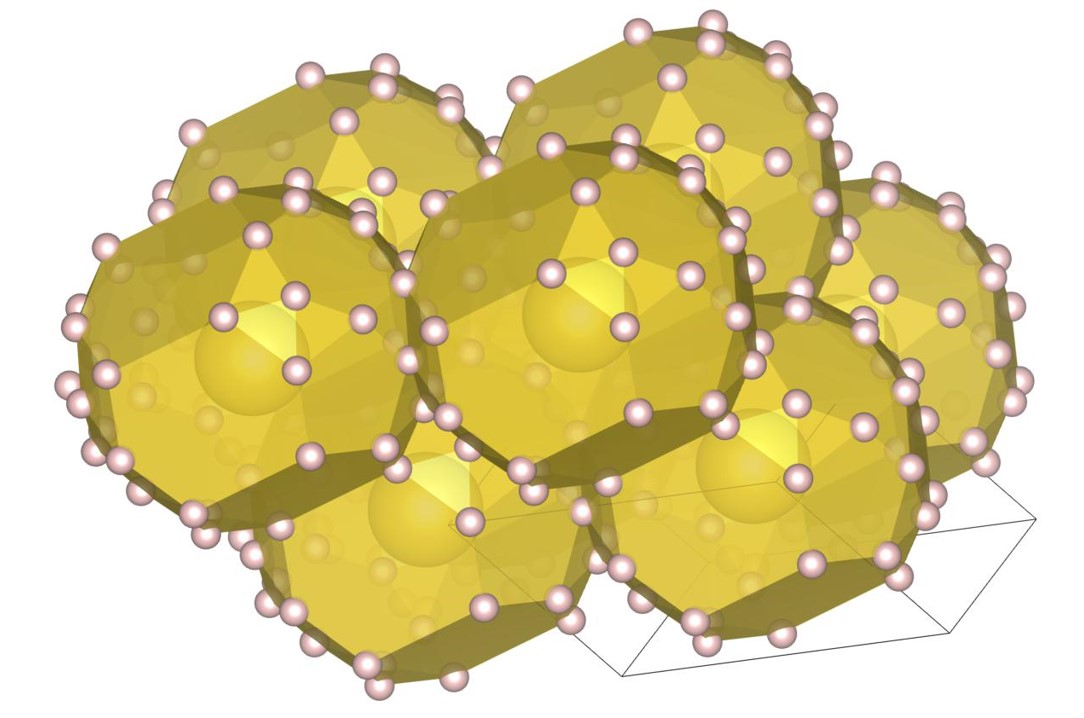}
\caption{(20-1) structure}
\label{fig:20-1}
\end{subfigure}
\begin{subfigure}{0.66\columnwidth}
\centering
\includegraphics[width=\columnwidth]{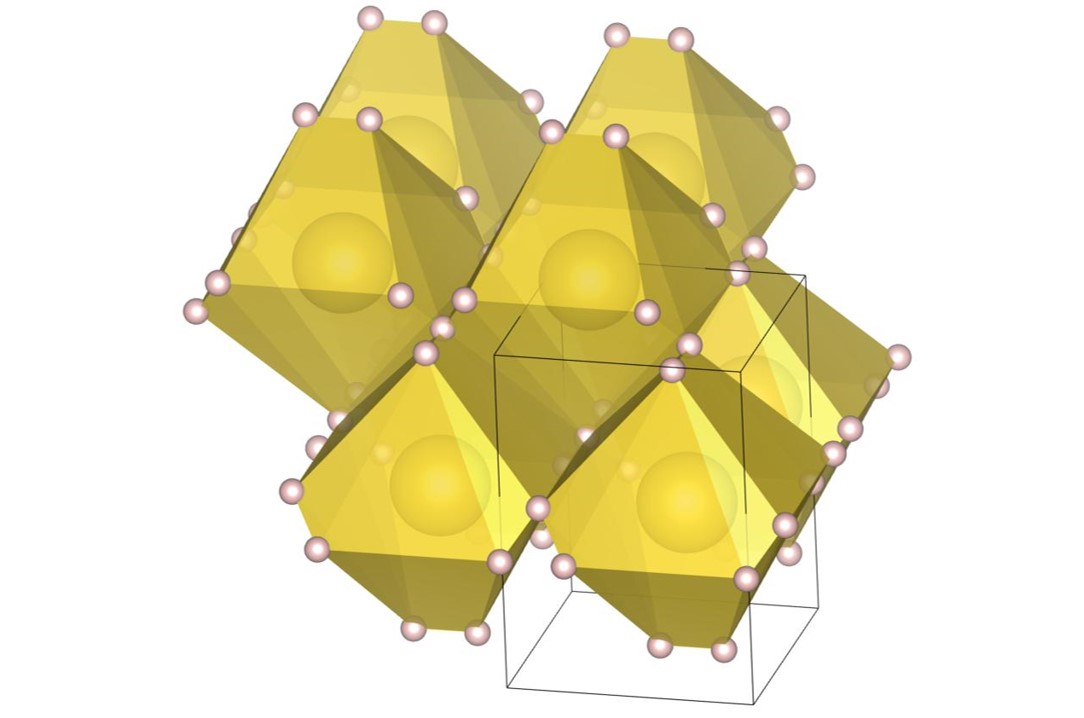}
\caption{(8-2) structure}
\label{fig:8-2}
\end{subfigure}
\begin{subfigure}{0.66\columnwidth}
\centering
\includegraphics[width=\columnwidth]{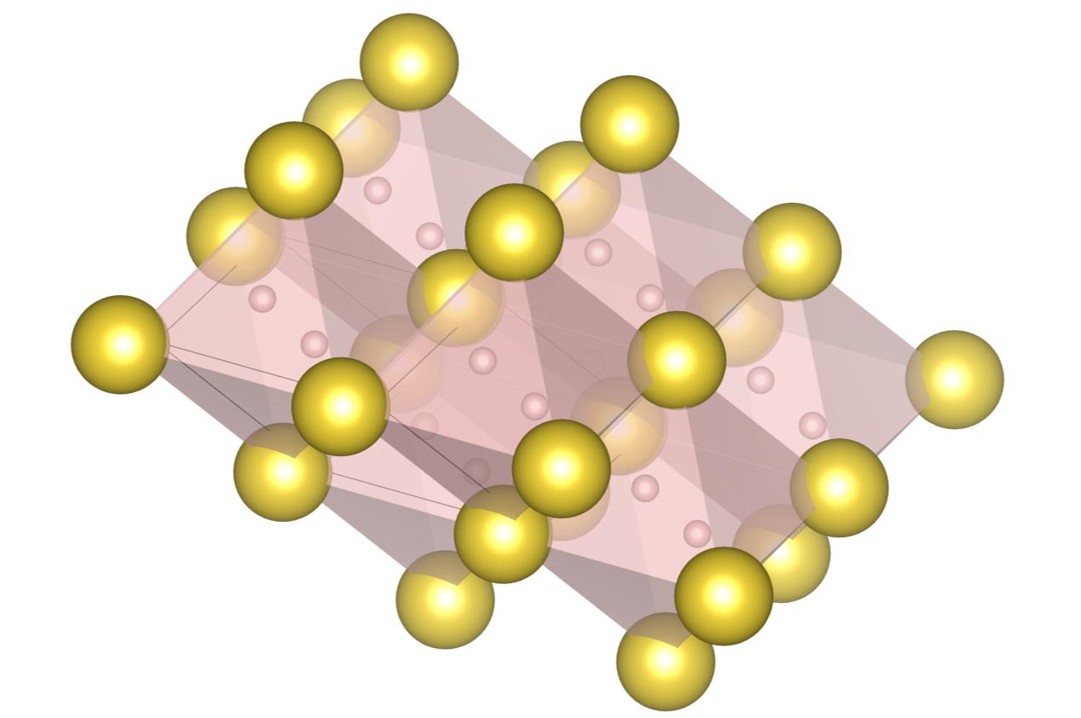}
\caption{(2-1) structure}
\label{fig:2-1}
\end{subfigure}
\caption{The six new DBSP in the radius ratio of $0.20 \le \alpha \le 0.40$. (a) The $\mathrm{XY_{12}}$ structure appears on the phase diagram at $\alpha=0.20$, and the packing fraction is $0.811567$. The unit cell contains 12 small spheres and 1 large sphere. (b) The (12-1) structure appears on the phase diagram at $\alpha=0.203$ and the packing fraction is $0.811932$. It can be understood as an extended $\mathrm{XY_{12}}$ structure. The unit cell contains 12 small spheres and 1 large sphere. It is not plotted on the phase diagram of Fig \ref{fig:smallPhaseDiagram}, because it appears in a very narrow region. (c) The (22-1) structure appears on the phase diagram at $\alpha=0.20$ and $0.203$. The packing fraction is $0.813313$ and $0.809182$, respectively. The unit cell contains 22 small spheres and 1 large sphere. (d) The (20-1) structure appears on the phase diagram at $\alpha=0.26$, and the packing fraction is $0.785154$. The unit cell contains 20 small spheres and 1 large sphere. (e) The (8-2) structure appears on the phase diagram at $\alpha = 0.26$, and the packing fraction is $0.788345$. The unit cell contains 8 small spheres and 2 large spheres. (f) The (2-1) structure appears on the phase diagram at $\alpha = 0.28$, and the packing fraction is $0.765223$. The unit cell contains 2 small spheres and 1 large sphere.}
\label{fig:smallNewDensestPackings}
\end{figure*}

For $0.20 \le \alpha \le 0.40$, the exhaustive search is conducted for 11 radius ratios of 0.20, 0.22, 0.24, 0.26, 0.28, 0.30, 0.32, 0.34, 0.36, 0.38, and 0.40. In order to confirm that the (12-1) structure appears on the phase diagram, the $\mathrm{XY_{12}}$ structure, the $\mathrm{XY_{11}}$ structure, and the (22-1) structure are re-optimized at $\alpha=0.203$. The number of spheres in the unit cell is set between 6 and 24. As a result, we have identified the new six putative DBSP: the $\mathrm{XY_{12}}$ structure, the (22-1) structure, the (12-1) structure, the (20-1) structure, the (8-2) structure, and the (2-1) structure. Thereby, we update the phase diagram. For the case of 12 or fewer spheres in the unit cell, our phase diagram is consistent with that of the previous study \cite{PhysRevE.85.021130} with a small correction. The putative DBSP already discovered are the $\mathrm{XY}$ structure~\cite{PhysRevE.85.021130}, the $\mathrm{XY_2}$ structure~\cite{PhysRevE.85.021130}, the $\mathrm{XY_4}$ structure~\cite{PhysRevE.85.021130}, the $\mathrm{XY_8}$ structure~\cite{PhysRevE.85.021130}, the $\mathrm{XY_{10}}$ structure~\cite{PhysRevE.85.021130}, the $\mathrm{XY_{11}}$ structure~\cite{PhysRevE.85.021130}, the (10-1) structure~\cite{PhysRevE.85.021130}, the (11-1) structure~\cite{PhysRevE.85.021130}, and the (6-1) structure~\cite{PhysRevE.85.021130}. On the phase diagram of Fig.~\ref{fig:smallPhaseDiagram}, the $\mathrm{XY_8}$ structure, the $\mathrm{XY_{11}}$ structure, and the (11-1) are not plotted, because they appear in the narrow region.

In this subsection, firstly we present the small modifications of the HST phase diagram for the case of 12 or fewer spheres in the unit cell. Secondly, we present the newly found six putative DBSP. The four of them have extended unit cells compared to the previous study~\cite{PhysRevE.85.021130}.

\subsubsection{Modifications of HST phase diagram}
In this radius range, if the number of small spheres is little enough, all of the small spheres can penetrate into the tetrahedral and octahedral sites in the fcc densest structure constituted by large spheres. If so, the packing fraction can be calculated as Eq.~(\ref{eq:analytical-solution-of-XYn}), as discussed in Sec.~\ref{sec:accuracy-of-filling-rate} and the phase separation into more than one $\mathrm{XY_n}$ structure is not intrinsic. Phase separation into several $\mathrm{XY_n}$ structures only complicates the phase diagram. Therefore, in the HST phase diagram, the densest phase separation is represented by the $\mathrm{XY_n}$ structure which contains the largest number of small spheres. In other words, at each $\left(\alpha, x \right)$ only the $\mathrm{XY_n}$ structure which contains the largest number of small spheres is written on the HST phase diagram. Nevertheless, the HST phase diagram shows that the phase separation into the $\mathrm{XY_{10}}$ structure and the $\mathrm{XY_4}$ structure is the densest in the area of $0.233 \le \alpha \le 0.245$ and $4/5 \le x \le 10/11$. The HST phase diagram also shows that the phase separation into the $\mathrm{XY_4}$ structure and the $\mathrm{XY_2}$ structure is the densest in the area of $0.258 \le \alpha \le 0.264$ and $2/3 \le x \le 4/5$. Furthermore, the HST phase diagram shows that the phase separation into the $\mathrm{XY_2}$ structure and the $\mathrm{XY}$ structure is the densest in the area of $0.275 \le \alpha \le 0.278$ and $1/2 \le x \le 2/3$. These phase separations are inconsistent with their representation rule because in their rule at each $\left(\alpha, x \right)$, only one $\mathrm{XY_n}$ structure which contains the largest number of small spheres is represented on the HST phase diagram.

Our calculation resolves these contradictions. At $\alpha=0.24$ and $0.26$, the (10-1) structure appears on our phase diagram. It is a distorted and expanded $\mathrm{XY_{10}}$ structure. Table~\ref{table:small-filling-factors} shows the packing fractions of the (10-1) structure at several radius ratios. We found that at $\alpha = 0.239$, $0.240$, and $0.242$, the (10-1) structure becomes an expanded $\mathrm{XY_{10}}$ structure whose symmetry is kept $Fm\overline{3}m$. In the radius range of $0.225 \le \alpha \le 0.233$ and $0.245 \le \alpha \le 0.270$, our result is consistent with that of the Hopkins and coworkers \cite{PhysRevE.85.021130} with a small correction. However, our result indicates that in the radius range of $0.233 \le \alpha \le 0.245$, the $\mathrm{XY_{10}}$ structure on the HST phase diagram has to be replaced for the (10-1) structure. At $\alpha=0.26$, the (8-2) structure appears on our phase diagram. It is a distorted and expanded $\mathrm{XY_{4}}$ structure. Table~\ref{table:small-filling-factors} also shows the packing fractions of the (8-2) structure at several radius ratios. The result indicates that in the radius range of $0.258 \le \alpha \le 0.264$, the $\mathrm{XY_{4}}$ structure on the HST phase diagram has to be replaced for the (8-2) structure. Finally, the (2-1) structure appears on our phase diagram at $\alpha=0.28$. It is a distorted and expanded $\mathrm{XY_{2}}$ structure. Table~\ref{table:small-filling-factors} also shows the packing fractions of the (2-1) structure at several radius ratios. The result indicates that in the radius range of $0.275 \le \alpha \le 0.278$, the $\mathrm{XY_{2}}$ structure on the HST phase diagram has to be replaced for the (2-1) structure. The appearance of the (2-1) structure at $\alpha=0.28$ is inconsistent with the HST phase diagram. The (8-2) structure and the (2-1) structure are the newly found structures, so they are discussed again in the next subsubsection.

\subsubsection{New putative densest sphere packings}

In this subsubsection, we introduce the newly found six putative DBSP. The four of them have extended unit cells compared to the previous study~\cite{PhysRevE.85.021130}.

The $\mathrm{XY_{12}}$ structure shown in Fig.~\ref{fig:smallNewDensestPackings}(a) appears on the phase diagram at $\alpha=0.20$, and the packing fraction is $0.811567$. All of the small spheres penetrate into the tetrahedral and octahedral sites in the fcc densest structure constituted by large spheres. An tetrahedral site is occupied by one small sphere and an octahedral site is occupied by ten small spheres.

The (12-1) structure shown in Fig.~\ref{fig:smallNewDensestPackings}(b) appears on the phase diagram at $\alpha=0.203$, and the packing fraction is $0.811932$. The structure can be understood as an extended $\mathrm{XY_{12}}$ structure in one direction. An extended octahedral site is occupied by a rectangle which consists of eight small spheres with two small spheres inside the sides of the rectangle. It is not plotted on the phase diagram of Fig.~\ref{fig:smallPhaseDiagram}, because it appears on the phase diagram in a very narrow region.

The (22-1) structure shown in Fig.~\ref{fig:smallNewDensestPackings}(c) appears on the phase diagram at $\alpha=0.20$ and $0.203$. The packing fraction is $0.813313$ and $0.809182$, respectively. The structure can be understood as a distorted and expanded $\mathrm{XY_{22}}$ structure. We have confirmed that at $\alpha=0.16$, the (22-1) structure becomes the $\mathrm{XY_{22}}$ structure.

The (20-1) structure shown in Fig.~\ref{fig:smallNewDensestPackings}(d) appears on the phase diagram at $\alpha=0.26$, and the packing fraction is $0.785154$. The large spheres constitute a distorted face-centered orthohombic lattice. The structure can be understood as a clathrate structure. In this radius range, most of the structures, e.g., $\mathrm{XY_{12}}$, (22-1), and (10-1), can be regarded as clathrate structure. 

The (8-2) structure shown in Fig.~\ref{fig:smallNewDensestPackings}(e) appears on the phase diagram at $\alpha = 0.26$, and the packing fraction is $0.788345$. The structure can be understood as a distorted $\mathrm{XY_{4}}$ structure. We also find the expanded $\mathrm{XY_{4}}$ structure whose framework is the fcc structure constituted by large spheres where they do not contact each other. However, the packing fraction of the (8-2) structure is $0.000023$ higher than that of the expanded $\mathrm{XY_{4}}$ structure.

The (2-1) structure shown in Fig.~\ref{fig:smallNewDensestPackings}(f) appears on the phase diagram at $\alpha = 0.28$, and the packing fraction is $0.765223$. The structure can be understood as a distorted $\mathrm{XY_{2}}$ structure.

\subsection{Densest Packings for $0.420 \le \alpha \le 0.500$}
\label{sec:medium-densest-packings}

\begin{figure}
\centering
\begin{subfigure}{0.49\columnwidth}
\centering
\includegraphics[width=\columnwidth]{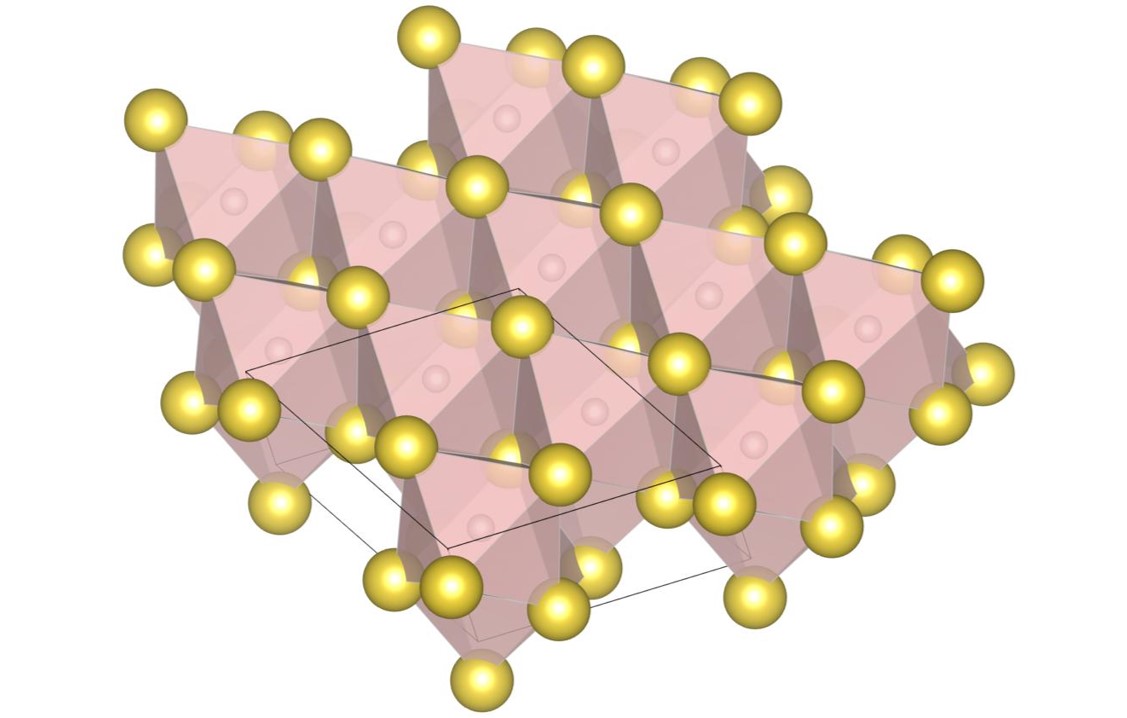}
\caption{(6-6) structure}
\label{fig:6-6}
\end{subfigure}
\begin{subfigure}{0.49\columnwidth}
\centering
\includegraphics[width=\columnwidth]{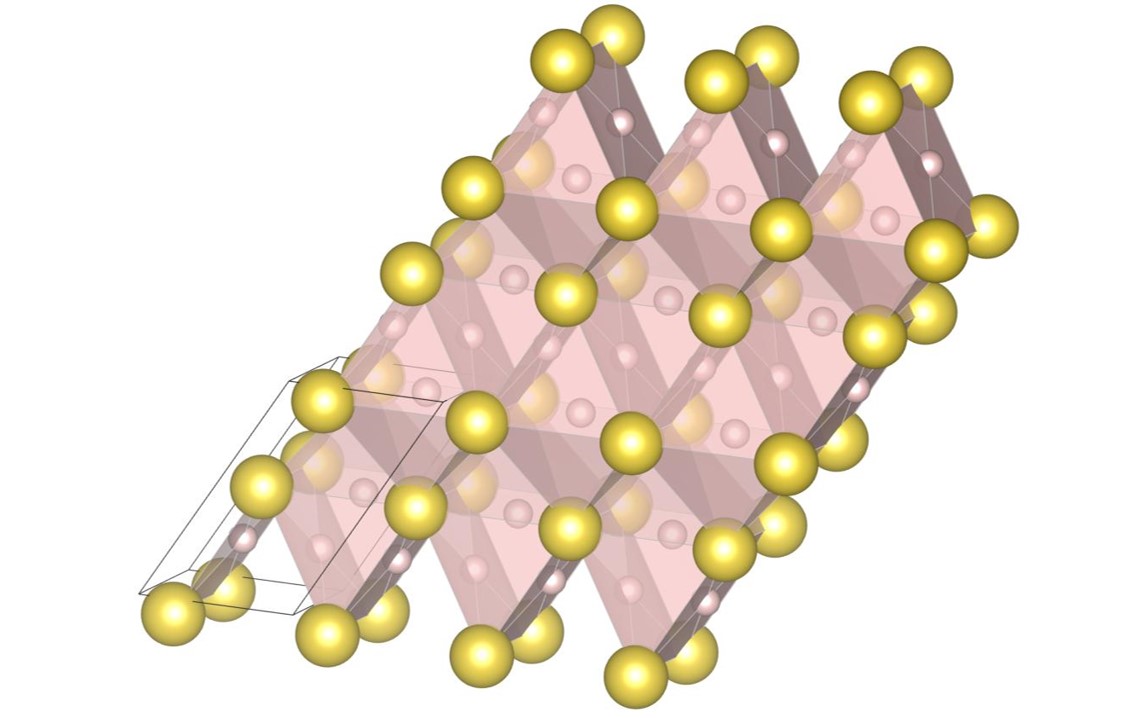}
\caption{$\mathrm{HgBr_2}$ structure}
\label{fig:HgBr2}
\end{subfigure}
\begin{subfigure}{0.49\columnwidth}
\centering
\includegraphics[width=\columnwidth]{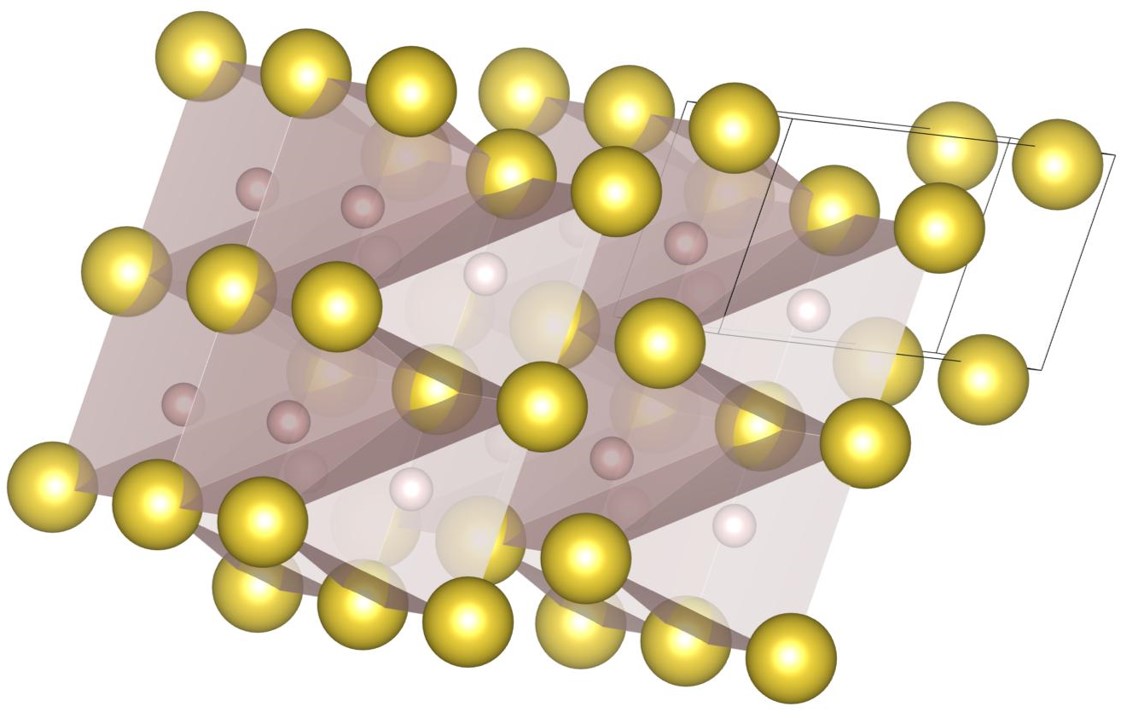}
\caption{(2-2) structure}
\label{fig:2-2}
\end{subfigure}
\begin{subfigure}{0.49\columnwidth}
\centering
\includegraphics[width=\columnwidth]{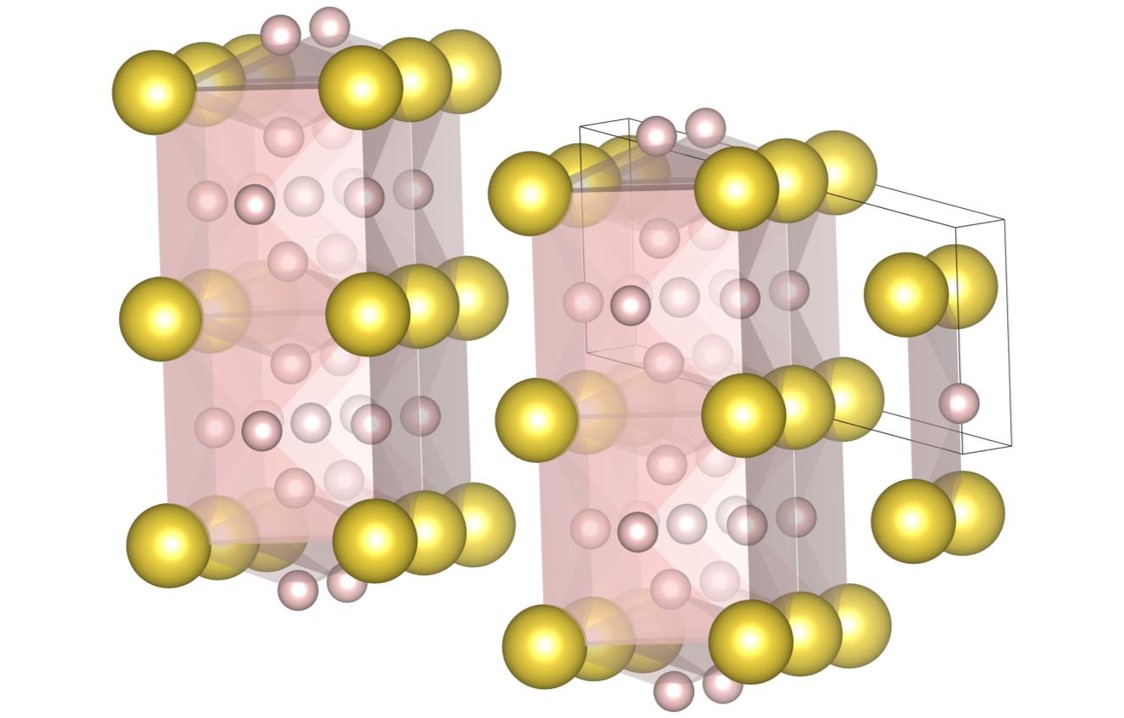}
\caption{(5-2) structure}
\label{fig:5-2}
\end{subfigure}
\begin{subfigure}{\columnwidth}
\centering
\includegraphics[width=0.65\columnwidth]{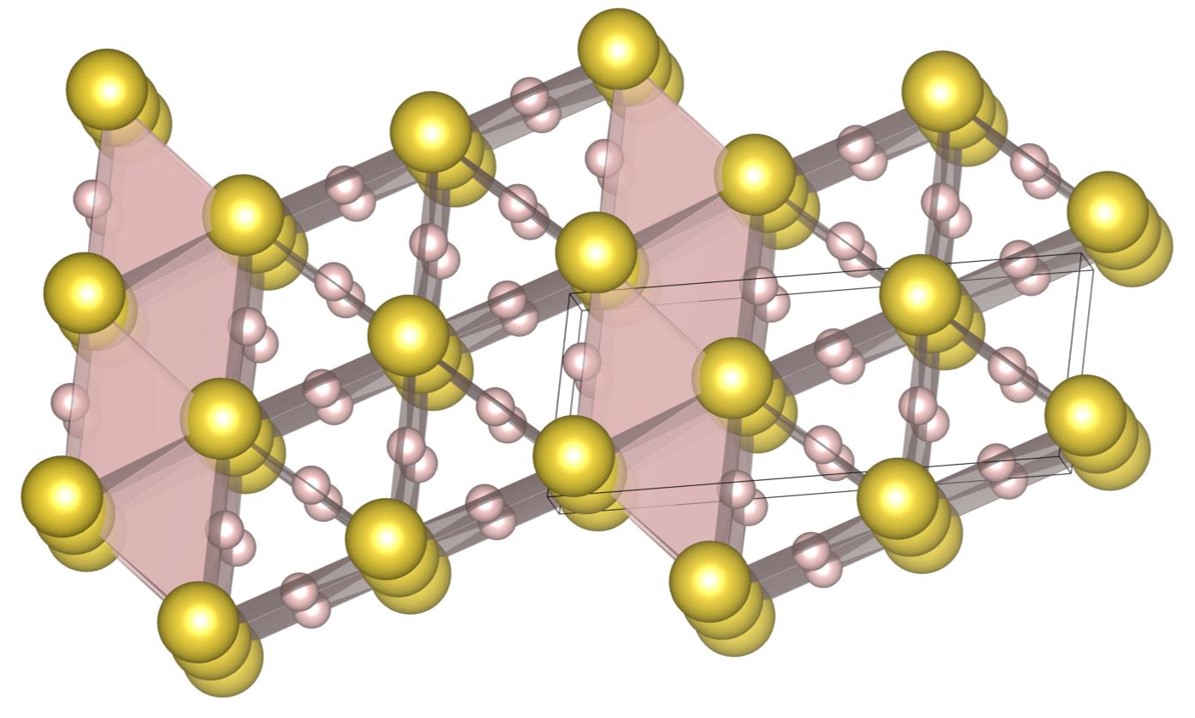}
\caption{(7-3) structure}
\label{fig:7-3}
\end{subfigure}
\caption{The five putative DBSP already identified on the HST phase diagram in the radius ratio of $0.420 \le \alpha \le 0.500$~\cite{PhysRevE.85.021130}. The other two putative DBSP are shown in Fig.~\ref{fig:AuTe2-and-4-2}.}
\label{fig:mediumKnownDensestStructures}
\end{figure}
\begin{figure}
\centering
\begin{subfigure}{0.49\columnwidth}
\centering
\includegraphics[width=\columnwidth]{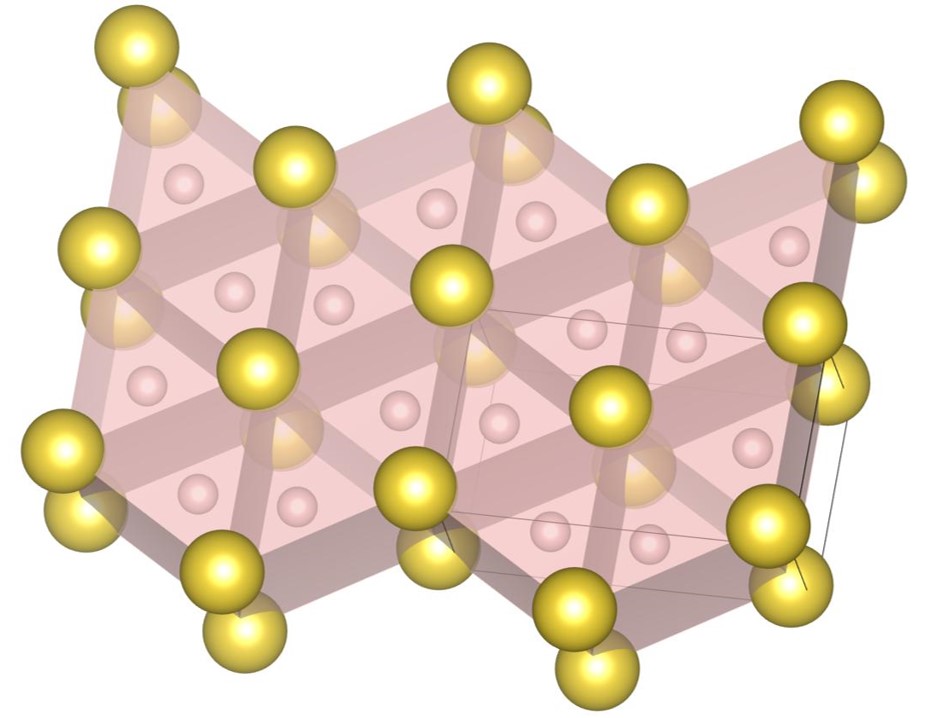}
\caption{}
\label{fig:0.48-AuTe2}
\end{subfigure}
\begin{subfigure}{0.49\columnwidth}
\centering
\includegraphics[width=\columnwidth]{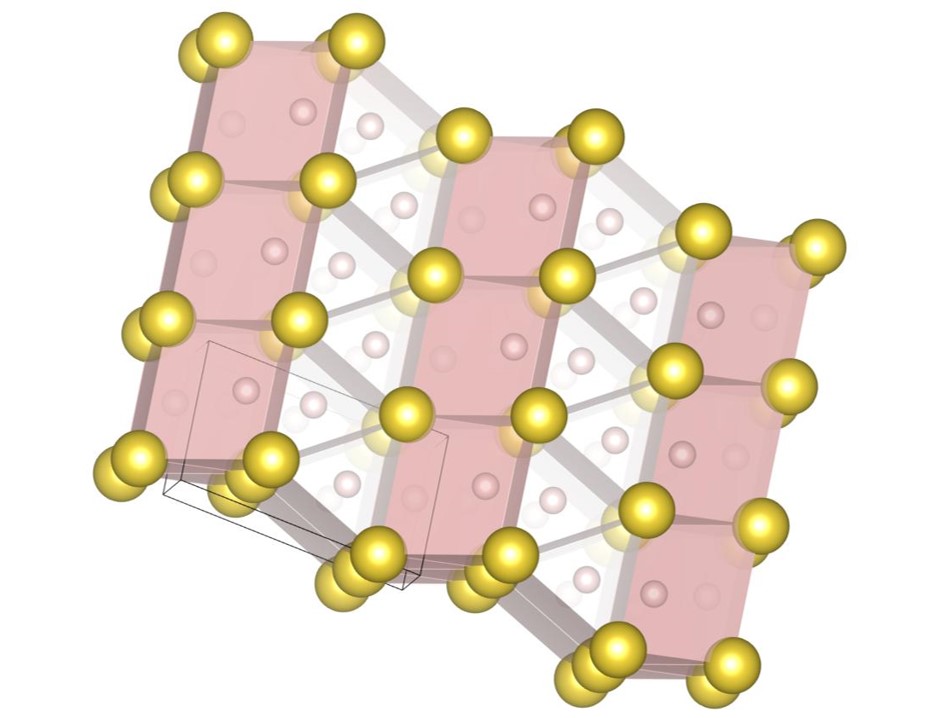}
\caption{}
\label{fig:0.48-4-2}
\end{subfigure}
\caption{(a) The $\mathrm{AuTe_2}$ structure~\cite{PhysRevE.79.046714, PhysRevE.85.021130} and (b) the (4-2) structure~\cite{PhysRevE.85.021130}. In our calculations, it turns out that the packing fractions of the $\mathrm{AuTe_2}$ structure is equal to that of the (4-2) structure.}
\label{fig:AuTe2-and-4-2}
\end{figure}
\begin{figure*}
\centering
\begin{subfigure}{0.66\columnwidth}
\includegraphics[width=\columnwidth]{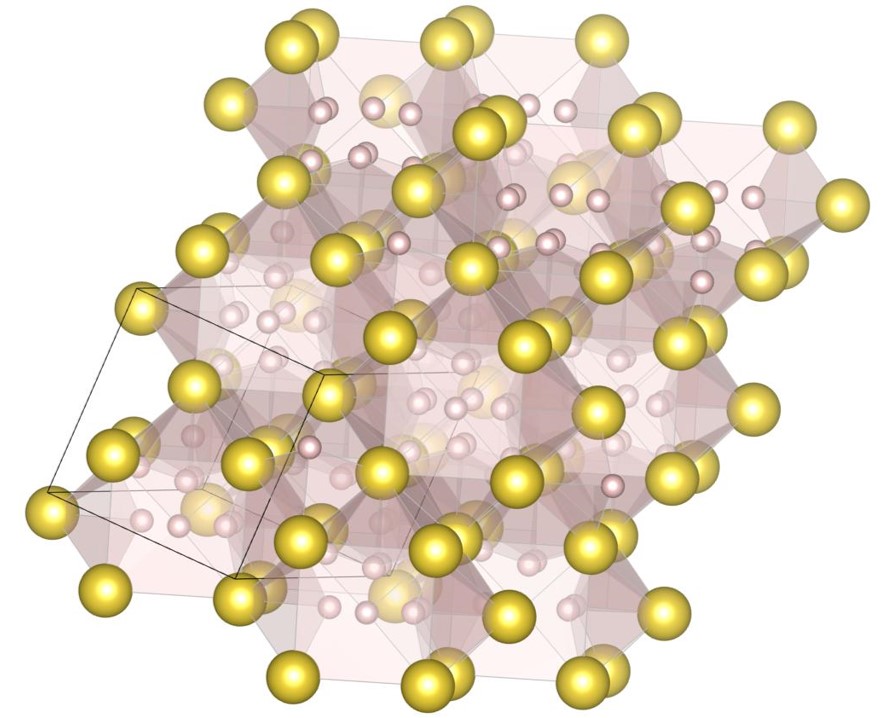}
\caption{(14-5) structure}
\label{fig:14-5}
\end{subfigure}
\begin{subfigure}{0.66\columnwidth}
\centering
\includegraphics[width=\columnwidth]{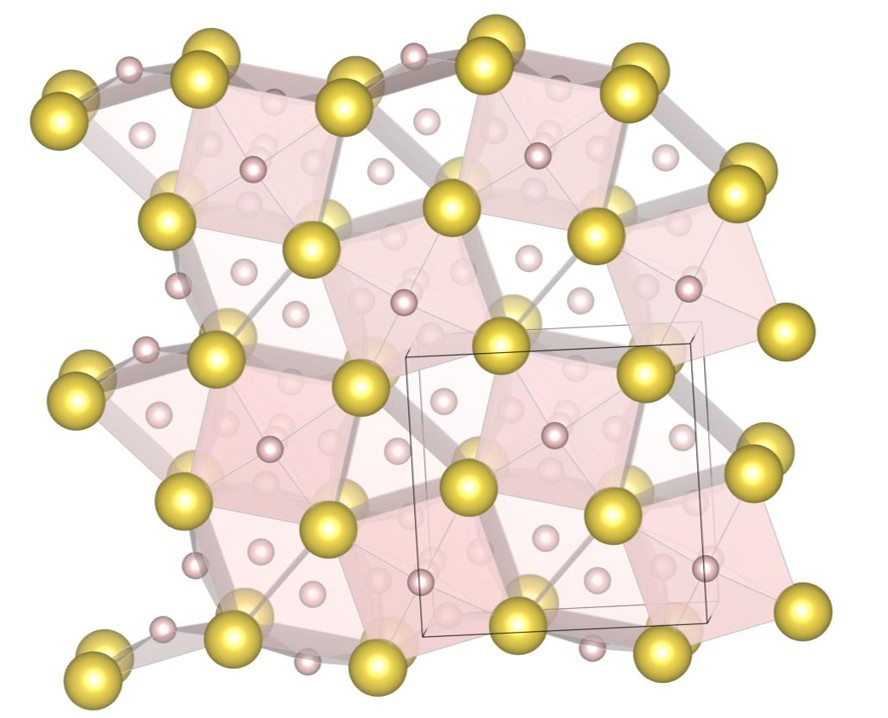}
\caption{(16-4) structure}
\label{fig:16-4}
\end{subfigure}
\begin{subfigure}{0.66\columnwidth}
\centering
\includegraphics[width=\columnwidth]{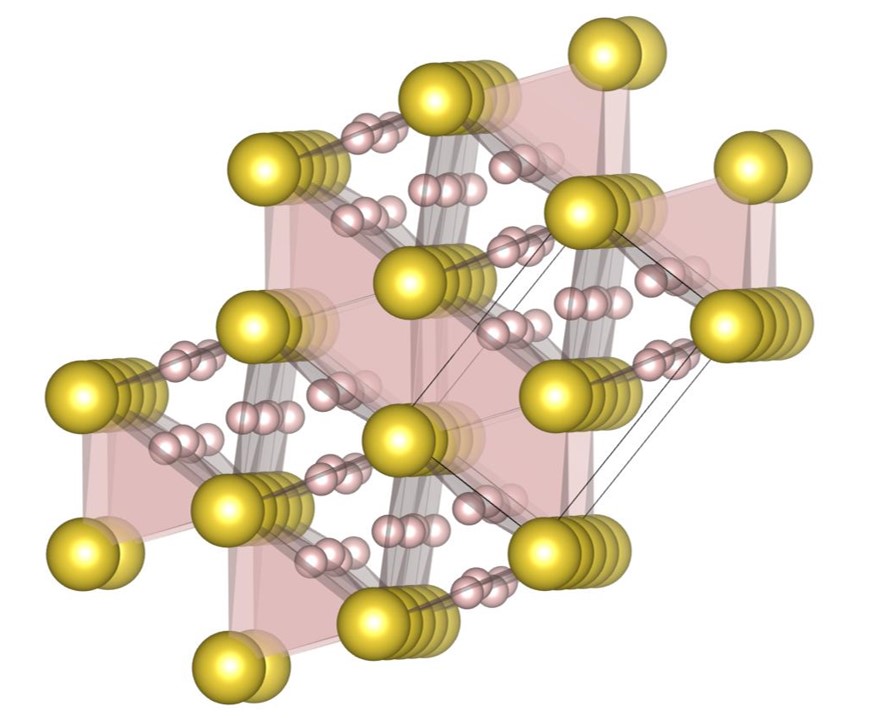}
\caption{(8-4) structure}
\label{fig:8-4}
\end{subfigure}
\begin{subfigure}{0.66\columnwidth}
\centering
\includegraphics[width=\columnwidth]{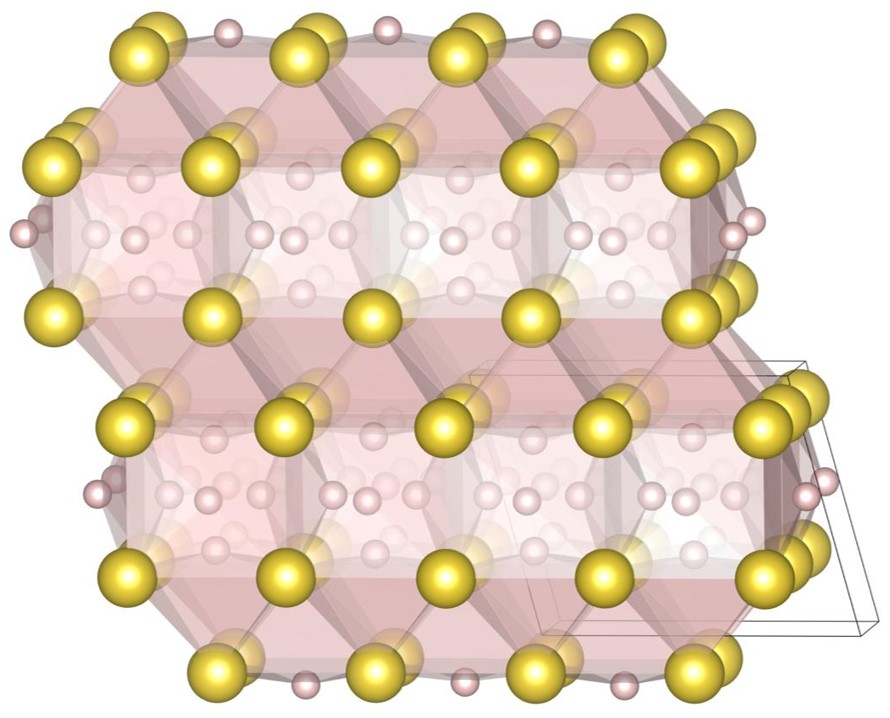}
\caption{(10-4) structure}
\label{fig:10-4}
\end{subfigure}
\begin{subfigure}{0.66\columnwidth}
\centering
\includegraphics[width=\columnwidth]{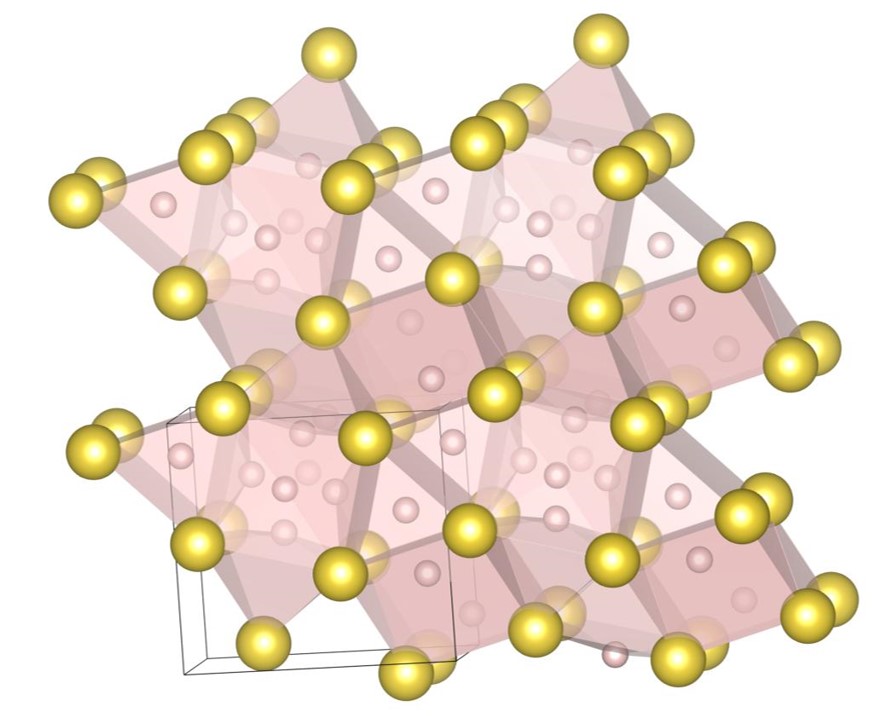}
\caption{(9-4) structure}
\label{fig:9-4}
\end{subfigure}
\caption{The five new putative DBSP for the radius ratio of $0.420 \le \alpha \le 0.500$. (a) The (14-5) structure appears on the phase diagram at the three radius ratios of $\alpha = 0.440$, $0.445$, and $0.450$. The packing fractions are $0.764247$, $0.765259$, and $0.759972$, respectively. The unit cell contains 14 small spheres and 5 large spheres. (b) The (16-4) structure appears on the phase diagram at the three radius ratios of $\alpha = 0.450$, $0.455$, and $0.460$. The packing fractions are $0.760246$, $0.759629$, and $0.759276$, respectively. The unit cell contains 16 small spheres and 4 large spheres. (c) The (8-4) structure appears on the phase diagram at $\alpha = 0.450$ and $0.465$. The packing fraction is $0.758885$ and $0.755714$, respectively. The unit cell contains 8 small spheres and 4 large spheres. The structure is isomorphic to the $\mathrm{HgBr_2}$ structure. (d) The (10-4) structure appears on the phase diagram at two radius ratios of $\alpha = 0.480$ and $0.481$. The packing fractions are $0.746216$ and $0.746246$, respectively. The unit cell contains 10 small spheres and 4 large spheres. The structure is isomorphic to the (5-2) structure. (e) The (9-4) structure appears on the phase diagram at the three radial ratios of $\alpha = 0.481$, $0.482$, and $0.483$. The packing fractions are $0.746093$, $0.746452$, and $0.746510$, respectively. The unit cell contains 9 small spheres and 4 large spheres. The structure does not appear on the phase diagram of Fig.~\ref{fig:mediumPhaseDiagram}, because it appears in the narrow region.}
\label{fig:mediumNewDensestPackings}
\end{figure*}

For $0.420 \le \alpha \le 0.500$, the exhaustive search is conducted for 17 radius ratios of 0.420, 0.425, 0.430, 0.435, 0.440, 0.445, 0.450, 0.455, 0.460, 0.465, 0.470, 0.475, 0.480, 0.485, 0.490, 0.495, and 0.500. In order to confirm that the (9-4) structure appears on the phase diagram, the (2-2) structure, the (10-4) structure, the $\mathrm{AuTe_2}$ structure and the (9-4) structure are re-optimized at $\alpha=0.481$, $0.482$, and $0.483$. The number of spheres in the unit cell is set between 12 and 32. As a result, we have identified 5 new putative DBSP: the (14-5) structure, the (16-4) structure, the (8-4) structure, the (10-4) structure, and the (9-4) structure. Thereby, we update the phase diagram. For the case of 12 or fewer spheres in the unit cell, our phase diagram is consistent with that of the previous study~\cite{PhysRevE.85.021130} with a small correction. The putative DBSP already discovered are the (6-6) structure~\cite{PhysRevE.85.021130}, the $\mathrm{HgBr_2}$ structure~\cite{PhysRevE.79.046714, PhysRevE.85.021130}, the (2-2) structure \cite{PhysRevE.85.021130}, the (5-2) structure~\cite{PhysRevE.85.021130}, and the (7-3) structure \cite{PhysRevE.85.021130} shown in Fig.~\ref{fig:mediumKnownDensestStructures}, and the $\mathrm{AuTe_2}$ structure~\cite{PhysRevE.79.046714, PhysRevE.85.021130} shown in Fig.~\ref{fig:AuTe2-and-4-2}.

In this subsection, firstly we present small modifications of the HST phase diagram for the case of 12 or fewer spheres in the unit cell. Secondly, we present the newly found five putative DBSP. The four of them have extended unit cells compared to the previous study~\cite{PhysRevE.85.021130}.

\subsubsection{Modifications of HST phase diagram}

In our calculations, it turns out that the packing fractions of the $\mathrm{AuTe_2}$ structure shown in Fig.~\ref{fig:AuTe2-and-4-2}(a) is equal to that of the (4-2) structure~\cite{PhysRevE.85.021130} shown in Fig.~\ref{fig:AuTe2-and-4-2}(b), at all of the radius ratio. The finding we have is inconsistent with the HST phase diagram. Here we provide a rational explanation below of why they should have the same packing fraction.

Both the $\mathrm{AuTe_2}$ structure and the (4-2) structure are composed of triangular prisms with one small sphere. In the (4-2) structure, the orientation of the triangular prisms changes alternately. The triangular prism is distorted in order to increase the packing fraction. If the small radius is less than 0.5 where the large radius is 1, there are two kinds of triangular prisms. In the (4-2) structure, different triangular prisms are pointing in a different direction. An interface between those two triangular prisms is a square made of large spheres; the length of one side of the square is 2. One of the two small spheres in those two triangular prisms contacts the four large spheres; the small sphere is placed at the hollow in the center of the square. In that case, the triangular prism including the small sphere has two degrees of freedom in its direction, because the small sphere is in the middle of the square. In other words, there is two freedom of the arrangement on how to place the other two large spheres which constitute the triangular prism. These two degrees of freedom correspond to the $\mathrm{AuTe_2}$ structure and the (4-2) structure. The other triangular prism has no freedom of the orientation if the small radius is less than 0.5 because the small sphere is not placed in the middle of the square. In conclusion, both the $\mathrm{AuTe_2}$ structure and the (4-2) structure consist of the same triangular prisms, so they have the same packing fraction.

Finally, on our phase diagram, the phase separation into the $\mathrm{HgBr_2}$ structure and the (7-3) structures appear at the radius ratio of $\alpha = 0.470$. The result is inconsistent with the HST phase diagram.

\subsubsection{New putative densest sphere packings}

In this subsubsection, we introduce the newly found five putative DBSP. The four of them have extended unit cells compared to the previous study~\cite{PhysRevE.85.021130}.

The (14-5) structure shown in Fig.~\ref{fig:mediumNewDensestPackings}(a) appears on the phase diagram at $\alpha = 0.440$, $0.445$, and $0.450$. The packing fractions are $0.764247$, $0.765259$, and $0.759972$, respectively. The structure contains the 14-oligomer structures constituted by small spheres. The local structures are embedded in the gap among large spheres. A 14-oligomer structure consists of a cubic constituted by eight small spheres and six small spheres attached to each side of the cubic.

The (16-4) structure shown in Fig.~\ref{fig:mediumNewDensestPackings}(b) appears on the phase diagram at $\alpha = 0.450$, $0.455$, and $0.460$. The packing fractions are $0.760246$, $0.759629$, and $0.759276$, respectively. The structure consists of $\mathrm{AlB_2}$-type local structure and cubic frameworks constituted by large spheres. A cubic framework is occupied by an octahedron consisting of small spheres.

The (8-4) structure shown in Fig.~\ref{fig:mediumNewDensestPackings}(c) appears on the phase diagram at $\alpha = 0.450$ and $0.465$. The packing fractions are $0.758885$ and $0.755714$, respectively. Table \ref{table:medium-filling-factors} shows the packing fractions of the structure at several radius ratios. The (8-4) structure is isomorphic to the $\mathrm{HgBr_2}$ structure, but as shown in Table~\ref{table:medium-filling-factors}, the packing fractions of the (8-4) structure are higher than those of the $\mathrm{HgBr_2}$ structure at some radius ratios: $\alpha=0.431$, $0.434$, $0.437$, $0.440$, $0.443$, $0.445$, $0.448$, $0.463$, and $0.465$. At $\alpha=0.450$, the packing fraction of the (8-4) structure is only 0.000002 higher than that of $\mathrm{HgBr_2}$; however in the radius range of $0.431 \le \alpha \le 0.450$, the (8-4) structure is denser than $\mathrm{HgBr_2}$ structure. Therefore, it is likely true that the (8-4) structure is denser than the $\mathrm{HgBr_2}$ structure at $\alpha = 0.450$.

The (10-4) structure shown in Fig.~\ref{fig:mediumNewDensestPackings}(d) appears on the phase diagram at $\alpha = 0.480$ and $0.481$. The packing fractions are $0.746216$ and $0.746246$, respectively. The structure is isomorphic to the (5-2) structure, but the packing fractions of (10-4) structure are higher than those of (5-2) structure.

Finally, the (9-4) structure shown in Fig.~\ref{fig:mediumNewDensestPackings}(e) appears on the phase diagram at the three radius ratios of $\alpha = 0.481$, $0.482$, and $0.483$. The packing fractions are $0.746093$, $0.746452$, and $0.746510$, respectively. The structure does not appear on the phase diagram of Fig.~\ref{fig:mediumPhaseDiagram}, because it appears in a very narrow region. The structure contains the (10-4)-type local structure.

\subsection{Densest Packings for $0.52 \le \alpha \le 0.64$}
\label{sec:large-densest-packings}

\begin{figure}
\centering
\begin{subfigure}{0.49\columnwidth}
\centering
\includegraphics[width=\columnwidth]{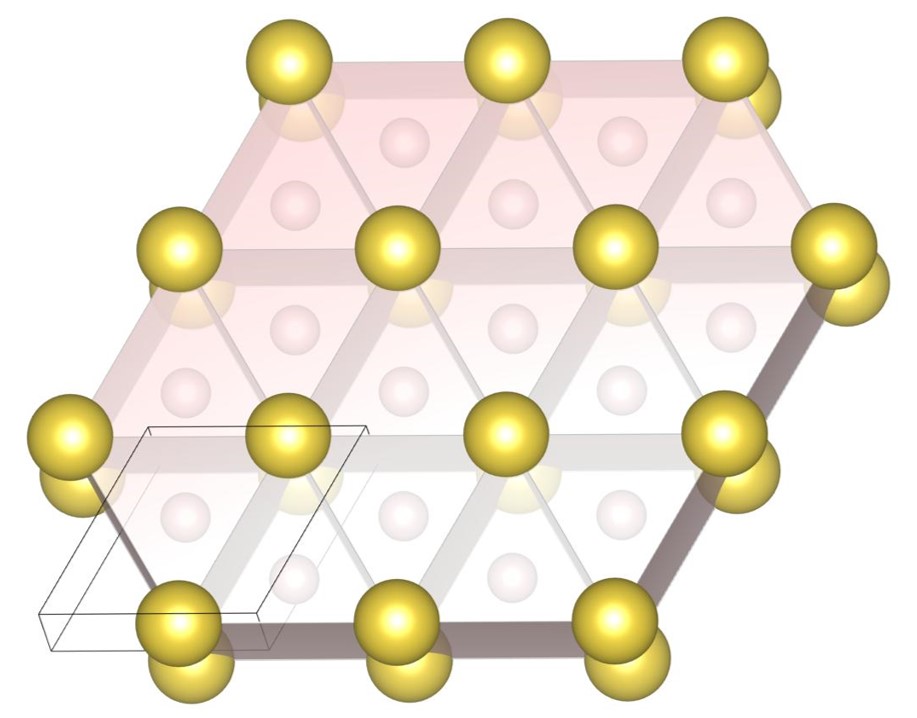}
\caption{$\mathrm{AlB_2}$ structure}
\label{fig:AlB2}
\end{subfigure}
\begin{subfigure}{0.49\columnwidth}
\centering
\includegraphics[width=\columnwidth]{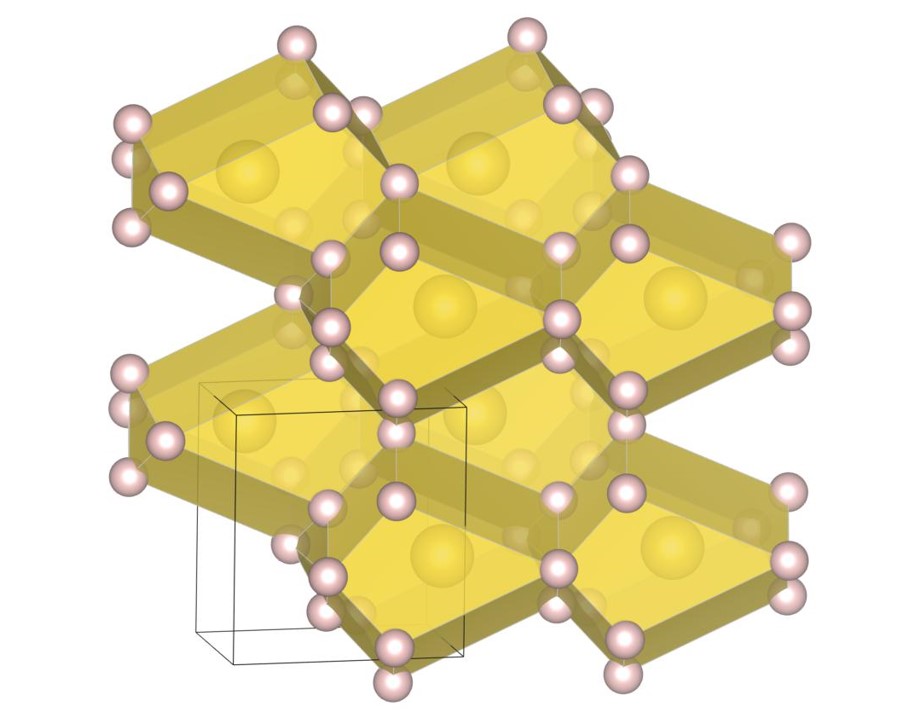}
\caption{$\mathrm{A_3B}$ structure}
\label{fig:A3B}
\end{subfigure}
\caption{The two of the three putative densest binary sphere packings already identified on the HST phase diagram in the radius ratio of $0.52 \le \alpha \le 0.64$~\cite{PhysRevE.85.021130}.}
\label{fig:largeKnownDensestStructures}
\end{figure}
\begin{figure} 
\centering
\includegraphics[width=0.75\columnwidth]{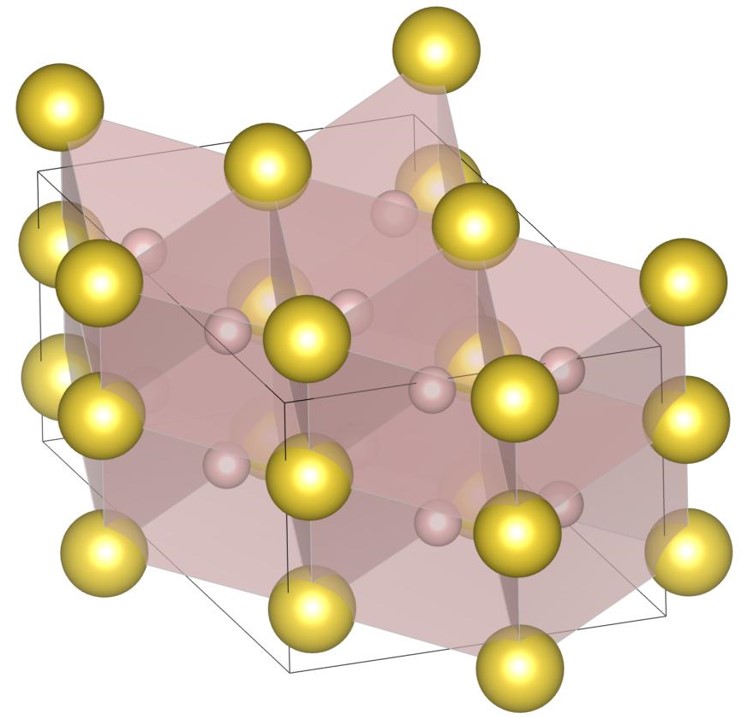}
\caption{The (12-6) structure appears on the phase diagram at $\alpha =0.54$ and $0.56$. The packing fractions are $0.780466$ and $0.779098$, respectively.  The unit cell contains 12 small spheres and 6 large spheres. The structure is a distorted $\mathrm{AlB_2}$ structure as well as the $\mathrm{AuTe_2}$ structure.}
\label{fig:12-6}
\end{figure}

For $0.52 \le \alpha \le 0.64$, the exhaustive search is conducted for 7 radius ratios of 0.52, 0.54, 0.56, 0.58, 0.60, 0.62, and 0.64. The number of spheres in the unit cell is set between 6 and 24. As a result, we have identified new one putative DBSP, named the (12-6) structure. The structure have an extended unit cell compared to the previous study~\cite{PhysRevE.85.021130}. Thereby, we update the phase diagram. For the case of 12 or fewer spheres in the unit cell, our phase diagram is completely consistent with that of the previous study~\cite{PhysRevE.85.021130}. The putative DBSP already discovered are the $\mathrm{AuTe_2}$ structure~\cite{PhysRevE.79.046714, PhysRevE.85.021130} shown in Fig.~\ref{fig:AuTe2-and-4-2}, and the $\mathrm{AlB_2}$ structure \cite{PhysRevE.79.046714, PhysRevE.85.021130} and $\mathrm{A_3B}$ structure~\cite{doi:10.1021/jp206115p, PhysRevE.85.021130} shown in Fig.~\ref{fig:largeKnownDensestStructures}. 

The (12-6) structure shown in Fig.~\ref{fig:12-6} appears on the phase diagram at $\alpha = 0.54$ and $0.56$. The packing fractions are $0.780466$ and $0.779098$, respectively. The structure can be understood as a distorted $\mathrm{AlB_2}$ structure as well as the $\mathrm{AuTe_2}$ structure. At these radius ratios, the $\mathrm{AuTe_2}$ structure and the $\mathrm{AlB_2}$ structure have the same packing fraction; the equivalence indicates that a large unit cell is necessary to realize an optimal distortion for higher packing fraction. At the three radius ratios of $\alpha = 0.58$, $0.60$, and $0.62$, the (12-6) structure and the $\mathrm{AlB_2}$ structure have the same packing fraction; however, if the unit cell is expanded, a distorted $\mathrm{AlB_2}$ structure might appear on the phase diagram. The further investigation will be in future work.

\subsection{Geometric features of densest packings}
\label{sec:Mono-phase_densest_sphere_packings}

The putative DBSP consist of local structures with high packing fraction. For example, $\mathrm{XY_n}$ structure is clearly made up of a combination of local structures with high packing fractions: the tetrahedron and octahedron constituted by large spheres. The same is true for the distorted $\mathrm{XY_n}$ structure such as the (2-1), (8-2), (10-1), (11-1), and (22-1) structures. The same is also true for the (6-6) structure in which small spheres penetrate into the octahedral sites in the distorted hcp densest structure constituted by large spheres. Similarly, the $\mathrm{AlB_2}$ structure, the $\mathrm{AuTe_2}$ structure, and the (12-6) structure consisting of the dense local structure in which a small sphere is embedded in the center of a triangular prism consisting of large spheres.

Some of the putative DBSP are complex structures, especially for $0.42 \le \alpha \le 0.50$. Both the $\mathrm{HgBr_2}$ structure and the (7-3) structure consist of equilateral triangular prisms and parallelepiped hexahedrons. The (10-4) structure consists of parallelepipedal hexahedrons and cubics constituted by large spheres with five small spheres. The (16-4) structure consists of $\mathrm{AlB_2}$-type local structure and cubic frameworks constituted by large spheres with an octahedron made of small spheres. In the (14-5) structure, 14-oligomer structures constituted by small spheres are embedded in the gap among the large spheres. As discussed below, these local structures also appear in the mono-phase densest binary sphere packings that have a lower packing fractions than the densest phase separations. The appearance shows that those local structures are dense.

When space is filled with two kinds of spheres under periodic boundary conditions, the highest packing fraction is achieved by phase separation into two or fewer packing structures. However, in the below discussion, we prohibit a phase separation. Under the restriction, a lot of mono-phase densest sphere packings are obtained.

\begin{figure}
\centering
\begin{subfigure}{0.49\columnwidth}
\includegraphics[width=\columnwidth]{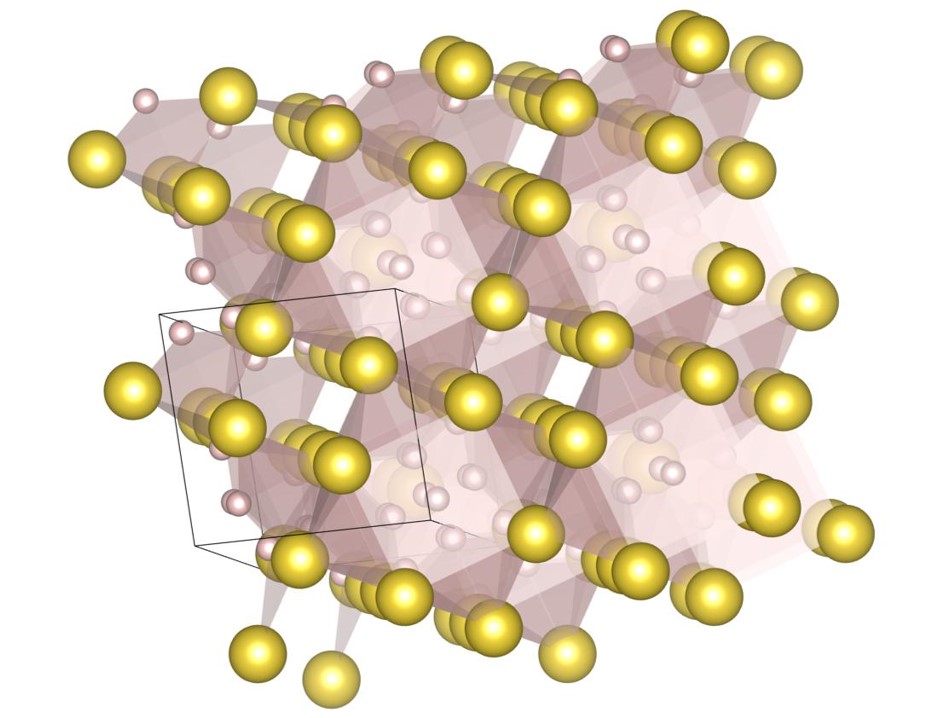}
\caption{(14-8) structure}
\label{fig:14-8}
\end{subfigure}
\begin{subfigure}{0.49\columnwidth}
\centering
\includegraphics[width=\columnwidth]{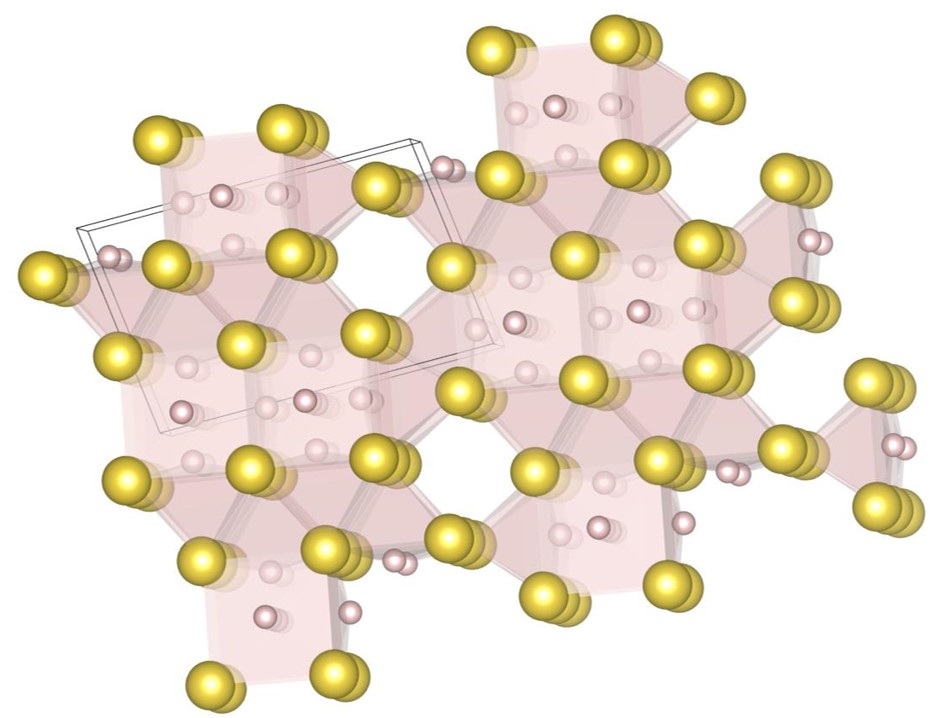}
\caption{(11-6) structure}
\label{fig:11-6}
\end{subfigure}
\caption{(a) The (14-8) structure at $\alpha = 0.44$. It contains local structures of 14-oligomer structure of small spheres embedded in the gap between the large spheres. Compared to the (14-5) structure, there are wasted gaps in it. (b) The (11-6) structure at $\alpha = 0.48$. it contains (10-4)-type local structure. Compared to the (10-4) structure, there are wasted gaps in it.}
\label{fig:14-8-and-11-6}
\end{figure}

Mono-phase densest binary sphere packings also consist of some of the dense local structures. The (14-8) structure is shown in Fig.~\ref{fig:14-8-and-11-6}(a). The structure is the densest at $\left(\alpha, x \right) = \left(0.44, 14/22 \right)$. The unit cell contains 14 small spheres and 8 large spheres. The structure contains 14-oligomer structures, which is the same local structure in the (14-5) structure. The (14-8) structure contains a wasteful gaps compared to the (14-5) structure, but the presence of the (14-5)-type local structure suggests that the local structure is dense. The (11-6) structure is shown in Fig.~\ref{fig:14-8-and-11-6}(b). The structure is the densest at $\left(\alpha, x \right) = \left(0.48, 11/17 \right)$. The unit cell contains 11 small sphere and 6 large spheres. The structure contains the (10-4)-type local structure. The (11-6) structure contains a wasteful gap compared to the (10-4) structure, but the presence of the (10-4)-type local structure indicates that the local structure is dense.

\begin{figure}
\centering
\includegraphics[width=\columnwidth]{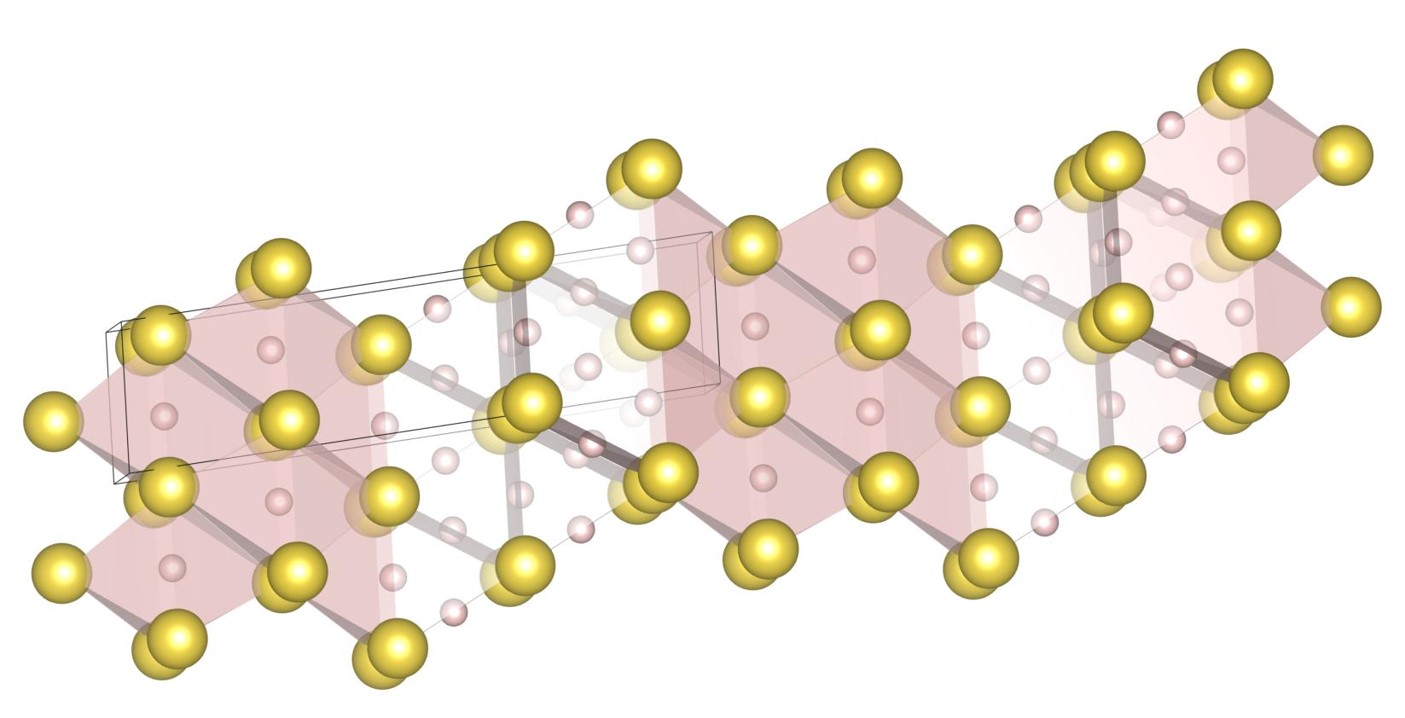}
\caption{The (9-5) structure at $\alpha = 0.46$. This structure consists of  $\mathrm{HgBr_2}$-type triangular prism phase and (6-6)-type stacking phase.}
\label{fig:9-5}
\end{figure}

\begin{figure}
\centering
\includegraphics[width=\columnwidth]{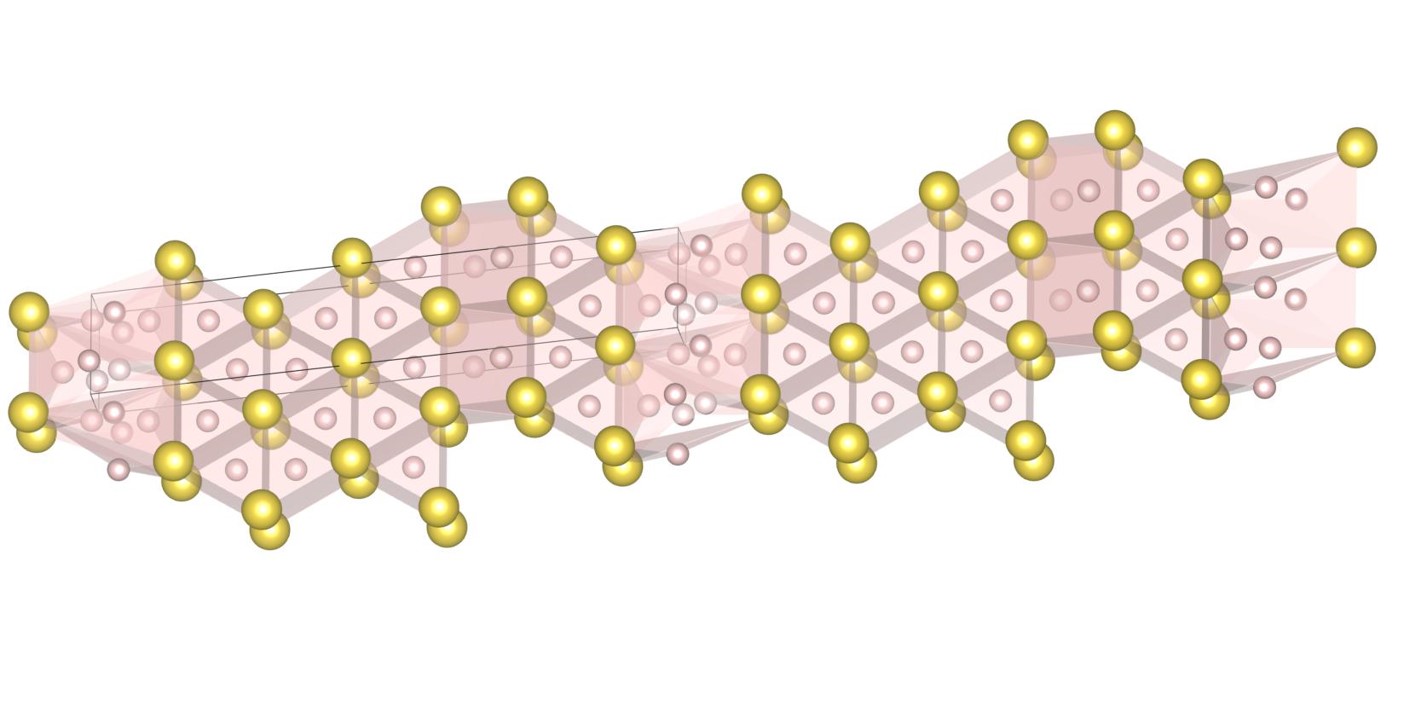}
\caption{The (16-6) structure at $\alpha = 0.56$. This structure consists of $\mathrm{AlB_2}$-type triangular prism phase and the remainder of a small spheres.}
\label{fig:16-6}
\end{figure}

Some of the mono-phase densest sphere packings are long-period structures; those structures are phase-separated into a few local structures in the unit cell. We discuss the two examples as such cases. Firstly, the (9-5) structure is shown in Fig.~\ref{fig:9-5}. The structure is the densest at $\left(\alpha, x \right) = \left(0.46, 9/14 \right)$. The unit cell contains 9 small spheres and 5 large spheres. It consists of the $\mathrm{HgBr_2}$-type triangular prism phase and the (6-6)-type stacking phase. The appearance of the $\mathrm{HgBr_2}$-type local structure indicates that the local structure is dense at $\alpha = 0.46$. Secondly, the (16-6) structure is shown in Fig.~\ref{fig:16-6}. The structure is the densest at $\left(\alpha, x \right) = \left(0.56, 16/22 \right)$. The unit cell contains 16 small spheres and 6 large spheres. It consists of the $\mathrm{AlB_2}$-type triangular prism phase and the remainder of small spheres. The appearance of the $\mathrm{AlB_2}$-type local structure indicates that the local structure is dense at $\alpha = 0.56$. The structural feature of the (16-6) structure indicates that the highest packing fraction at $\left(\alpha, x \right) = \left(0.56, 16/22 \right)$ is achieved by phase separation into the $\mathrm{AlB_2}$ structure and the fcc densest structure constituted by small spheres. In fact, our phase diagram shows that the highest packing fraction at $\left(\alpha, x \right) = \left(0.56, 16/22 \right)$ is achieved by the phase separation into the (12-6) structure and fcc densest packing of small spheres.

Those results show that all of the mono-phase densest binary sphere packings are made of a few local structures with high packing fractions. The concept of dense local structures are very similar to the compact packing; the concept is defined for disc packings \cite{Fernique2020}. The compact packing corresponds to a structure in which all the circles are in contact with the perimeter and the gaps consist only of curvilinear triangles. It has been proved analytically that there are 164 ternary compact packings~\cite{Fernique2020}. A local structure with a high packing fraction in the three-dimensional DBSP may be regarded as an extended concept of the compact packing for three dimensions.

\section{Discussion}
\label{sec:discussion}

In this section, we detail the complexity of local minima, the effectiveness of our method, and the relationship between crystals and densest packings.

\subsection{Complexity of local minima}
\label{sec:difficulty-of-determining-filling-rate}

\begin{figure}
\centering
\begin{subfigure}{0.45\columnwidth}
\includegraphics[width=\columnwidth]{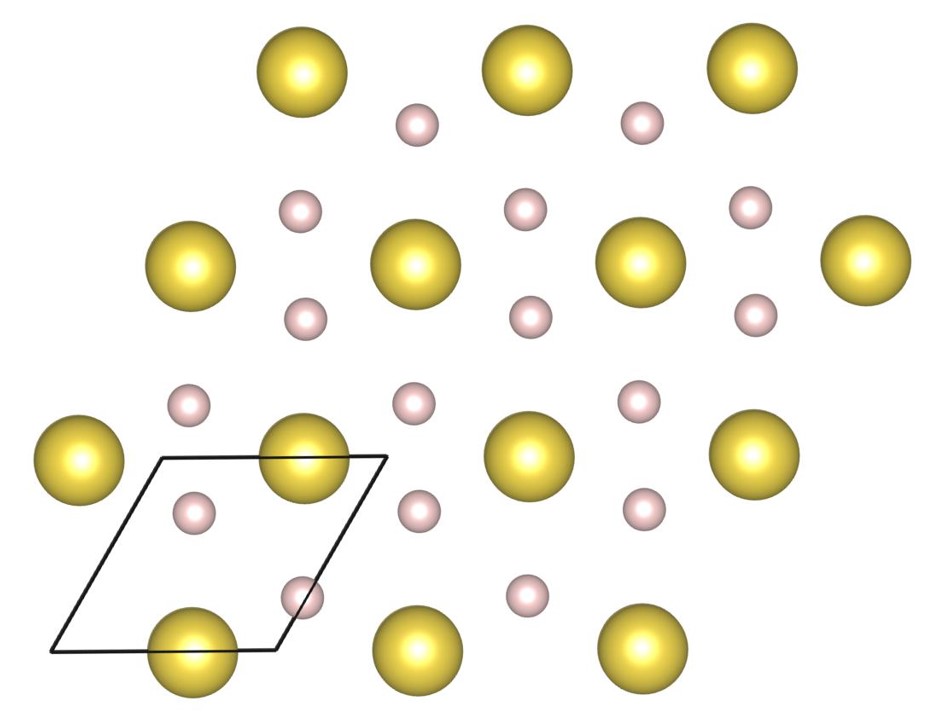}
\caption{The unitcell of $\mathrm{AlB_2}$ structure}
\label{fig:unitcell-AlB2}
\end{subfigure}
\begin{subfigure}{0.45\columnwidth}
\centering
\includegraphics[width=\columnwidth]{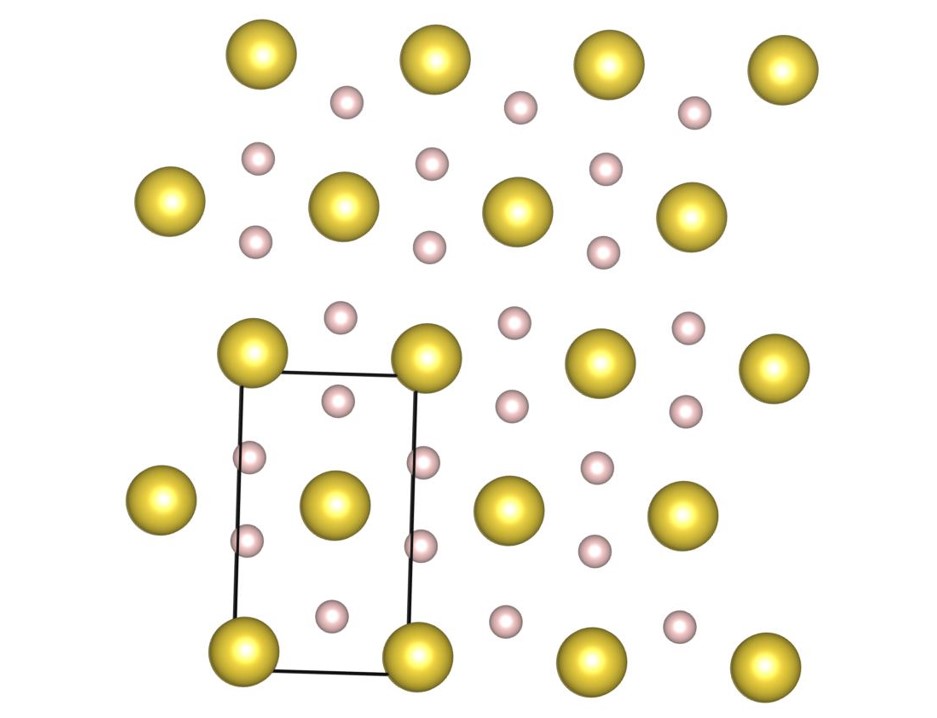}
\caption{The unitcell of $\mathrm{AuTe_2}$ structure}
\label{fig:unitcell-AuTe2}
\end{subfigure}
\caption{A representative example that the unit cell can be reduced by adjusting the distortion. (a) The unit cell of the $\mathrm{AlB_2}$ structure. (b) The unit cell of the $\mathrm{AuTe_2}$ structure. The two large spheres in the unit cell are not in the same plane. The unit cell of the $\mathrm{AuTe_2}$ is consistent with that of $\mathrm{AlB_2}$ structure when the distortion is adjusted.}
\label{fig:unitcell-AlB2-and-AuTe2}
\end{figure}

Sometimes the number of local minima becomes large. In this section, we discuss some representative examples, where the complexity of local minima can be understood. 

Firstly, we analytically compute the packing fraction of the $\mathrm{HgBr_2}$ structure~\cite{PhysRevE.79.046714, PhysRevE.85.021130} shown in Fig.~\ref{fig:mediumKnownDensestStructures}(b). Hereafter, the large radius is set to 1 and the small radius is set to $r$. The structure contains equilateral triangular prisms constituted by large spheres; the length of one side of the equilateral triangle is 2 and the height of the sides of the triangular prism is $h$. Small spheres are placed in the center of each side of the triangular prism and one of them is inserted inside until it touches the other two small spheres. We name the side as $K$. In that case, $h$ can be calculated as
\begin{equation}
h = 2 \sqrt{r^2 + 2r}.
\end{equation}
The $\mathrm{HgBr_2}$ structure also contains parallelepipedal hexahedrons. The bottom surface of the hexahedron is side $K$. One of the large spheres contacts the four large spheres which constitute the side $K$; here we assume that the large sphere does not overlap with the inserted small sphere. In that case, the height of parallelepiped hexahedron $l$ can be calculated as
\begin{equation}
l = \sqrt{3 - \frac{h^2}{4}}.
\end{equation}
The distance $d$ between the large sphere and small sphere can be calculated as
\begin{equation}
d = \frac{\sqrt{3}}{2} - \sqrt{4r^2 - \frac{1}{4}}+ l.
\end{equation}
The $\mathrm{HgBr_2}$ structure consists of two equilateral triangular prism and one parallelepiped hexahedron, so the packing fraction can be calculated as
\begin{equation}
\phi = \frac{2\pi \left(2 + 4 r^3 \right)}{3h \left(\sqrt{3} + \sqrt{3 - \frac{h^2}{4}} \right)}.
\end{equation}
If we set the small sphere radius $r$ as $0.46$, $d$ and $\phi$ are calculated as
\begin{align}
d &\simeq 1.4606503, \\
\phi &\simeq 0.75900904.
\end{align}
On the other hand, the packing fraction calculated by \textit{\textbf{SAMLAI}} is 0.759011. This small difference comes from the slight gap between small spheres; $d$ is slightly larger than 1.46. The gap allows the two small spheres placed in the center of the sides of the equilateral triangular prism to penetrate slightly into the interior of the triangular prism; the distortion allows a slight increase in the packing fraction.

Secondly, the $\mathrm{HgBr_2}$ structure and the (7-3) structure~\cite{PhysRevE.85.021130} shown in Fig.~\ref{fig:mediumKnownDensestStructures}(e) are very similar. These two structures consist of equilateral triangular prisms and parallelepiped hexahedrons. The packing fraction is increased by small spheres penetrating slightly into the interior of triangular prisms. The parallelepiped hexahedron consisting of large spheres is necessary for periodic penetrating. In the region where the radius is getting larger, the (7-3) structure appears on the phase diagram instead of the $\mathrm{HgBr_2}$ structure. This is because the (7-3) structure has twice as many triangular prisms as $\mathrm{HgBr_2}$ structure; the connection of many triangular prisms is necessary for small spheres to penetrate into gaps in triangular prisms without making wasted void between large spheres.

Thirdly, the (10-1) structure~\cite{PhysRevE.85.021130} appears on the phase diagram at the two radius ratios of $\alpha = 0.240$ and $0.260$. Table~\ref{table:small-filling-factors} shows the packing fractions of the (10-1) structure at several radius ratios. The (10-1) structure is an expanded or a distorted $\mathrm{XY_{10}}$ structure so that small spheres can penetrate into the gap constituted by large spheres. The distortion pattern is varied according to the radius ratio. At most radius ratios, the (10-1) structure is distorted, but at $\alpha = 0.239$, $0.240$, and $0.242$, it becomes an enlarged $\mathrm{XY_{10}}$ structure whose symmetry is kept $Fm\overline{3}m$. We are surprised to find that the highly symmetric structure appears again at those radius ratios. This suggests that a delicate balance among the numerous distortion patterns determines which one has the highest packing fraction.

As we have discussed, the packing fraction is increased by distortions. Determinating the maximum packing fraction becomes more difficult when the number of distortion pattern is larger. Besides, the unit cell is also related to the difficulty in determining the highest packing fraction.

Most of the densest packings are distorted; however, if the distortions are adjusted, some structures become highly symmetric and their unit cell could be reduced. For example, the unit cell of the $\mathrm{AuTe_2}$ structure can be reduced to that of the $\mathrm{AlB_2}$ structure \cite{PhysRevE.79.046714, PhysRevE.85.021130} if the distortion of the $\mathrm{AuTe_2}$ structure is adjusted. The unit cell of the $\mathrm{AlB_2}$ structure and the $\mathrm{AuTe_2}$ structure are shown in Fig.~\ref{fig:unitcell-AlB2-and-AuTe2}(a) and Fig.~\ref{fig:unitcell-AlB2-and-AuTe2}(b), respectively. In the $\mathrm{AuTe_2}$ structure, the two large spheres in the unit cell are not in the same plane, so the highest packing fraction of the $\mathrm{AlB_2}$ structure cannot be larger than that of the $\mathrm{AuTe_2}$ structure. Furthermore, even if the unit cell of the $\mathrm{AlB_2}$ structure is expanded, the highest packing fraction of the $\mathrm{AuTe_2}$ structure is not always achievable. This is a representative example that unit cell places a limit on the maximum packing fraction that can be reached. Hence, several isomorphic structures consisting of different unit cells need to be optimized to determine the maximum packing fraction. This is one of the main reasons why it is difficult to determine the maximum packing fraction.

\subsection{Effectiveness of \textit{iterative balance method}}

As discussed in the previous section, it is difficult to determine the maximum packing fraction. However, not only almost all the packing fractions calculated by our method are the same as those shown in the previous study~\cite{PhysRevE.85.021130} by more than three decimal points, but also some of the packing fractions have successfully been updated. The results indicate that the iterative balance method can find the maximum packing fraction. Besides, the computational time for the structural optimization is very short as discussed in Sec.~\ref{sec:speed-of-structure-generation}. The effectiveness enables us to find the densest packing from the vast coordination space. Finally, as discussed in Sec.~\ref{sec:local_optimization}, the implementation is straightforward since it requires only simple calculations of forces and stress at each steepest decent step. Those validity indicates that the iterative balance method might also be useful for the optimization of higher-dimensional sphere packings.

\subsection{Effectiveness of \textit{piling up method}}

As discussed in Sec.~\ref{sec:Distribution_and_update_history_of_packing_fractions}, our method can create diverse packing structures, so our method can discover a wide variety of mono-phase densest packings from the vast coordination space, as discussed in Sec.~\ref{sec:Mono-phase_densest_sphere_packings}. The validity indicates that the piling up method is an effective way to generate initial structures for predicting densest packings.

\subsection{Public release of structural data and \textbf{\textit{SAMLAI}}}

Three-dimensional data of the DBSP and the \textbf{\textit{SAMLAI}} package, in which our methods are implemented, are available on the website~\cite{samlai}. The distribution of the program package and the source codes follow the practice of the GNU General Public License version 3 (GPLv3).

\subsection{Crystal structures and densest sphere packings}

\begin{table*}
\caption{Correspondence between crystals and densest sphere packings}
\begin{tabular}{cccc} \toprule
Densest sphere packing type$\,\,\,\,$ & Crystal structure type$\,\,\,\,$ & space group$\,\,\,\,$ & Material example \\ \midrule
$\mathrm{XY}$ & $\mathrm{NaCl}$ & $Fm\overline{3}m$ & $\mathrm{NaCl}$, $\mathrm{BaO}$, $\mathrm{CeN}$ \\
$\mathrm{XY_{2}}$, (2-1) & $\mathrm{KO_2}$ & $I4/mmm$ & $\mathrm{KO_2}$, $\mathrm{CaO_2}$, $\mathrm{BaO_2}$ \\
$\mathrm{XY_{10}}$, (10-1) & - & $Fm\overline{3}m$ & $\mathrm{LaH_{10}}$~\cite{doi:10.1002/anie.201709970}\\
(6-1) & - & $Im\overline{3}m$ & $\mathrm{YH_{6}}$~\cite{PhysRevLett.119.107001} \\
(6-6) & $\mathrm{NiAs}$ & $P6_3/mmc$ & $\mathrm{NiAs}$, $\mathrm{FeSe}$, $\mathrm{VP}$ \\
(16-4) & $\mathrm{UB_{4}}$ & $P 4/mbm$ & $\mathrm{UB_4}$, $\mathrm{YB_4}$, $\mathrm{LaB_4}$ \\
(4-2) & $\mathrm{ThSi_2}$ & $I4_1/amd$ & $\mathrm{ThSi_2}$, $\mathrm{BeGe_2}$, $\mathrm{SrGe_{2-\delta}}$~\cite{doi:10.1021/acs.inorgchem.7b01446} \\
(12-6), $\mathrm{AuTe_2}$, $\mathrm{AlB_2}$ & $\mathrm{AlB_2}$ & $P6/mmm$ & $\mathrm{AlB_2}$, $\mathrm{ThSi_2}$, $\mathrm{SrGa_2}$ \\
(2-2) & $\mathrm{CaSi}$\#$\mathrm{AlTh}$ & $Cmcm$ & $\mathrm{CaSi}$, $\mathrm{AlTh}$, $\mathrm{YSi}$ \\
$\mathrm{A_3 B}$ & $\mathrm{Cu_3 Ti}$ & $Pmmn$ & $\mathrm{Cu_3 Ti}$, $\mathrm{Ni_3 Nb}$, $\mathrm{Au_3 Sm}$ \\ \bottomrule
\end{tabular}
\label{table:Torquato-densest-packing-and-crystal}
\end{table*}

In some cases, atoms in crystals are approximated as spheres: In ion-bonded materials such as NaCl~\cite{villars}, atoms are often spherically symmetrical because atoms have closed-shell structures due to the charge transfer. In intermetallic compounds such as AgCu~\cite{10.1007/BF02652162}, atoms are also sometimes spherically symmetrical due to the bonds formed by electrons populating in \textit{s}-orbitals. In materials under high pressure, distances between atoms become so close that the directional orientation of the bond is weakened due to the strong repulsive force by Pauli's exclusion principle~\cite{doi:10.1021/jacs.9b02634}. Therefore, we can assume that many crystals can be understood as densest sphere packings~\cite{villars}. Then, we investigated the correspondence of DBSP with the crystal structures with reference to the space groups. The Spglib~\cite{togo2018textttspglib} is used for determining the space group of a densest sphere packing. The distortions of DBSP are corrected by the Spglib. As a result, we have succeeded in finding many crystals corresponding to DBSP, e.g., $\mathrm{LaH_{10}}$~\cite{doi:10.1002/anie.201709970} and $\mathrm{SrGe_{2-\delta}}$ \cite{doi:10.1021/acs.inorgchem.7b01446} synthesized under high pressure, $\mathrm{YH_6}$~\cite{PhysRevLett.119.107001} predicted theoretically under high pressure, $\mathrm{NaCl}$; their crystal structures correspond to the (4-2) structure~\cite{PhysRevE.85.021130}, the $\mathrm{XY_{10}}$ structure~\cite{PhysRevE.85.021130}, the (6-1) structure~\cite{PhysRevE.85.021130}, and the $\mathrm{XY}$ structure~\cite{PhysRevE.85.021130}, respectively. Besides, the crystal structure of $\mathrm{UB_{4}}$ corresponds to the (16-4) structure. All of the correspondence we find are shown in Table~\ref{table:Torquato-densest-packing-and-crystal}. These correspondences indicate that the densest sphere packings can be used effectively as structural prototypes for searching complex crystal structures, especially for high-pressure phases.

On the other hand, we could not find the correspondence between crystals and the 15 densest binary sphere packings: $\mathrm{XY_{4}}$, $\mathrm{XY_{8}}$, $\mathrm{XY_{11}}$, $\mathrm{XY_{12}}$, (8-2), (11-1), (12-1), (20-1), (22-1), (14-5), (8-4), $\mathrm{HgBr_{2}}$, (7-3), (10-4), and (9-4). However, we consider that these structure can be realized by crystals, e.g., the (20-1) structure and the (22-1) structure may be realized by hydrides. As discussed in Sec.~\ref{sec:Mono-phase_densest_sphere_packings}, the DBSP consist of a few local structures with high packing fractions. This is a different principle compared to crystals, but it is apparently true that the volume of a crystal structure is decreased as much as possible under high pressure, so many of the densest packings may correspond to crystals especially for high-pressure phases.

\section{Conclusions}
\label{sec:conclusions}

In the present research, we revisit the densest binary sphere packings under the periodic boundary conditions. To efficiently explore the densest sphere packings, firstly we invent the \textit{piling up method} to generate initial structures in an unbiased way; secondly, we develop the \textit{iterative balance method} to optimize the volume of a unit cell while keeping the overlap of hard spheres minimized. The piling up method is developed based on the idea of stacking spheres randomly one by one on top of a randomly generated first layer. It enables us to search the densest packings unbiasedly from the vast coordination space. The iterative balance method is developed based on the idea of repeating collision and repulsion among spheres under pressure while the maximum displacement in position vectors and lattice vectors is gradually decreased. The method not only generates a dense, periodic packing of nonoverlapping spheres but also predicts the maximum packing fraction with high accuracy. Those two methods are implemented in our open source program package \textit{\textbf{SAMLAI}} (Structure search Alchemy for MateriaL Artificial Invention).

With the \textbf{\textit{SAMLAI}}, we exhaustively search the DBSP with extending the unit cell compared to the previous study~\cite{PhysRevE.85.021130} and as a result we have discovered 12 putative DBSP, named $\mathrm{XY_{12}}$, (12-1), (22-1), (20-1), (8-2), (2-1), (14-5), (16-4), (8-4), (10-4), (9-4), and (12-6), shown in Figs.~\ref{fig:smallNewDensestPackings}, \ref{fig:mediumNewDensestPackings}, and \ref{fig:12-6}. Acccordingly, we have updated the phase diagram over the $x - \alpha$ plane, shown in Figs.~\ref{fig:smallPhaseDiagram}, \ref{fig:mediumPhaseDiagram}, and \ref{fig:largePhaseDiagram}. For the case of 12 or fewer spheres in the unit cell, our phase diagram is consistent with that of the previous study~\cite{PhysRevE.85.021130} with a small correction. Through the exhaustive search, diverse mono-phase densest binary sphere packings have been discovered and accordingly we have found that high packing fractions are achieved by introducing a distortion and/or combining a few local dense structural units. Three-dimensional data of DBSP and the \textbf{\textit{SAMLAI}} package are available on the website~\cite{samlai}.

If the $\mathrm{XY_n}$ structures are excluded, there are 21 putative DBSP. Their packing fractions are shown in Tables~\ref{table:small-filling-factors}, \ref{table:medium-filling-factors}, and \ref{table:large-filling-factors}. In some cases, it is difficult to determine the maximum packing fraction due to the large number of local minima corresponding to the large number of distortion patterns. However, comparing our results with those of the previous study~\cite{PhysRevE.85.021130}, it is found that in most cases the packing fractions are consistent by more than three decimal points. Furthermore, some of the packing fractions are 0.001 larger. The results indicate that our method can identify the maximum packing fraction.

We examined the distribution and the update history of highest packing fraction during the exhaustive search. As a result, we have confirmed that a diverse packing structures are created and our method can find the densest packings from diverse structures. 

Furthermore, we have investigated the correspondence of the DBSP with crystals based on the space group. The result shows that many structural units in real crystals, e.g., $\mathrm{LaH_{10}}$~\cite{doi:10.1002/anie.201709970} and $\mathrm{SrGe_{2-\delta}}$ \cite{doi:10.1021/acs.inorgchem.7b01446} synthesized under high pressure, can be understood as DBSP. The correspondence implies that the densest sphere packings can be used effectively as structural prototypes for searching complex crystal structures, especially for high-pressure phases.

\begin{acknowledgments}

All of the figures of packing structures are generated by VESTA~\cite{Momma:db5098}.

\end{acknowledgments}

\appendix

\section{Phase separation}
\label{sec:phase_separation}

As Hopkins and coworkers have shown, the highest packing fraction is achieved by phase separation~\cite{PhysRevE.85.021130}. In this appendix, we prove that the highest packing fractions for the binary system are achieved by phase separation into two or fewer densest packings.

In the proof, the small radius is fixed. Suppose that there are $n$ kinds of periodic dense packing structures, and they contain the fcc densest structure consisting of the large spheres and the fcc densest structure consisting of the small spheres. The composition ratio of small spheres in each structure $i$ is defined as $x_i$.

Generally, phase separation into several structures is necessary to realize a composition ratio $x$. Any composition ratio $x$ can be achieved because the $n$ candidates contain the fcc densest structure consisting of the large spheres or small spheres. To realize a certain composition $x$, three equations have to be satisfied as
\begin{align}
1 &= \sum_{i=1}^n y_i, \label{eq:condition1} \\
0 &= \sum_{i=1}^n a_i y_i, \label{eq:condition2}\\
0 & \le y_i, \label{eq:condition3}
\end{align}
where $y_i$ is defined as the composition ratio of each phase $i$. The $a_i$ is defined as
\begin{equation}
a_i \equiv \left(S_i + s_i \right) \left(x - x_i \right).
\end{equation}
where $S_i$ ($s_i$) is the number of large (small) spheres in the unit cell of the phase $i$. The packing fraction $\phi$ can be calculated as
\begin{equation}
\phi = \frac{\sum_{i=1}^n y_i v_i \phi_i}{\sum_{i=1}^n y_i v_i}, \label{eq:totalFillingFactor}
\end{equation}
where $v_i$ is defined as the volume of unit cell of phase $i$.

Now we have to find the group of $\{y_i \}$ that maximizes the packing fraction with the conditional formula (\ref{eq:condition1}), (\ref{eq:condition2}), and (\ref{eq:condition3}). First, any $\left(y_1, y_2, \cdots y_n \right)$ satisfying the conditional formula (\ref{eq:condition1}), (\ref{eq:condition2}), and (\ref{eq:condition3}), is a point on the hyperplane of ($n-2$) dimension. The hyperplane is specified by ($n-2$) linearly independent vectors and one point on the hyperplane. In other words, the hyperplane of ($n-2$) dimension is determined by the linear combination of the coordinates of ($n-1$) different points where no three points of them do not exist on the same line. In our case, a point on the hyperplane corresponds to a certain phase separation, therefore any phase separation can be represented as a linear combination of ($n-1$) kinds of phase separations, chosen to include all $n$-type structures.

As ($n-1$) different points, we can choose ($n-1$) kind of phase separations which consist of two or fewer densest packings. All of the $n$ candidates have to be chosen. For any $x$, there is at least one structure that satisfies $x_i < x$ and similarly there is at least one structure that satisfies $x < x_i$, so there are at least ($n-1$) kinds of phase separations whose composition ratio is $x$. The number is sufficient to specify the hyperplane. The total packing fraction $\phi$ of any phase separation are calculated as
\begin{equation}
\phi = \frac{\sum_{i=1}^n y_i \tilde{v_i} \tilde{\phi_i}}{\sum_{i=1}^n y_i \tilde{v_i}} = \tilde{\phi}_1 - \frac{\sum_{i=2}^n y_i \tilde{v_i} \left(\tilde{\phi_1}- \tilde{\phi_i} \right)}{\sum_{i=2}^n y_i \tilde{v_i}}, \label{eq:phaseSeparatedFillingRate}
\end{equation}
where $\tilde{\phi}_i$ is defined as the packing fraction of each phase $i$. The $\tilde{\phi}_1$ is chosen to be the highest packing fraction. Equation (\ref{eq:phaseSeparatedFillingRate}) shows that the supremum of the total packing fraction is $\tilde{\phi}_1$. The result shows the highest packing fraction can be achieved by a phase separation into two or fewer densest packings.

\section{Packing Fractions}

\begin{table*}
\caption{The packing fractions of the DBSP in the radius ratio of $0.200 \le \alpha \le 0.352$. The $\mathrm{XY_{n}}$-type structures are excluded. Compared to the HST phase diagram~\cite{PhysRevE.85.021130}, the larger packing fractions are shown in bold and the smaller ones are shown in italics.}
\begin{ruledtabular}
\begin{tabular}{ccccccccc}
$\alpha$ & (22-1) & (12-1) & (11-1) & (10-1) & (20-1) & (6-1) & (8-2) & (2-1) \\ \midrule
0.200 & 0.813313 & & & & & & & \\
0.203 & 0.809182 & 0.811932 & & & & & & \\
0.206 & 0.805783 & 0.807009 & & & & & & \\
0.209 & 0.801278 & 0.802364 & & & & & & \\
0.212 & 0.797876 & 0.799775 & & & & & & \\
0.214 & 0.796231 & 0.799418 & & & & & & \\
0.217 & 0.794616 & 0.796801 & 0.822630 & & & & & \\
0.220 & 0.792056 & 0.794140 & 0.817957 & & & & & \\
0.223 & 0.790229 & 0.791817 & 0.814383 & & & & & \\
0.225 & & & 0.811122 & \textit{0.824311} & & & & \\
0.228 & & & & 0.821822 & & & & \\
0.230 & & & & 0.820323 & & & & \\
0.233 & & & & 0.817962 & & & & \\
0.236 & & & & 0.815926 & & & & \\
0.239 & & & & 0.812289 & & & & \\
0.242 & & & & 0.807668 & 0.782166 & & & \\
0.245 & & & & \textit{0.803451} & 0.781966 & & & \\
0.247 & & & & \textbf{0.800974} & 0.782001 & & & \\
0.250 & & & & \textbf{0.797510} & 0.782303 & & & \\
0.253 & & & & 0.794335 & 0.782890 & & & \\
0.256 & & & & 0.791435 & 0.783738 & & & \\
0.258 & & & & \textbf{0.789649} & 0.784441 & & & \\
0.261 & & & & 0.787181 & 0.785285 & & 0.786449 & \\
0.264 & & & & 0.784959 & 0.783329 & \textit{0.780418} & 0.780930 & \\
0.267 & & & & 0.782973 & 0.778084 & 0.781892 & 0.775655 & \\
0.270 & & & & 0.781213 & 0.773025 & 0.783601 & 0.770617 & \\
0.273 & & & & & 0.768679 & \textbf{0.785533} & 0.768293 & \\
0.275 & & & & & 0.766132 & 0.786941 & 0.767796 & \\
0.278 & & & & & 0.762136 & 0.789225 & 0.767489 & 0.767544 \\
0.281 & & & & & & 0.791710 & & 0.764182 \\
0.284 & & & & & & 0.794388 & & 0.761459 \\
0.287 & & & & & & 0.797252 & & 0.759235 \\
0.289 & & & & & & \textit{0.799263} & & 0.757982 \\
0.292 & & & & & & 0.800869 & & 0.756394 \\
0.295 & & & & & & 0.799598 & & 0.755110 \\
0.298 & & & & & & 0.798549 & & 0.754086 \\
0.301 & & & & & & 0.797714 & & \\
0.304 & & & & & & 0.797085 & & \\
0.307 & & & & & & 0.796654 & & \\
0.309 & & & & & & 0.796068 & & \\
0.312 & & & & & & 0.794900 & & \\
0.315 & & & & & & 0.793889 & & \\
0.318 & & & & & & 0.793034 & & \\
0.321 & & & & & & 0.791582 & & \\
0.324 & & & & & & 0.788239 & & \\
0.326 & & & & & & 0.786121 & & \\
0.329 & & & & & & 0.783107 & & \\
0.332 & & & & & & 0.780285 & & \\
0.335 & & & & & & 0.777650 & & \\
0.338 & & & & & & 0.775199 & & \\
0.341 & & & & & & 0.772928 & & \\
0.343 & & & & & & \textbf{0.771512} & & \\
0.346 & & & & & & \textbf{0.769533} & & \\
0.349 & & & & & & 0.767724 & & \\
0.352 & & & & & & 0.766084 & & \\
\end{tabular}
\label{table:small-filling-factors}
\end{ruledtabular}
\end{table*}
\begin{table*}
\caption{The packing fractions of the DBSP in the radius ratio of $0.414 \le \alpha \le 0.500$. Compared to the HST phase diagram~\cite{PhysRevE.85.021130}, the larger packing fractions are shown in bold and the smaller ones are shown in italics while the packing fractions of the (7-3) structure are shown nomally.}
\begin{ruledtabular}
\begin{tabular}{ccccccccccc}
$\alpha$ & (16-4) & (14-5) & (10-4) & (9-4) & (8-4) & $\mathrm{HgBr_2}$ & (7-3) & $\mathrm{AuTe_2}$ & (6-6) & (2-2) \\ \midrule
0.414 & & & & & & & & & 0.793023 & \\
0.417 & & & & & & & & & \textbf{0.789534} & \\
0.420 & & & & & & & & & 0.785872 & \\
0.423 & & & & & & & & & \textit{0.782347} & \\
0.426 & & & & & & & & & 0.778968 & \\
0.428 & & & & & & & & & \textbf{0.776794} & \\
0.431 & & 0.755399 & & & 0.760930 & 0.760315 & & & \textbf{0.773646} & \\
0.434 & & 0.758308 & & & 0.760710 & 0.759992 & & & 0.770629 & \\
0.437 & & 0.761257 & & & 0.760455 & 0.759726 & & & 0.767739 & \\
0.440 & & 0.764247 & & & 0.760144 & 0.759518 & & & 0.764971 & \\
0.443 & 0.757311 & 0.767278 & & & 0.759747 & 0.759365 & & & \textit{0.762320} & \\
0.445 & 0.759701 & 0.765259 & & & 0.759348 & 0.759294 & & & \textbf{0.760616} & \\
0.448 & 0.760568 & 0.762049 & & & 0.759041 & 0.759035 & & & 0.758151 & \\
0.451 & 0.760101 & 0.758952 & & & 0.758830 & 0.758830 & & & 0.755793 & \\
0.454 & 0.759731 & 0.755967 & & & 0.758762 & 0.758762 & & & 0.753537 & \\
0.457 & 0.759456 & 0.753092 & & & 0.758825 & 0.758825 & & & 0.751380 & \\
0.460 & 0.759276 & 0.750298 & & & 0.759011 & 0.759011 & & & 0.749319 & \\
0.463 & 0.754469 & & & & 0.757833 & 0.757814 & 0.750988 & & & \\
0.465 & 0.750732 & & & & 0.755714 & 0.755701 & 0.751180 & & & \\
0.468 & 0.745076 & & & & 0.752668 & 0.752668 & 0.751510 & & & \\
0.471 & & & 0.746408 & 0.743894 & 0.749796 & 0.749796 & 0.750015 & & & 0.741836 \\
0.474 & & & 0.746249 & 0.744177 & 0.747076 & 0.747076 & 0.747831 & 0.740802 & & 0.742356 \\
0.477 & & & 0.746185 & 0.744854 & 0.744506 & 0.744506 & 0.745776 & 0.742471 & & 0.742966 \\
0.480 & & & 0.746216 & 0.745757 & & & 0.743866 & 0.744233 & & 0.743661 \\
0.481 & & & 0.746246 & 0.746093 & & & 0.743257 & 0.744840 & & 0.743911 \\
0.482 & & & 0.746287 & 0.746452 & & & 0.742661 & 0.745457 & & 0.744170 \\
0.483 & & & 0.746337 & 0.746510 & & & 0.742080 & 0.746084 & & 0.744437 \\
0.485 & & & 0.746468 & 0.746117 & & & 0.740956 & \textit{0.747366} & & 0.744997 \\
0.488 & & & 0.746727 & 0.745626 & & & & \textit{0.749358} & & 0.745898 \\
0.491 & & & & & & & & 0.751431 & & 0.746869 \\
0.494 & & & & & & & & 0.753583 & & 0.747907 \\
0.497 & & & & & & & & 0.755810 & & 0.749010 \\
0.500 & & & & & & & & 0.758114 & & 0.750174 \\
\end{tabular}
\label{table:medium-filling-factors}
\end{ruledtabular}
\end{table*}
\begin{table*}
\caption{The packing fractions of the DBSP in the radius ratio of $0.537 \le \alpha \le 0.64$. Compared to the HST phase diagram~\cite{PhysRevE.85.021130}, the larger packing fractions are shown in bold and the smaller ones are shown in italics.}
\begin{ruledtabular}
\begin{tabular}{ccccccccccc}
$\alpha$ & $\mathrm{AuTe_2}$ & $\mathrm{AlB_2}$ & (12-6) & $\mathrm{A_3 B}$ \\ \midrule
0.537 & 0.780598 & 0.780598 & 0.780743 & \\
0.540 & 0.780217 & 0.780217 & 0.780466 & \\
0.543 & 0.779883 & 0.779883 & 0.780262 & \\
0.546 & 0.779595 & 0.779595 & 0.780130 & \\
0.549 & 0.779352 & 0.779352 & 0.780067 & \\
0.552 & & 0.779154 & 0.779719 & \\
0.554 & & 0.779046 & 0.779522 & \\
0.557 & & 0.778922 & 0.779280 & \\
0.560 & & 0.778841 & 0.779098 & \\
0.563 & & 0.778804 & 0.778977 & \\
0.566 & & 0.778808 & 0.778916 & \\
0.569 & & 0.778856 & 0.778913 & \\
0.572 & & 0.778944 & 0.778968 & \\
0.574 & & 0.779027 & 0.779036 & \\
0.577 & & 0.779184 & 0.779184 & \\
0.580 & & 0.776382 & 0.776382 & \\
$\cdots$ & & & & \\
0.612 & & 0.748108 & & 0.743387 \\
0.614 & & 0.746679 & & 0.743268 \\
0.617 & & 0.744608 & & 0.743174 \\
0.620 & & 0.742622 & & 0.743180 \\
0.623 & & 0.740720 & & 0.743285 \\
$\cdots$ & & & & \\
0.640 & & & & 0.745690 \\
\end{tabular}
\label{table:large-filling-factors}
\end{ruledtabular}
\end{table*}

If the $\mathrm{XY_n}$ structures are excluded, there are 21 putative DBSP. Their packing fractions are shown in Tables~\ref{table:small-filling-factors}, \ref{table:medium-filling-factors}, and \ref{table:large-filling-factors}. The tables include all the radius ratio shown in Table I in Ref~\cite{PhysRevE.85.021130}. Comparing our results with those of the previous study~\cite{PhysRevE.85.021130}, it is found that in most cases the packing fractions are consistent by more than three decimal points. Some of the packing fractions are 0.001 larger; they are shown in bold. On the other hand, a few packing fractions are 0.001 smaller; they are shown in italics.

The packing fractions shown in the tables are determined by re-optimization. When the number of local minima is small, almost all the re-optimized structures converge to the densest packing. On the other hand, when the number of local minima is large, we obtain many packing fractions corresponding to diverse distortion patterns in the re-optimization process. This is because many structures are trapped at local minima. In that case, the large number of re-optimization is necessary to determine the highest packing fraction. In some cases, tens of thousand re-optimizations are necessary. 

We consider that the packing fractions of the (7-3) structure shown in the previous study~\cite{PhysRevE.85.021130} are misprints. According to the HST phase diagram, at $\alpha = 0.480$ the (7-3) structure has the same packing fraction as the phase separation into the (5-2) structure and the (4-2) structure. Therefore, the packing fraction of the (7-3) structure is less than that of (5-2) structure or (4-2) structure at $\alpha = 0.480$. However, in Table I in Ref~\cite{PhysRevE.85.021130}, the packing fraction of the (7-3) structure is obviously higher than that of (5-2) structure or (4-2) structure. The contradiction indicates that the packing fraction of the (7-3) structure shown in Table I in Ref \cite{PhysRevE.85.021130} is a misprint. Therefore, Table \ref{table:medium-filling-factors} shows the packing fractions of the (7-3) structure nomally.

\section{Derivatives of the enthalpy}
\label{sec:force-and-stress}

In our method, lattice vectors and the position vectors are simultaneously optimized by using the steepest descent method. To apply the steepest descent method, we need to calculate the derivatives of the enthalpy. In this section, we explain how to derive the derivatives analytically.

At first, we define the matrix $A$ as
\begin{equation}
A \equiv\begin{pmatrix}
a_{11} & a_{21} & a_{31} \\
a_{12} & a_{22} & a_{32} \\
a_{13} & a_{23} & a_{33}
\end{pmatrix},
\end{equation}
where $\bm{a}_1$, $\bm{a}_2$, and $\bm{a}_3$ are the lattice vectors. The $x$, $y$, and $z$ components of the vector $\bm{a}_i$ is written as $a_{i1}$, $a_{i2}$, and $a_{i3}$, respectively. The matrix $C$ is defined as
\begin{equation}
C \equiv \begin{pmatrix}
\bm{a}_1 \cdot \bm{a}_1 & \bm{a}_2 \cdot \bm{a}_1 & \bm{a}_3 \cdot \bm{a}_1 \\
\bm{a}_1 \cdot \bm{a}_2 & \bm{a}_2 \cdot \bm{a}_2 & \bm{a}_3 \cdot \bm{a}_2 \\
\bm{a}_1 \cdot \bm{a}_3 & \bm{a}_2 \cdot \bm{a}_3 & \bm{a}_3 \cdot \bm{a}_3
\end{pmatrix}.
\end{equation}
Hereafter, the position of sphere $i$ is represented by $\bm{r}_i$ and the fractional coordinates are represented by $\bm{q}_ {i} = \left(q _{i 1}, q _{i 2}, q _{i 3} \right)$. The relative fractional coordinate $\bm{q}_{i j} ^{\left(\bm{Q}\right)}$ is defined as
\begin{equation}
\bm{q}_{i j} ^{\left(\bm{Q}\right)} \equiv \bm{q} _{j} + \bm{Q} - \bm{q}_{i},
\end{equation}
where $\bm{Q}$ is defined as interger vector. The relative vector $\bm{r}_{i j} ^{\left(\bm{Q}\right)}$ is defined as
\begin{equation}
\bm{r}_{i j} ^{\left(\bm{Q}\right)} \equiv \bm{r} _{j} + \bm{T} - \bm{r}_{i} = A \bm{q}_{i j} ^{\left(\bm{Q}\right)}, \label{eq:relative-vector}
\end{equation}
where $\bm{T}$ are translational vectors that satisfy the equation as $\bm{T} = A \bm{Q}$. Taking the absolute value of Eq. (\ref{eq:relative-vector}) leads to
\begin{equation}
r_{i j} ^{\left(\bm{Q}\right)} \equiv \left| \bm{r} _{j} + \bm{T} - \bm{r}_{i} \right| = \sqrt{{^{t} \bm{q}_{i j} ^{\left(\bm{Q}\right)}} C \bm{q}_{i j} ^{\left(\bm{Q}\right)}}.
\end{equation}
Finally, we define $z _{i j} ^{\left(\bm{Q}\right)}$ as
\begin{equation}
z _{i j} ^{\left(\bm{Q}\right)} \equiv r_{i j} ^{\left(\bm{Q}\right)} - \left(c _i + c _j \right),
\end{equation}
where $c_i$ is defined as the radius of the sphere $i$.

The potential working between spheres is defined as Eq.~(\ref{eq:potential}) and the total energy per unit cell $E$ is given by Eq.~(\ref{eq:energy}). The enthalpy per unitcell $H$ is given by
\begin{equation}
H = \frac{1}{2} \sum _{\bm{Q}} \sum _{i = 1}^{N} \sum _{j = 1}^{N} U \left(z _{i j} ^{\left(\bm{Q}\right)} \right) + PV. \label{eq:enthalpy}
\end{equation}
The volume $V$ can be calculated as
\begin{equation}
V = \left|\mathrm{det} A \right| = \left|\bm{a}_1 \cdot \left(\bm{a}_2 \times \bm{a}_3 \right) \right|.
\end{equation}
If we take the lattice vectors to a right-handed system, $V$ can be calculated as $V = \mathrm{det} A$. We set the initial lattice vector to the right-handed system.

The derivative of the enthalpy with respect to $q_{ik}$ can be calculated as
\begin{equation}
\frac{\partial E}{\partial q _{ik}} = \sum _{\bm{Q}} \sum _{j = 1}^{N} \frac{\partial r _{i j} ^{\left(\bm{Q}\right)}}{\partial q _{ik}} \cdot f \left(z _{i j} ^{\left(\bm{Q}\right)} \right), \label{eq:dE-dq-kIi}
\end{equation}
where $f \left(z _{i j} ^{\left(\bm{Q}\right)} \right)$ is defined as
\begin{equation}
f \left(z _{i j} ^{\left(\bm{Q}\right)} \right) \equiv \frac{\partial U \left(z _{i j} ^{\left(\bm{Q}\right)} \right)}{\partial z_{i j} ^{\left(\bm{Q}\right)}} = \begin{cases}
-1 & \text{$z _{i j} ^{\left(\bm{T}\right)} \le 0$} \\
0 & \text{$0 < z _{i j} ^{\left(\bm{T}\right)}$} \end{cases}.
\end{equation}
The derivative of $r _{i j} ^{\left(\bm{Q}\right)}$ with respect to $q_{ik}$ can be calculated as
\begin{equation}
\frac{\partial r_{i j} ^{\left(\bm{Q}\right)}}{\partial q _{ik}} = -\frac{1}{r_{i j} ^{\left(\bm{Q}\right)}} \sum_{t=1} ^{3} c_{kt} \, q_{i j,t} ^{\left(\bm{Q}\right)}.
\end{equation}

The derivative of the enthalpy with respect to $a_{mn}$ can be calculated as
\begin{equation}
\frac{\partial H}{\partial a _{mn}} = \frac{\partial E}{\partial a _{mn}} + P \frac{\partial V}{\partial a _{mn}}. \label{eq:dH_damn}
\end{equation}
The first term of Eq. (\ref{eq:dH_damn}) can be calculated as
\begin{equation}
\frac{\partial E}{\partial a_{mn}} = \frac{1}{2} \sum _{\bm{Q}} \sum _{i = 1}^{N} \sum _{j = 1}^{N} \frac{\partial r _{i j} ^{\left(\bm{Q}\right)}}{\partial a_{mn}} \cdot f \left(z _{i j} ^{\left(\bm{Q}\right)} \right),
\end{equation}
where the derivative of $r _{i j} ^{\left(\bm{Q}\right)}$ with respect to $a_{mn}$ can be calculated as
\begin{equation}
\frac{\partial r_{i j} ^{\left(\bm{Q}\right)}}{\partial a_{mn}} = \frac{1}{r_{i j} ^{\left(\bm{Q}\right)}} \sum_{s=1} ^{3} a_{sn} \, q_{i j,s} ^{\left(\bm{Q}\right)} \, q_{i j,m} ^{\left(\bm{Q}\right)}.
\end{equation}
The second term of Eq.~(\ref{eq:dH_damn}) can be calculated as
\begin{equation}
P \, \frac{\partial V}{\partial a_{mn}} = \frac{1}{2}\, P \, \sum_{i,j} \sum_{k,l} \epsilon_{mij} \, \epsilon_{nkl} \, a_{ik} \, a_{jl},
\end{equation}
where $\epsilon_{ijk}$ is the Levi-Civita symbol.

Sometimes the lattice vectors change from a right-handed system to a left-handed system due to the spheres slipping through each other. The slipping is a fatal accident for structural optimization. The change causes a negative value of $\Omega = \mathrm{det} A$, so we can monitor the appropriate structural optimization by confirming that the sign of $V$ is positive.

\bibliographystyle{apsrev4-2}
\bibliography{draft4}

\end{document}